\documentclass[acmsmall,screen,table]{acmart}
\usepackage{tabularx}  
\usepackage{xcolor}
\usepackage{graphicx}
\usepackage{subfigure}
\usepackage{subcaption}
\usepackage{multirow}
\usepackage{multicol}
\usepackage{rotating}
\usepackage{lscape}
\usepackage{placeins}
\usepackage[most]{tcolorbox}

\newtheorem{definition}{\textbf{Definition}}[section]
\newtheorem{example}{\textbf{Example}}

\usepackage{tcolorbox}

\newcommand{\finding}[2]{%
\begin{tcolorbox}[
  breakable,
  enhanced,
  width=\linewidth,
  boxrule=0.7pt,          
  colframe=black, 
  colback=blue!5,         
  top=0.5pt,bottom=0.5pt,left=0.5pt,right=0.5pt,
  before skip=2.3pt, after skip=2.3pt,
  pad at break*=1mm,
]
\textbf{\textit{Summary of Answer to RQ#1:}} #2
\end{tcolorbox}%
}

\usepackage[pagewise]{lineno}

\usepackage{xcolor}

\newcommand{\textbg}[2]{%
  \begingroup
  \setlength{\fboxsep}{1pt}%
  \smash{\colorbox{#1}{\rule[-0.25ex]{0pt}{2.2ex}#2}}%
  \endgroup
}

\newcommand{\greenbg}[1]{\textbg{green!30}{#1}}
\newcommand{\yellowbg}[1]{\textbg{yellow!50}{#1}}
\newcommand{\redbg}[1]{\textbg{magenta!20}{#1}}
\newcommand{\purplebg}[1]{\textbg{blue!20}{#1}}

\newcounter{ChenhuiCommentCounter}
\newcommand{\ddu}[1]{
    \stepcounter{ChenhuiCommentCounter}
    \textcolor{blue}{\textit{/**Chenhui's comment [\arabic{ChenhuiCommentCounter}]: I don't understand the intended meaning in the next sentence. Please revise/delete/explain. **/}}
}

\newcommand{\dns}[1]{
    \stepcounter{ChenhuiCommentCounter}
    \textcolor{blue}{\textit{/**Chenhui's comment [\arabic{ChenhuiCommentCounter}]: I'm not sure that I have captured the intended meaning in the next sentence. Please check/confirm. **/}}
}

\newcounter{LisaCommentCounter}
\AtBeginDocument{%
  }

\setcopyright{acmlicensed}
\copyrightyear{2018}
\acmYear{2018}
\acmDOI{XXXXXXX.XXXXXXX}

\acmJournal{JACM}
\acmVolume{37}
\acmNumber{4}
\acmArticle{111}
\acmMonth{8}

\begin{document}

\title{An Extensive Empirical Study on Evaluation Metrics for Combinatorial Interaction Testing}

\author{Lisha Qin}
\email{2250002450@student.must.edu.mo}
\orcid{0009-0008-4594-2890}
\affiliation{
  \institution{School of Computer Science and Engineering, Macau University of Science and Technology}
  \city{Macao SAR}
  \postcode{999078}
  \country{China}  
  }

\author{Chenhui Cui}
\orcid{0009-0004-8746-316X}
\email{3230002105@student.must.edu.mo}
\affiliation{
  \institution{School of Computer Science and Engineering, Macau University of Science and Technology}
  \city{Macao SAR}
  \country{China}
  \postcode{999078}
  }

\author{Tao Li}
\email{3220007015@student.must.edu.mo}
\orcid{0009-0001-7413-9692}
\affiliation{
  \institution{School of Computer Science and Engineering, Macau University of Science and Technology}
  \city{Macao SAR}
  \country{China}
  \postcode{999078}
  }

\author{Rubing Huang}
\orcid{0000-0002-1769-6126}
\email{rbhuang@must.edu.mo}
\affiliation{
  \institution{School of Computer Science and Engineering, Macau University of Science and Technology}
  \city{Macao SAR}
  \postcode{999078}
  \country{China}
  }
  \affiliation{
  \institution{Macau University of Science and Technology Zhuhai MUST Science and Technology Research Institute}
  \city{Zhuhai}
  \state{Guangdong}
  \postcode{519099}
  \country{China}
}


\author{Shikai Guo}
\email{shikai.guo@dlmu.edu.cn}
\orcid{0000-0002-8554-6365}
\affiliation{
  \institution{School of Information Science and Technology, Dalian Maritime University}
  \city{Dalian}
  \state{Liaoning}
  \postcode{116026}
   \country{China}
}

\author{Lei Ma}
\email{ma.lei@acm.org}
\orcid{0000-0002-8621-2420}
\affiliation{
  \institution{School of Science, University of Tokyo}
  \city{Tokyo}
  \postcode{113-0033}
  \country{Japan}
}
\affiliation{
  \institution{Department of Electrical and Computer Engineering, University of Alberta}
  \city{Edmonton}
  \postcode{T6G 2R3}
  \country{Canada}
}


\begin{abstract}
Combinatorial interaction testing (CIT) is a black-box testing method that has received extensive attention in both research and practice over recent years. Its primary objective is to construct an effective combinatorial test suite that detects software failures caused by parameter interactions. As a fundamental component of the CIT testing process, the evaluation metric plays a critical role in assessing and comparing combinatorial test suites, as well as in evaluating various test generation techniques. For CIT practitioners, selecting an appropriate evaluation metric is both important and challenging, given the wide variety of available options. Nevertheless, no prior work has systematically addressed this problem. To fill this gap, this paper first provides a comprehensive survey of black-box evaluation metrics for combinatorial test suites, offering rigorous definitions, clear classifications, illustrative examples, and complexity analyses. We then conduct an extensive empirical study involving eight open-source projects, encompassing 32 test scenarios and 295,624 combinatorial test suites. In this study, we examine the correlation between each static evaluation metric and fault-detection effectiveness using two correlation measures. Experimental results show that the Value Combination Coverage (VCC) metric serves as a valid predictor for test-suite evaluation. However, distribution-based metrics generally incur lower computational costs than interaction coverage-based ones. The choice of an appropriate metric should also account for the test suite's inherent properties, as these characteristics can substantially influence the effectiveness of the evaluation. Finally, we provide practical guidelines to assist CIT practitioners in selecting suitable evaluation metrics for assessing or comparing combinatorial test suites.

\end{abstract}



 \received{24 January 2026}

\maketitle

\section{Introduction}
As the software under test (SUT) grows increasingly configurable, it becomes affected by a wide range of parameters (or factors), including user inputs, configuration options, and external events. To ensure system quality, it would be ideal to test every possible combination of parameter values—that is, to conduct exhaustive testing. In practice, however, this approach is often infeasible due to the considerable resources and effort it demands.

Combinatorial Interaction Testing (CIT), also known as combinatorial testing (CT)~\cite{Nie2011}, is a well-established black-box testing method designed to generate effective test suites for revealing software failures caused by parameter interactions. CIT has been widely applied in practice, especially in highly configurable systems~\cite{Kuhn2004, Yilmaz2014, patil2026combinatorial} and software product lines~\cite{Oster2010, Henard2014, chen2026act, balouchi2025exploratory, 10962496}. Recently, CIT has also been adopted for systematic and comprehensive assessments of Large Language Models (LLMs)~\cite{chandrasekaran2025evaluating, perko5782210evaluating} and Machine Learning models~\cite{10962484}.

CIT achieves a trade-off between testing effectiveness and efficiency by requiring coverage of only selected combinations of parameter values, rather than all possible ones. For example, 2-wise CIT (or pairwise testing) examines every possible 2-wise parameter-value tuple, but not higher-order tuples. Empirical evidence supports the effectiveness of this approach. Kuhn et al.~\cite{Kuhn2002, Kuhn2004, Kuhn2006} analyzed fault reports from several software projects and found that over 70\% of faults could be detected through 2-wise testing, and approximately 90\% through 3-wise testing. In contrast, a similar study~\cite{Bell2005} reported that nearly all faults were identifiable with 6-wise testing.

Researchers have continually explored ways to better leverage CIT's potential for enhancing test quality. For instance, Lin et al.~\cite{lin2025t} proposed an advanced evolutionary heuristic algorithm to guide test case generation, while Muazu et al.~\cite{muazu2025enhancing} introduced an improved version of the Harmony Search algorithm for generating CIT test suites. Furthermore, investigations into the effectiveness of CIT in test automation~\cite{daoud2025effectiveness} provide valuable insights into its practical implications in real-world software products.

\begin{figure}[!b]
\graphicspath{{Figures/}}
\centering
\includegraphics[width=3.5in]{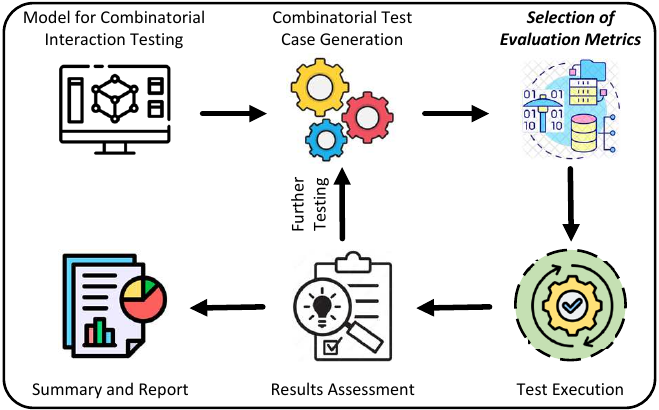}
\caption{The general testing procedure for CIT.}
\label{Figure1}
\end{figure}

As an indispensable part of the CIT testing procedure~\cite{Nie2011}, the evaluation metric plays a critical role in assessing CIT. Figure~\ref{Figure1} illustrates the basic CIT testing procedure, in which the stage of ``\textit{Selection of Evaluation Metrics}'' determines how the effectiveness or efficiency of CIT is judged, and this stage helps testers or users select a test suite from a set of candidates based on the chosen metric. According to Nie's classification~\cite{Nie2011}, evaluation in CIT falls into two categories: (1) measuring and evaluating CIT itself, and (2) measuring and evaluating the quality of the SUT after CIT. This paper focuses on the former category—specifically, we evaluate different CIT techniques by measuring their respective combinatorial test suites. Intuitively, different types of software testing may adopt different evaluation metrics. For instance, in white-box testing, common metrics for evaluating test suites include statement, branch, and method coverage, and these metrics can also be applied to combinatorial test suites when the SUT's source code is accessible. However, since CIT is a black-box testing method where inputs are specified and expected outputs are derived from a specification, obtaining the program source code is often difficult; the SUT is therefore treated as a simple processor that accepts inputs and produces outputs, with no prior knowledge of its internal workings expected. Consequently, this paper concentrates on evaluation metrics that do not require source code knowledge—namely, black-box evaluation metrics.

As is well known, various evaluation metrics have been developed for combinatorial test suites; however, it remains a critical issue how to select metrics when assessing or selecting candidate test suites. Unfortunately, no previous study has investigated this challenge. Motivated by this fact, this paper presents an empirical investigation of evaluation metrics for CIT. More specifically, we first survey common metrics for measuring combinatorial test suites, including categories, descriptions, examples, and complexity analyses. We then conduct an extensive empirical study across 8 open-source projects, using various test scenarios and a large-scale collection of combinatorial test suites to evaluate each evaluation metric. Finally, we provide practical guidelines for testers when selecting metrics to evaluate combinatorial test suites.

The main contributions of this study are listed as follows:
\begin{itemize}
\item We surveyed more than 10 well-known black-box evaluation metrics for combinatorial test suites and analyzed their space and time complexities.
\item We conducted an extensive empirical study on nine open-source projects, involving 32 test scenarios and 295,624 test suites.
\item We investigated the correlation between each static evaluation metric and fault-detection effectiveness using two correlation measures.
\item We provided some practical guidelines to help CIT testers select evaluation metrics.
\end{itemize}

The remainder of this paper is organized as follows:
Section \ref{SECTION:Background} briefly introduces some background on CIT.
Section \ref{SECTION:Metrics} surveys and summarizes several evaluation metrics for combinatorial test suites, including their definitions, examples, and complexity analyses.
Section \ref{SECTION:Design} describes the experimental methodology.
Section \ref{SECTION:Results} presents and analyzes the experimental results, and discusses some threats to validity.
Section \ref{SECTION:Guidelines} provides practical guidelines for testers.
Section \ref{SECTION:Conclusions} concludes the paper and outlines potential directions for future work.

\section{Background}
\label{SECTION:Background}
In this section, we provide background information on the input parameter model, the combinatorial test suite, and the test sequence.

\subsection{Input Parameter Model}

Assume that the software under test (SUT) is influenced by $k$ parameters $p_1, p_2, \cdots, p_k$ that form the set $\mathcal{P}$, and each parameter $p_i$ has a finite set of distinct values (or levels) $\mathcal{V}_i$, where $i=1,2,\cdots,k$. In practice, these parameters may correspond to configuration options, components, or user inputs of the SUT.

\begin{definition}[\textbf{$\lambda$-wise Parameter Combination}]
A $\lambda$-wise parameter combination is a $\lambda$-tuple, denoted as $\phi_\lambda=(p_{i_1},p_{i_2},\cdots,p_{i_\lambda})$, where $1\leq \lambda \leq k$, $p_{i_j} \in \mathcal{P}~(j=1,2,\cdots,\lambda)$, and $1\leq i_1 \leq i_2 \leq \cdots \leq i_\lambda \leq k$.
\end{definition}

For ease of discussion, we use $\Phi_\lambda$ to denote the set of all possible $\lambda$-wise parameter combinations derived from an input parameter model. Intuitively, the total number of $\lambda$-wise parameter combinations for $k$ parameters is $$|\Phi_\lambda|=\mathrm{C}(k,\lambda).$$

\begin{definition}[\textbf{$\lambda$-wise Value Combination}]
A $\lambda$-wise value combination is a $\lambda$-tuple, denoted as $\sigma_\lambda=(v_{i_1},v_{i_2},\cdots,v_{i_\lambda})$, which is derived from a $\lambda$-wise parameter combination $\phi_\lambda=(p_{i_1}, p_{i_2}, \cdots, p_{i_\lambda})$, where $1 \leq \lambda \leq k$, $1 \leq i_1 \leq i_2 \leq \cdots \leq i_\lambda \leq k$, $p_{i_j}\in \mathcal{P}$, and $v_{i_j} \in \mathcal{V}_{i_j}~(j=1,2,\cdots,\lambda)$—that is, each parameter $p_{i_j}$ is assigned the value $v_{i_j}$.
\end{definition}

Intuitively, a $\gamma$-wise value combination covers $\mathrm{C}(\gamma, \lambda)$ $\lambda$-wise value combinations, where $\lambda \leq \gamma \leq k$. For ease of description, we define a function $\varphi(\lambda,\sigma_{\gamma})$ that returns the set of all $\lambda$-wise value combinations covered by a $\gamma$-wise ($\gamma \geq \lambda$) value combination. In addition, when $\lambda = k$, a $\lambda$-wise value combination becomes a combinatorial test case.

\begin{definition}[\textbf{Combinatorial Test Case}]
A combinatorial test case $t$ is a $k$-wise value combination $\sigma_k$, i.e., a specific assignment of a value to each parameter.
\end{definition}

Due to certain constraints on parameter values in the SUT, each $\lambda$-wise value combination may be either \textit{valid} (i.e., it satisfies the constraints) or \textit{invalid} (i.e., it violates the constraints). The definition of constraints is given in \cite{Zhang2014}.
\begin{definition}[\textbf{Constraint}]
A constraint $c(\sigma_j)$ is a predicate on the $j$-wise value combination $\sigma_j$ for the corresponding parameter combination $\phi_j$, where $1 \leq j \leq k$. A combinatorial test case is valid if and only if the assignment of those parameters in $\phi_j$ makes the predicate true.
\end{definition}

For the SUT, the parameters, their corresponding values, and the constraints together constitute the input parameter model, which is formally defined as follows.
\begin{definition} [\textbf{Input Parameter Model}]
An input parameter model (IPM) for the SUT can be represented as a $3$-tuple $IPM(\mathcal{P},\mathcal{V},\mathcal{R})$, where $\mathcal{P}=\{p_1,p_2,\cdots,p_k\}$ is a set of $k$ parameters, $\mathcal{V}=\{\mathcal{V}_1,\mathcal{V}_2,\cdots,\mathcal{V}_k\}$ denotes the set of possible parameter-value sets, with each $\mathcal{V}_i$ being a finite set of values for parameter $p_i~(i=1,2,\cdots,k)$, and $\mathcal{R}$ is a set of constraints among parameter values in the SUT (if any).
\end{definition}

\begin{example}[\textbf{IPM Example}]
   Consider an example of a component-based system, as shown in Table~\ref{Table1}, with some configurations provided. The IPM for this system can be expressed as $IPM(\mathcal{P},\mathcal{V},\mathcal{R})$, where each element is described as follows: $\mathcal{P}=\{p_1,p_2,p_3,p_4\}$, $\mathcal{V}=\{\mathcal{V}_1,\mathcal{V}_2,\mathcal{V}_3,\mathcal{V}_4\}$ with $\mathcal{V}_1=\{\textrm{\texttt{Windows}}, \textrm{\texttt{Linux}}, \textrm{\texttt{Solaris}}\}$, $\mathcal{V}_2=\{\textrm{\texttt{IE}}, \textrm{\texttt{Google}}, \textrm{\texttt{Firefox}}\}$, $\mathcal{V}_3=\{\textrm{\texttt{LAN}}, \textrm{\texttt{VPN}}\}$, and $\mathcal{V}_4=\{\textrm{\texttt{DB/2}}, \textrm{\texttt{Oracle}}\}$, and $\mathcal{R}=\{r_1\}$ (since VPN can only be deployed on IE under Linux).
\end{example}

Since the specific values of each parameter do not affect the IPM, we can, without loss of generality, assign increasing index numbers to the parameter values (see the index numbers for each value in Table~\ref{Table1}). In addition, we can assume that $|\mathcal{V}_1| \geq |\mathcal{V}_2| \geq \cdots \geq |\mathcal{V}_k|$ for the $IPM(\mathcal{P},\mathcal{V},\mathcal{R} )$, i.e., ordering all parameters in descending order according to the number of parameter values. After that, the IPM can be abbreviated as $IPM(|\mathcal{V}_1||\mathcal{V}_2|\cdots|\mathcal{V}_k|,{\mathcal{R}})$. The example above can therefore be represented as $IPM(3^22^2,\mathcal{R}=\{r_1\})$.

\begin{table}[!t]
\centering
\footnotesize
 \caption{Configurations of A Component-Based System}
  \label{Table1}
  \setlength{\tabcolsep}{3.3mm}{
  \begin{tabular}{rrrr}
    \hline
    \textbf{$p_1$: \texttt{OS}}      & \textbf{$p_2$: \texttt{Browser}}      &\textbf{$p_3$: \texttt{Network}}         &\textbf{$p_4$: \texttt{Database}} \\\hline

    \texttt{Windows} ($0$)     &\texttt{IE} ($3$)      &\texttt{LAN} ($6$)        &\texttt{DB/2} ($8$)   \\

    \texttt{Linux} ($1$)    &\texttt{Firefox} ($4$)   &\texttt{VPN} ($7$)        &\texttt{Oracle} ($9$)  \\

    \texttt{Solaris} ($2$)     &\texttt{Google} ($5$)    &--     &--\\     \hline
   \multicolumn{4}{l}{Constraint $r_1$: $(p_1=\textrm{\texttt{Linux}}) \wedge (p_2=\textrm{\texttt{IE}}) \Rightarrow (p_3 \neq \textrm{\texttt{LAN}})$.}

  \end{tabular}}
\end{table}

For the purpose of clear description, let $\Psi_\lambda$ be the set of all possible $\lambda$-wise value combinations ($1\leq \lambda \leq k$) from $IPM(\mathcal{P},\mathcal{V},\mathcal{R})$, and let $\tilde{\Psi}_\lambda$ be the set of all valid $\lambda$-wise value combinations under the constraint set $\mathcal{R}$.

\begin{definition}[\textbf{Combinatorial Test Space}]
Given an $IPM(\mathcal{P},\mathcal{V},\mathcal{R})$, the combinatorial test space $T_{\textit{\textrm{valid}}}$ is the set of all valid test cases, i.e., $T_{\textit{\textrm{valid}}}=\{t \mid t \in \tilde{\Psi}_k\}$ (or $T_{\textit{\textrm{valid}}} = \tilde{\Psi}_k$).
\end{definition}

For the $IPM(\mathcal{P},\mathcal{V},\mathcal{R})$, the number of all possible combinatorial test cases is $|\Psi_k|=\prod_{i=1}^k{|\mathcal{V}_i|}$. However, the size of the combinatorial test space is $|\tilde{\Psi}_k|~(|\tilde{\Psi}_k|\leq |{\Psi}_k|)$, because some combinatorial test cases are invalid under the constraint set $\mathcal{R}$. Consider the IPM shown in Table~\ref{Table1}; this example has $3^2\times 2^2 =36$ possible combinatorial test cases, while its combinatorial test space contains $(36-2)=34$ valid combinatorial test cases.

\subsection{Combinatorial Test Suite and Sequence}

\subsubsection{Combinatorial Test Suite}
A combinatorial test suite is a finite set of combinatorial test cases, which can be defined as follows:

\begin{definition}[\textbf{Combinatorial Test Suite}]
\textit{A combinatorial test suite $T$ is a subset of $\tilde{\Psi}_k$, i.e., $T \subseteq \tilde{\Psi}_k$.}
\end{definition}

Intuitively, any subset of $\tilde{\Psi}_k$ can serve as a combinatorial test suite. In other words, a combinatorial test suite consists of valid combinatorial test cases that satisfy the constraint set $\mathcal{R}$. An intuitive example is a random combinatorial test suite constructed via random testing (RT)~\cite{Hamlet2002,Myers2004}, in which each combinatorial test case is chosen at random~\cite{Arcuri2012}.

In the field of CIT, the most commonly used type of combinatorial test suite is the $\lambda$-wise covering array (CA)~\cite{Cohen1997,Cohen2003}, which is specifically designed to cover all $\lambda$-wise value combinations. Previous studies have shown that the number of parameters involved in software failures is typically 1 or 2, and generally fewer than 6~\cite{Kuhn2002,Kuhn2004,Kuhn2006}.

\begin{definition}[\textbf{$\lambda$-wise Covering Array}] Given an $IPM(\mathcal{P},\mathcal{V},\mathcal{R})$,
a $\lambda$-wise covering array, denoted as $T_\lambda={\textrm{CA}}(N;\lambda,IPM(\mathcal{P},\mathcal{V},\mathcal{R}))$, is a set of combinatorial test cases that satisfy the constraints $\mathcal{R}$, such that each valid $\lambda$-wise value combination is covered by at least one valid combinatorial test case, where $1 \leq \lambda \leq k$, i.e., $\forall \sigma_\lambda \in \tilde{\Psi}_\lambda, \exists t\in T_\lambda, \sigma_\lambda \in \varphi(\lambda,t)$.
\end{definition}

Intuitively, $T_\lambda \subseteq \tilde{\Psi}_k$, and $|T_\lambda| \geq \prod_{i=1}^\lambda{|\mathcal{V}_i|}$ due to the constraints. In general, $\lambda$ is referred to as the strength of the CA.

\begin{example}[\textbf{CA Example}]
   As shown in Table~\ref{Table2}, a covering array with strength $\lambda=2$ is derived from the IPM shown in Table~\ref{Table1}. This covering array can be written as $\textit{\textrm{CA}}(9;2,IPM(3^22^2,\mathcal{R}=\{r_1\}))$. It can be seen that only $9$ of the $34$ valid combinatorial test cases are sufficient to cover all valid 2-wise value combinations.
\end{example}

\begin{table}[!t]
\centering
\footnotesize
\caption{An Example of A $2$-wise CA for $IPM(\mathcal{P},\mathcal{V},\mathcal{R})$ Shown in Table~\ref{Table1}}
\label{Table2}
\setlength{\tabcolsep}{2mm}{\begin{tabular}{rrrrr} \hline

\textbf{No.}  &\textbf{$p_1$: \texttt{OS}}      & \textbf{$p_2$: \texttt{Browser}}      &\textbf{$p_3$: \texttt{Network}}         &\textbf{$p_4$: \texttt{Database}} \\ \hline
$t_1$ &\texttt{Windows} ($0$) &\texttt{IE} ($3$)      &\texttt{LAN} ($6$) &\texttt{Oracle} ($9$)\\
$t_2$ &\texttt{Linux} ($1$)   &\texttt{Firefox} ($4$) &\texttt{LAN} ($6$) &\texttt{DB/2} ($8$)\\
$t_3$ &\texttt{Solaris} ($2$) &\texttt{Google} ($5$)  &\texttt{VPN} ($7$) &\texttt{Oracle} ($9$)\\
$t_4$ &\texttt{Windows} ($0$) &\texttt{Google} ($5$)  &\texttt{LAN} ($6$) &\texttt{Oracle} ($9$)\\
$t_5$ &\texttt{Linux} ($1$)   &\texttt{IE} ($3$)      &\texttt{VPN} ($7$) &\texttt{Oracle} ($9$)\\
$t_6$ &\texttt{Solaris} ($2$) &\texttt{IE} ($3$)      &\texttt{LAN} ($6$) &\texttt{DB/2} ($8$)\\
$t_7$ &\texttt{Windows} ($0$) &\texttt{Firefox} ($4$) &\texttt{VPN} ($7$) &\texttt{DB/2} ($8$)\\
$t_8$ &\texttt{Linux} ($1$)  &\texttt{Google} ($5$)  &\texttt{VPN} ($7$) &\texttt{DB/2} ($8$)\\
$t_9$ &\texttt{Solaris} ($2$) &\texttt{Firefox} ($4$) &\texttt{VPN} ($7$) &\texttt{Oracle} ($9$)\\\hline
\end{tabular}
}
\end{table}

To support covering array generation, many algorithms and tools have been proposed and developed, such as the Advanced Combinatorial Testing System (ACTS)~\cite{Lei1998,Lei2008}, the Automatic Efficient Test Generator (AETG)~\cite{Cohen1997}, Practical Independent Combinatorial Testing (PICT)~\cite{Czerwonka2006}, Covering Arrays by Simulated Annealing (CASA)~\cite{Garvin2009,Garvin2011}, and others. All these algorithms and tools are presented in a comprehensive survey~\cite{Khalsa2014}.\setlength{\parskip}{0em}

\subsubsection{Combinatorial Test Sequence}
When the order of a combinatorial test suite $T$ is taken into account, the position of each element in $T$ must be fixed. In such cases, a combinatorial test sequence is required.

\begin{definition}[\textbf{Combinatorial Test Sequence}]
\textit{Given a combinatorial test suite $T=\{t_1,t_2,\cdots,t_n\}$ of size $n$, a combinatorial test sequence $S$ is a sequence obtained by fixing the position of each combinatorial test case in $T$, i.e., $S=\langle t_{i_1},t_{i_2},\cdots,t_{i_n} \rangle~(t_{i_j} \in T, j=1,2,\cdots,n)$.}
\end{definition}

Combinatorial test sequences have been widely used in the CIT field~\cite{Nie2011}, especially when testing resources are limited and only a subset of combinatorial test cases in a test suite can be executed. In such scenarios, the potentially failure-revealing combinatorial test cases should be executed as early as possible. In other words, a well-designed combinatorial test sequence may be more likely to detect failures earlier, thereby enabling earlier fault characterization, diagnosis, and revision~\cite{Nie2011}. The process of determining the execution order of test cases in a test suite is referred to as test case prioritization (TCP)~\cite{Nie2011,Rothermel2001,Elbaum2002}, which ensures that higher-priority test cases are executed earlier based on specific metrics.

To prioritize combinatorial test cases, many techniques have been studied and can be generally divided into two categories~\cite{Qu2007}: 1) \textit{Pure prioritization}, which re-prioritizes combinatorial test cases within an existing combinatorial test suite; and 2) \textit{Re-generation prioritization}, which takes prioritization into account during the generation of combinatorial test suites.

\textit{For ease of discussion, in the remainder of this study, we will refer to a combinatorial test case as a test case; a combinatorial test suite as a test suite; a combinatorial test sequence as a test sequence; and the combinatorial test space as the test space. In other words, the term ``combinatorial'' is omitted to simplify descriptions, unless explicitly stated otherwise. In addition, only valid combinatorial test cases are considered in the following sections.}

\section{Black-box Metrics for Evaluating Test Suites}
\label{SECTION:Metrics}

In this section, we summarize the evaluation metrics used to assess test suites and sequences, including their rigorous definitions, usage, and classification. We also provide numerous examples to illustrate these metrics along with detailed explanations. Since various metrics have been proposed by other researchers, we also provide references to their work.

It is reasonable to categorize these evaluation metrics according to whether they are applied to test suites or test sequences, given that test case order is considered in test sequences but not in test suites. However, it should be noted that there is a ``\textit{one-to-one relationship}'' between evaluation metrics for test suites and those for test sequences. Specifically, when a test sequence needs to be evaluated with respect to a certain property derived from the test cases, that same property can also be used to assess test suites, and vice versa.

\begin{figure*}[!t]
\graphicspath{{Figures/}}
  \centering
  \subfigure[Evaluation for test suites.]{
  \label{Figure2a}
  \includegraphics[width=0.49\linewidth]{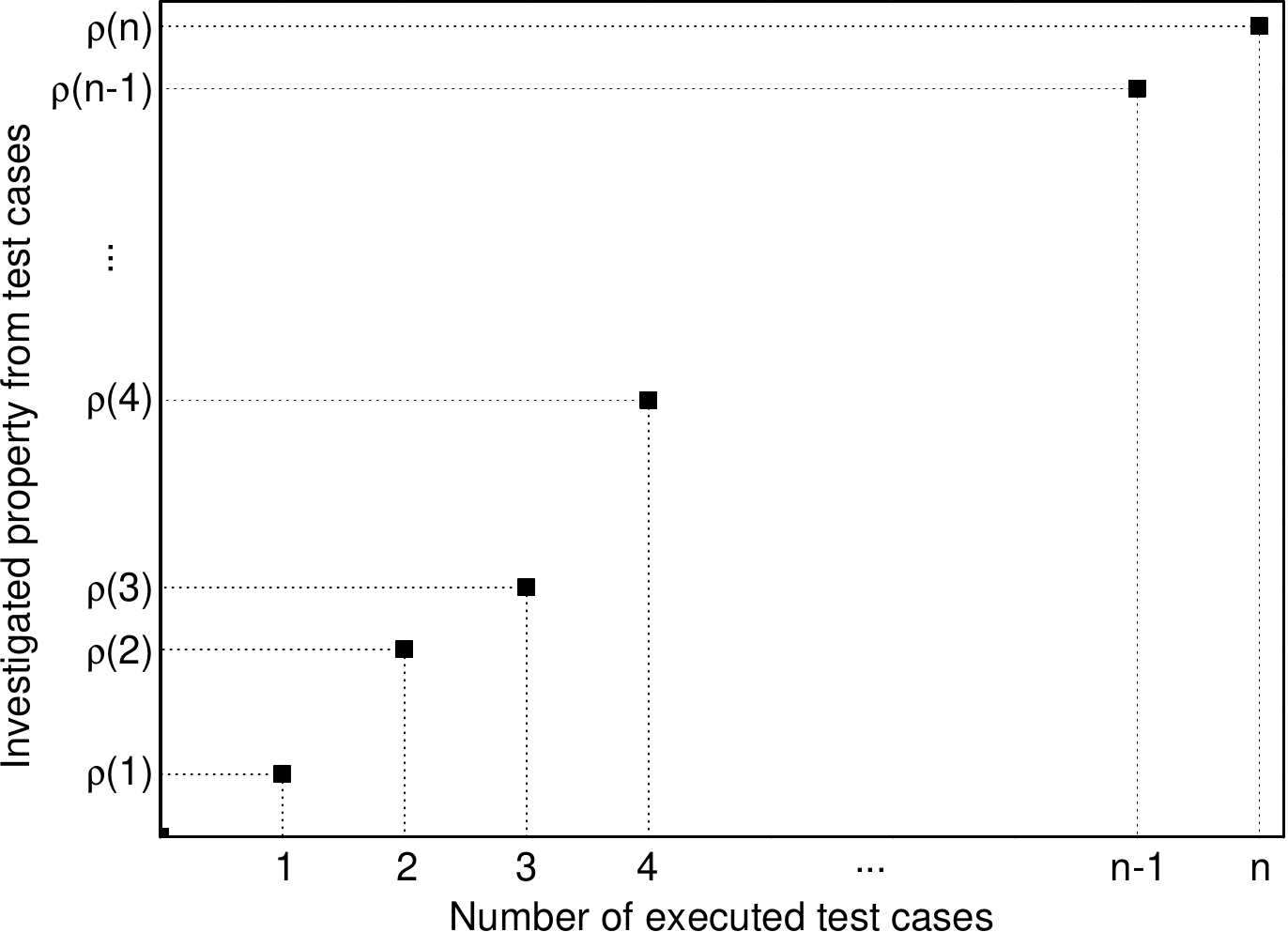}}
  \subfigure[Evaluation for test sequences.]{
  \label{Figure2b}
  \includegraphics[width=0.49\linewidth]{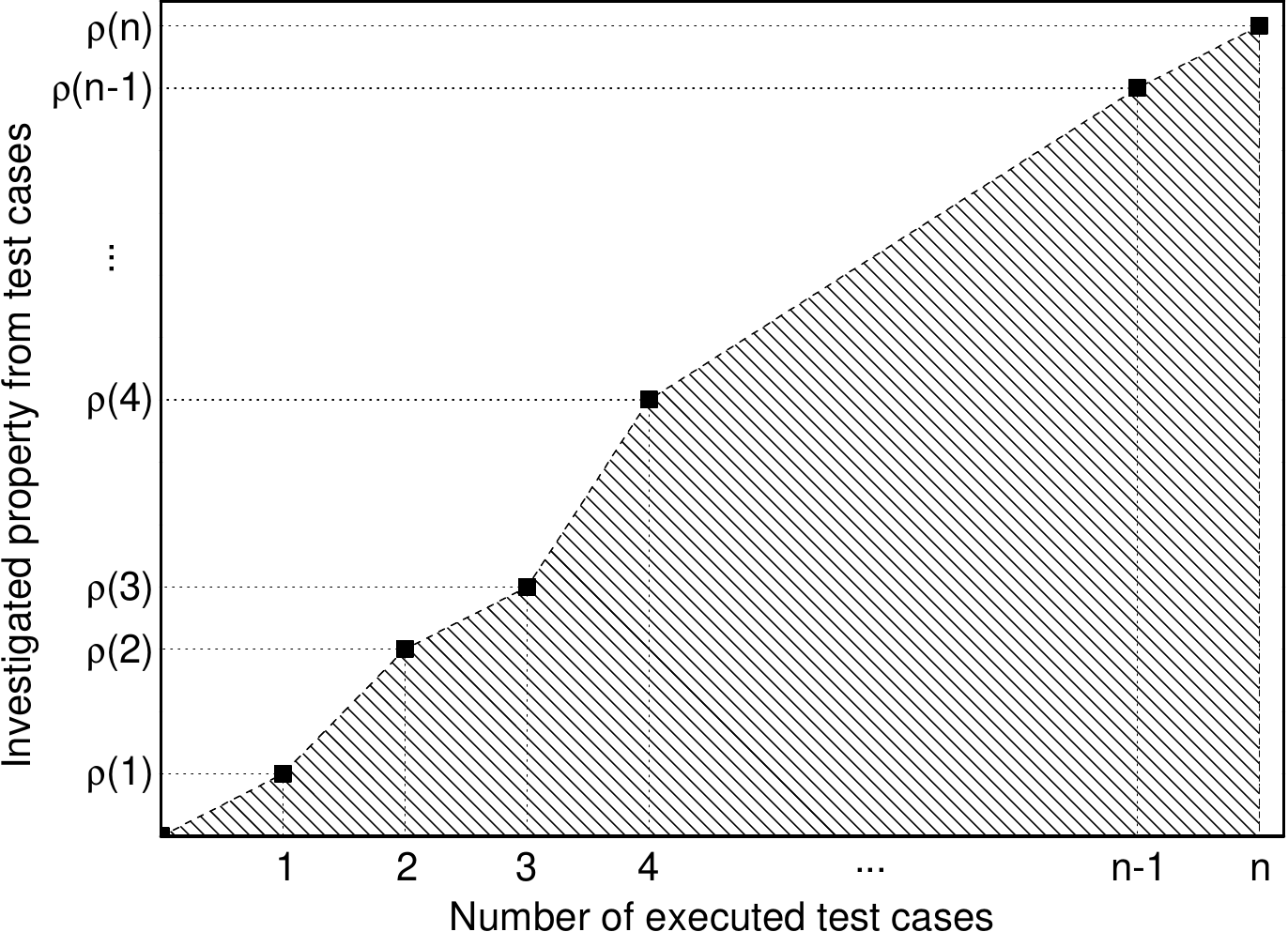}}
  \caption{Evaluation for test suites and sequences.}
  \label{Figure2}
\end{figure*}

Consider a test suite $T=\{t_1,t_2,\cdots,t_n\}$ of size $n$. Let $g(t)$ be a function that returns a set of elements according to some property after executing test case $t$; for example, $g(t)$ can be defined as $\varphi(\lambda,t)$, where each element in $g(t)$ corresponds to a $\lambda$-wise value combination. Let $\rho(x)=\big|\bigcup_{i=1}^{x}g(t_i)\big|$, where $1 \leq x \leq n$. Thus, $\rho(x)$ denotes the number of elements contributed by the first $x$ test cases in $T$, i.e., $T'=\{t_1,t_2,\cdots,t_x\}$. Intuitively, we have $\rho(0) = 0 \leq \rho(1) \leq \rho(2) \leq \cdots \leq \rho(n)$. Figure~\ref{Figure2} provides an illustrative example for evaluating test suites and sequences. When evaluating a test suite $T' \subseteq T$, as shown in Fig.~\ref{Figure2a}, $\rho(x)$ can be directly used as an evaluation metric to measure the property of the $x$ executed test cases in $T'$. When evaluating a test sequence $S=\langle t_1,t_2,\cdots,t_x \rangle$, as shown in Fig.~\ref{Figure2b}, we use $\int_{0}^{x}\rho(x)\,dx$ (i.e., the shaded region) to measure the rate of the property accumulated from the ordered $x$ test cases in $S$. In this example (as shown in Fig.~\ref{Figure2b}), a linear function is used to connect the points of $\rho(x)$; however, other functions, such as a step function, could also be used. Nevertheless, the choice of the connecting function does not affect the evaluation of test sequences\footnote{Specifically, if an evaluation metric using a linear function to connect $\rho(x)$ determines that test sequence $S_1$ is better than test sequence $S_2$, another metric using a step function for the same connection will yield the same determination, and vice versa.}. Therefore, in this paper, we adopt a linear function to connect $\rho(x)$ when evaluating test sequences, unless explicitly stated otherwise.

As discussed previously, given an evaluation metric $\rho(x)$ for evaluating $x$ test cases in a test suite, we can use a function to connect $\rho(x)$ for assessing the corresponding test sequence. Consequently, in this paper, we focus on the details of each evaluation metric for test suites rather than for test sequences.

Depending on the context, the underlying evaluation metrics can be classified into three categories: \textit{interaction coverage-based metrics}, \textit{distribution-based metrics}, and \textit{mutation-based metrics}.

\subsection{Interaction Coverage-Based Metrics}

In this section, we review evaluation metrics related to value combination coverage of test cases, namely \textit{interaction coverage-based metrics} (ICBM), including $\lambda$\textit{-wise value combination coverage}~\cite{Dalal1998,Williams2001,Kuhn2010}, $\lambda$-wise value interaction diversity~\cite{Dalal1998}, $(\lambda+a)$\textit{-wise value combination coverage}~\cite{Maximoff2010,Kuhn2010,Kuhn2013}, \textit{tuple density}~\cite{Chen2011}, \textit{weighted multiple value combination coverage}~\cite{Huang2014}, $\lambda$\textit{-wise parameter-value configuration coverage}~\cite{Maximoff2010,Kuhn2010,Kuhn2013}, \textit{simple $\lambda$-wise value combination coverage}~\cite{Maximoff2010,Kuhn2010,Kuhn2013}, and $(q,\lambda)$\textit{-completeness}~\cite{Maximoff2010,Kuhn2010,Kuhn2013}.

\subsubsection{$\lambda$-wise Value Combination Coverage}

In~\cite{Dalal1998}, Dalal et al. first proposed $\lambda$-wise value combination coverage (VCC), also referred to as $\lambda$-coverage~\cite{Dalal1998}, interaction coverage of degree $\lambda$~\cite{Williams2001}, or $\lambda$-wise combination coverage~\cite{Kuhn2010}. This metric measures the proportion of all possible valid $\lambda$-wise value combinations covered by a given test suite.

\begin{definition}[\textbf{$\lambda$-wise Value Combination Coverage}]
Given a test suite $T$ and a strength $\lambda~(1 \leq \lambda \leq k)$, the $\lambda$-wise value combination coverage (VCC) of $T$ is defined as the ratio of the number of $\lambda$-wise value combinations covered by test cases in $T$ to the total number of possible valid $\lambda$-wise value combinations, i.e.,
\begin{equation}
\label{Equation3}
VCC(\lambda,T)=\frac{\big|\bigcup_{t \in T}{\varphi(\lambda,t)}\big|}{\big|\bigcup_{t \in \tilde{\Psi}_k}{\varphi(\lambda,t)}\big|},
\end{equation}
where $\bigcup_{t \in \tilde{\Psi}_k}{\varphi(\lambda,t)}$ denotes the set of all possible valid $\lambda$-wise value combinations derived from the IPM.
\end{definition}

\subsubsection{$\lambda$-wise Value Interaction Diversity}
Dalal et al.~\cite{Dalal1998} first proposed $\lambda$-wise value interaction diversity (VID), also referred to as $\lambda$-diversity~\cite{Dalal1998}, to measure the extent to which the test design avoids replicating each $\lambda$-wise value combination. Its definition is very similar to that of VCC.

\begin{definition}[\textbf{$\lambda$-wise Value Interaction Diversity}]
\textit{Given a test suite $T$ and a strength $\lambda~(1 \leq \lambda \leq k)$, the $\lambda$-wise value interaction diversity (VID) of $T$ is defined as the percentage of the number of $\lambda$-wise value combinations covered by test cases in $T$ relative to the maximum number of $\lambda$-wise value combinations that could be covered by $|T|$ test cases, i.e.,}
\begin{equation}
\label{Equation10}
VID(\lambda,T)=\frac{\big|\bigcup_{t \in T}{\varphi(\lambda,t)}\big|}{\mathrm{C}(k,\lambda)\times |T|}.
\end{equation}
\end{definition}

As we know, when two test suites have the same VCC metric value at strength $\lambda$, for example, two $\lambda$-wise CAs denoted as $T_\lambda^1$ and $T_\lambda^2$ (both having VCC values equal to $100.00\%$), the $\lambda$-wise VCC metric fails to differentiate between them. In such cases, the VID metric can be used for further comparison. If the size of $T_\lambda^1$ is smaller than that of $T_\lambda^2$, we consider $T_\lambda^1$ to be better than $T_\lambda^2$; otherwise, $T_\lambda^2$ is considered better. However, the VID metric faces the same limitation when $T_\lambda^1$ and $T_\lambda^2$ have the same size, as it cannot determine which CA is superior. Therefore, in the following sections, we review other metrics for evaluating CAs, such as \textit{$(\lambda+a)$-wise value combination coverage}~\cite{Maximoff2010,Kuhn2010,Kuhn2013} and \textit{tuple density}~\cite{Chen2011}.

\subsubsection{$(\lambda+a)$-wise Value Combination Coverage}

The \textit{$(\lambda+b)$-wise value combination coverage} (BVCC) is defined similarly to the VCC metric; however, it evaluates a $\lambda$-wise CA at a strength higher than $\lambda$ (i.e., $\lambda+b$, where $1 \leq b \leq k-\lambda$). This metric is particularly useful when two test suites are $\lambda$-wise CAs of the same size.

\begin{definition}[\textbf{$(\lambda+b)$-wise Value Combination Coverage}]
\textit{Given a $\lambda$-wise CA denoted as $T_\lambda=CA(N;\lambda,IPM(\mathcal{P},\mathcal{V},\mathcal{R}))$, the $(\lambda+b)$-wise value combination coverage (BVCC) of $T_\lambda$ is defined as:}
\begin{equation}
BVCC(b,T_\lambda)=VCC(\lambda+b,T_\lambda).
\end{equation}
\end{definition}

\subsubsection{Tuple Density}

\textit{Tuple density} (TD)~\cite{Chen2011} incorporates the information of $(\lambda+b)$-wise value combination coverage with $b=1$. In other words, for a $\lambda$-wise CA $T_\lambda$, TD evaluates $T_\lambda$ based on $BVCC(1,T_\lambda)$. As discussed in~\cite{Chen2011}, the use of $BVCC(1, T_\lambda)$ is practical for the following reasons: 1) Since the number of $\lambda$-wise value combinations to be covered grows exponentially with the increase of $\lambda$, the coverage of $\gamma$-wise value combinations becomes much lower when $\gamma > \lambda+1$; 2) the $(\lambda+1)$-wise value combination coverage also provides some indication of coverage at higher strengths $\gamma > \lambda+1$; and 3) the number of additional faults triggered by $\lambda$-wise value combinations drops rapidly for $\lambda>2$~\cite{Kuhn2004}. The definition of TD is given as follows:

\begin{definition}[\textbf{Tuple Density}]
\textit{Given a $\lambda$-wise CA denoted as $T_\lambda=CA(N;\lambda,IPM(\mathcal{P},\mathcal{V},\mathcal{R}))$, the tuple density (TD) of $T_\lambda$ is defined as:}
\begin{align}
TD(T_\lambda) & = \lambda + BVCC(1,T_\lambda) \nonumber\\
&= \lambda + VCC(\lambda+1,T_\lambda).
\end{align}
\end{definition}

In effect, the TD of $T_\lambda$ ranges in $(\lambda, \lambda+1.0]$: a) since $T_\lambda$ covers some $(\lambda+1)$-wise value combinations, $BVCC(1,T_\lambda) > 0.0$, and hence $TD(T_\lambda) > \lambda$; and b) when $T_\lambda$ is also a $(\lambda+1)$-wise CA, its TD metric value reaches the maximum, which is equal to $\lambda+1.0$.

Similar to BVCC, TD can only be used to measure CAs rather than general test suites (that are not CAs), because both metrics require $\lambda$ obtained from the CA $T_\lambda$ as a parameter. To measure general test suites, Huang et al.~\cite{Huang2014c} proposed a new metric, namely \textit{weighted multiple value combination coverage} (WMVCC).

\subsubsection{Weighted Multiple Value Combination Coverage}
As discussed previously, prior metrics use a fixed strength to evaluate test suites, which may not provide sufficient information. To address this limitation, the weighted multiple value combination coverage (WMVCC) metric comprehensively accounts for varying strengths.

\begin{definition}[\textbf{Weighted Multiple Value Combination Coverage}]
\textit{Given a test suite $T$ and a strength $\lambda~(1 < \lambda \leq k)$, the weighted multiple value combination coverage (WMVCC) of $T$ is defined as:}
\begin{equation}
WMVCC(\lambda,T)=\sum_{i=1}^{\lambda}{(\omega_i \cdot VCC(i,T))},
\label{Eq.WMVCC}
\end{equation}
\textit{where $\omega_i~(i=1,2,\cdots,\lambda)$ is a constant weight for each strength $i$ that must be assigned in advance, satisfying the following two requirements: (1) $0 \leq \omega_i \leq 1.0$; and (2) $\sum_{i=1}^\lambda\omega_i=1.0$.}
\end{definition}

As shown in Eq.~(\ref{Eq.WMVCC}), the weight $\omega_i~(i=1,2,\cdots,\lambda)$ must be assigned before evaluation. As investigated in~\cite{Huang2014c,Huang2015}, there are three weight distribution schemes: (1) Equal weight assignment: each $\omega_i$ is assigned the same value, i.e., $\omega_1=\omega_2=\cdots=\omega_\lambda=1/\lambda$; (2) Random weight assignment: each $\omega_i$ is assigned randomly according to a uniform distribution; and (3) Empirical weight assignment: since previous studies (e.g., \cite{Kuhn2002,Kuhn2004,Kuhn2006}) have reported that $1$-wise parameter interactions account for 29\% to 82\% of faults; $2$-wise for 6\% to 47\%; $3$-wise for 2\% to 19\%; $4$-wise for 1\% to 7\%; and even less for interactions higher than $4$-wise, we assign each weight as follows:
$\omega_1=\omega$, and $\omega_{i+1}=\frac{1}{2}\omega_i$, where $i=1,2,\cdots,\lambda-1$.
For example, if $\lambda=2$, then we have $\omega_1=0.67$ and $\omega_2=0.33$;
if $\lambda=3$, then we have $\omega_1=0.57$, $\omega_2=0.29$, and $\omega_3=0.14$.

\subsubsection{$\lambda$-wise Parameter-value Configuration Coverage}
As discussed previously, the above metrics based on value-combination coverage consider all $\lambda$-wise ($1\leq \lambda \leq k$) value combinations, regardless of the $\lambda$-wise parameter combinations from which they are derived. In other words, they do not take into account the underlying $\lambda$-wise parameter combinations. In the following sections, we present metrics for test suites that are based on $\lambda$-wise parameter combinations.

As discussed earlier, given a $\lambda$-wise parameter combination $\phi_\lambda=(p_{i_1},p_{i_2},\cdots,p_{i_\lambda})$, where $1\leq\lambda\leq k$, it can derive $(\prod_{j=1}^\lambda|\mathcal{V}_{i_j}|)$ possible $\lambda$-wise value combinations, which form a set denoted as $M_{\phi_\lambda}=\{\sigma_\lambda \mid \sigma_\lambda=(v_{i_1,j_1},v_{i_2,j_2},\cdots,v_{i_\lambda,j_\lambda}), v_{i_a,j_a}\in \mathcal{V}_{i_a}, 1 \leq j_a \leq |\mathcal{V}_{i_a}|, a=1,2,\cdots, \lambda\}$. To check whether all $\lambda$-wise value combinations derived from $\phi_\lambda$ have been covered by a test suite, a new metric was proposed, namely \textit{$\lambda$-wise parameter-value configuration coverage} (PCC)~\cite{Maximoff2010,Kuhn2010,Kuhn2013}.

\begin{definition}[\textbf{$\lambda$-wise Parameter-value Configuration Coverage}]
\textit{Given a test suite $T$, and a $\lambda$-wise parameter combination $\phi_\lambda \in \Phi_\lambda~(1\leq\lambda\leq k)$, the $\lambda$-wise parameter-value configuration coverage (PCC) of $T$ is defined as the percentage of the $\lambda$-wise value combinations from $\phi_\lambda$ covered by $T$ relative to the total valid $\lambda$-wise value combinations derived from $\phi_\lambda$, i.e.,}
\begin{equation}
PCC(\phi_\lambda,T)=\frac{\big|M_{\phi_\lambda} \bigcap (\bigcup_{t \in T}{\varphi(\lambda,t)})\big|}{\big|\{\sigma_\lambda \mid \sigma_\lambda \in M_{\phi_\lambda} \wedge isValid(\sigma_\lambda,\mathcal{R})\}\big|}.
\label{Eq.PCC}
\end{equation}
\end{definition}

Intuitively, the PCC metric ranges from $0$ to $100\%$, with higher values indicating better test suites.

\subsubsection{Simple $\lambda$-wise Value Combination Coverage}
Different from the PCC metric, the \textit{simple $\lambda$-wise value combination coverage} (SVCC) considers all $\lambda$-wise parameter combinations by checking whether all $\lambda$-wise value combinations of each parameter combination have been covered. Its definition is given in~\cite{Maximoff2010,Kuhn2010,Kuhn2013}.

\begin{definition}[\textbf{Simple $\lambda$-wise Value Combination Coverage}]
\textit{Given a test suite $T$ and a strength $\lambda~(1 \leq \lambda \leq k)$, the simple $\lambda$-wise value combination coverage (SVCC) of $T$ is defined as the proportion of $\lambda$-wise parameter combinations $\phi_\lambda \in \Phi_\lambda$ such that all possible valid $\lambda$-wise value combinations derived from $\phi_\lambda$ are covered, i.e.,}
\begin{equation}
SVCC(\lambda,T)=\frac{\big|\{\phi_\lambda \mid \phi_\lambda \in \Phi_\lambda, PCC(\phi_\lambda,T) = 1.0\}\big|}{\mathrm{C}(k,\lambda)}.
\label{Eq.SVCC}
\end{equation}
\end{definition}

\subsubsection{$(q,\lambda)$-completeness}
The \textit{$(q,\lambda)$-completeness} can be viewed as a generalized version of SVCC, as it checks whether the $\lambda$-wise PCC of each parameter combination meets a specified threshold $q$. It is clear that when $q=1.0$, $(q,\lambda)$-completeness reduces to SVCC. Its detailed definition is provided in~\cite{Maximoff2010,Kuhn2010,Kuhn2013}.

\begin{definition}[\textbf{$(q,\lambda)$-completeness}]
\textit{Given a test suite $T$ and a strength $\lambda~(1 \leq \lambda \leq k)$, the $(q,\lambda)$-completeness of $T$ is defined as the proportion of $\lambda$-wise parameter combinations $\phi_\lambda \in \Phi_\lambda$ with $PCC(\phi_\lambda,T)$ greater than or equal to $q$, i.e.,}
\begin{equation}
Completeness(q,\lambda,T)=\frac{\big|\{\phi_\lambda \mid \phi_\lambda \in \Phi_\lambda, PCC(\phi_\lambda,T) \geq q\}\big|}{\mathrm{C}(k,\lambda)}.
\label{Eq.Completeness}
\end{equation}
\end{definition}

In effect, the $(q,\lambda)$-completeness metric checks whether the PCC value of each $\lambda$-wise parameter combination is at least $q$.

\subsection{Distribution-Based Metrics}
The \textit{distribution-based metric} (DBM) measures the test case distribution of a given test suite by utilizing the locations of test cases or the occurrences of parameter values. To date, the following six well-studied metrics based on test case distribution have been proposed: \textit{discrepancy}~\cite{Chen2009b}, \textit{dispersion}~\cite{Chen2009b}, \textit{diversity}~\cite{Bueno2014}, \textit{similarity}~\cite{Henard2014}, \textit{novelty score}~\cite{Xiang2022}, and \textit{divergence}~\cite{Choi2016}.

\subsubsection{Discrepancy}

The \textit{discrepancy}~\cite{Chen2009b} measures the density degree of sample test cases located in different subdomains of a domain (or test space). A smaller discrepancy value indicates that the set of test cases is more evenly distributed. Its definition is given in~\cite{Chen2009b}.

\begin{definition}[\textbf{Discrepancy}]
Given a test suite $T$, the discrepancy of $T$ is defined as:
\begin{equation}
\textit{\textrm{Discrepancy}}(T)=\max_{i=1,2,\cdots,m}\Big| \frac{|T_i|}{|T|} - \frac{|\mathcal{D}_i|}{|\mathcal{D}|} \Big|,
\label{Eq.Discrepancy}
\end{equation}
where
\begin{itemize}
    \item $\mathcal{D}$ is the test space of $T$, i.e., $\mathcal{D}=\tilde{\Psi}_k$ in this paper,
    \item $\mathcal{D}_i~(i = 1,2,\cdots,m)$ is a randomly defined subdomain of $\mathcal{D}$, i.e., $\mathcal{D}_i \subseteq \mathcal{D}$,
    \item $T_i$ is the subset of $T$ whose test cases are located in $\mathcal{D}_i$, i.e., $T_i = T \cap \mathcal{D}_i$, and
    \item $m$ is a constant.
\end{itemize}
\end{definition}

\subsubsection{Dispersion}

The \textit{dispersion}~\cite{Chen2009b} indicates whether a large empty spherical region surrounds any test case in $T$. A smaller dispersion value indicates that $T$ is reasonably equidistributed. Its definition is given as follows.

\begin{definition}[\textbf{Dispersion}]
Given a test suite $T$, the dispersion of $T$ is defined as:
\begin{equation}
Dispersion(T)=\max_{t \in T}{\min_{t' \in T,t' \neq t}dist(t, t')},
\label{Eq.Dispersion}
\end{equation}
where $dist(t,t')$ is a distance (or dissimilarity) function that measures the difference between test cases $t$ and $t'$.
\end{definition}

As we know, there are many distance measures to evaluate the difference between two test cases, such as \textit{Hamming distance} and \textit{Euclidean distance}. For ease of description, in this paper we adopt Hamming distance to evaluate the difference between test cases. Specifically, given two test cases $t_1=(v_1,v_2,\cdots,v_k)$ and $t_2=(u_1,u_2,\cdots,u_k)$, where $v_i,u_i\in \mathcal{V}_i~(1 \leq i \leq k)$, the distance between $t_1$ and $t_2$ is given as follows:
\begin{equation}
dist(t_1,t_2)= \frac{\big|\{i \mid v_i \neq u_i, 1\leq i \leq k\}\big|}{k}.
\end{equation}

\subsubsection{Diversity}

The \textit{diversity}~\cite{Bueno2014} is a variant of the dispersion metric; however, it uses the sum of the distance values between each element and its nearest neighbor rather than the maximum value as the evaluation criterion. Similarly, the higher the diversity of a test suite, the better its performance.

\begin{definition}[\textbf{Diversity}]
Given a test suite $T$, the diversity of $T$ is defined as:
\begin{equation}
Diversity(T)=\frac{1}{|T|}\sum_{t \in T}{\min_{t' \in T,t' \neq t}dist(t, t')}.
\label{Eq.Diversity}
\end{equation}
\end{definition}

\subsubsection{Similarity}

The \textit{similarity}~\cite{Henard2014} is a generalized similarity measure for a test set, which aggregates all distance values between any pair of distinct test cases. A larger similarity value indicates a better test suite. Here, we provide a normalized version of the similarity metric.

\begin{definition}[\textbf{Similarity}]
Given a test suite $T$, the similarity of $T$ is defined as:
\begin{equation}
Similarity(T)=\frac{1}{|T|\times(|T|-1)}\sum_{t\in T}\sum_{t' \in T, t' \neq t}dist(t, t').
\label{Eq.Similarity}
\end{equation}
\end{definition}

\subsubsection{Novelty Score}

The \textit{novelty score}~\cite{Xiang2022} is used to measure the sparseness of a test suite. It calculates the average distance from each test case in $T$ to its $\varepsilon$-nearest neighbors (i.e., the top-$\varepsilon$ nearest elements to each test case), where $\varepsilon$ is a constant integer. A higher novelty score indicates a better test suite.

\begin{definition}[\textbf{Novelty Score}]
Given a test suite $T$, the novelty score of $T$ is defined as:
\begin{equation}
    Novelty(T) = \frac{1}{|T|} \cdot \frac{1}{\varepsilon} \sum_{t \in T} \sum_{j=1}^{\varepsilon}dist(t,t_j),
\end{equation}
where $\varepsilon$ is a pre-defined constant integer ($1 \leq \varepsilon \leq |T|-1$), and $t_j$ is the $j$-th nearest neighbor of $t$ within $T$.
\end{definition}

Intuitively, the novelty score of $T$ can be viewed as a normalized version of the similarity and diversity of $T$, i.e., similarity and diversity are two special cases of the novelty score. The main reason is that when $\varepsilon = |T|-1$, the novelty score of $T$ is equivalent to the similarity of $T$; however, when $\varepsilon = 1$, the novelty score reduces to the diversity metric.

\subsubsection{Divergence}

The \textit{divergence}~\cite{Choi2016} measures the distribution of values for each parameter against the ideal situation under the assumption that the number of occurrences (or frequencies) of each value for a given parameter should be identical across a test suite, to spread parameter values as evenly as possible. A larger divergence value indicates a more effective test suite.

There are many types of metrics to measure the divergence of probability distributions, such as $\chi^2$-divergence~\cite{Pearson1900}, Jensen-Shannon divergence~\cite{Endres2003}, and Kullback-Leibler divergence~\cite{Endres2003}. Following previous studies~\cite{Choi2016}, we adopt the $\chi^2$-divergence in this study.

\begin{definition}[\textbf{Divergence}]
Given an $IPM(\mathcal{P},\mathcal{V},\mathcal{R})$ and a test suite $T=\{t_1,t_2,\cdots,t_n\}$, the divergence of $T$ is defined as:
\begin{align}
   Divergence(T) &= \chi^2(U(T) \parallel U_{IPM}) \nonumber\\
    &= \frac{1}{2}\sum_{1 \leq i \leq |\mathcal{P}|}{\sum_{v \in \mathcal{V}_i}{\frac{\big(O_v(T) - O_v^{IPM}\big)^2}{O_v(T) + O_v^{IPM}}}},
    \label{Eq.Divergence}
\end{align}
where
\begin{itemize}
    \item $U(T)$ represents the probability distribution of value occurrences in the test suite $T$,
    \item $U_{IPM}$ represents the ideal probability distribution of occurrences of parameter values in the IPM of the SUT,
    \item $O_v(T)$ is the occurrence probability of each parameter value $v$ in the test suite $T$, i.e., $O_v(T)$ is calculated as the ratio of the total number of occurrences of $v$ in $T$ to the size of $T$, and
    \item $O_v^{IPM}$ is the occurrence probability of each parameter value $v$ in the IPM, i.e., for each value $v \in \mathcal{V}_i$ of parameter $p_i$, $O_v^{IPM}=1/|\mathcal{V}_i|$ in $U_{IPM}$.
\end{itemize}
\end{definition}

\begin{figure*}[!b]
\graphicspath{{Figures/}}
\centering
\includegraphics[width=\textwidth]{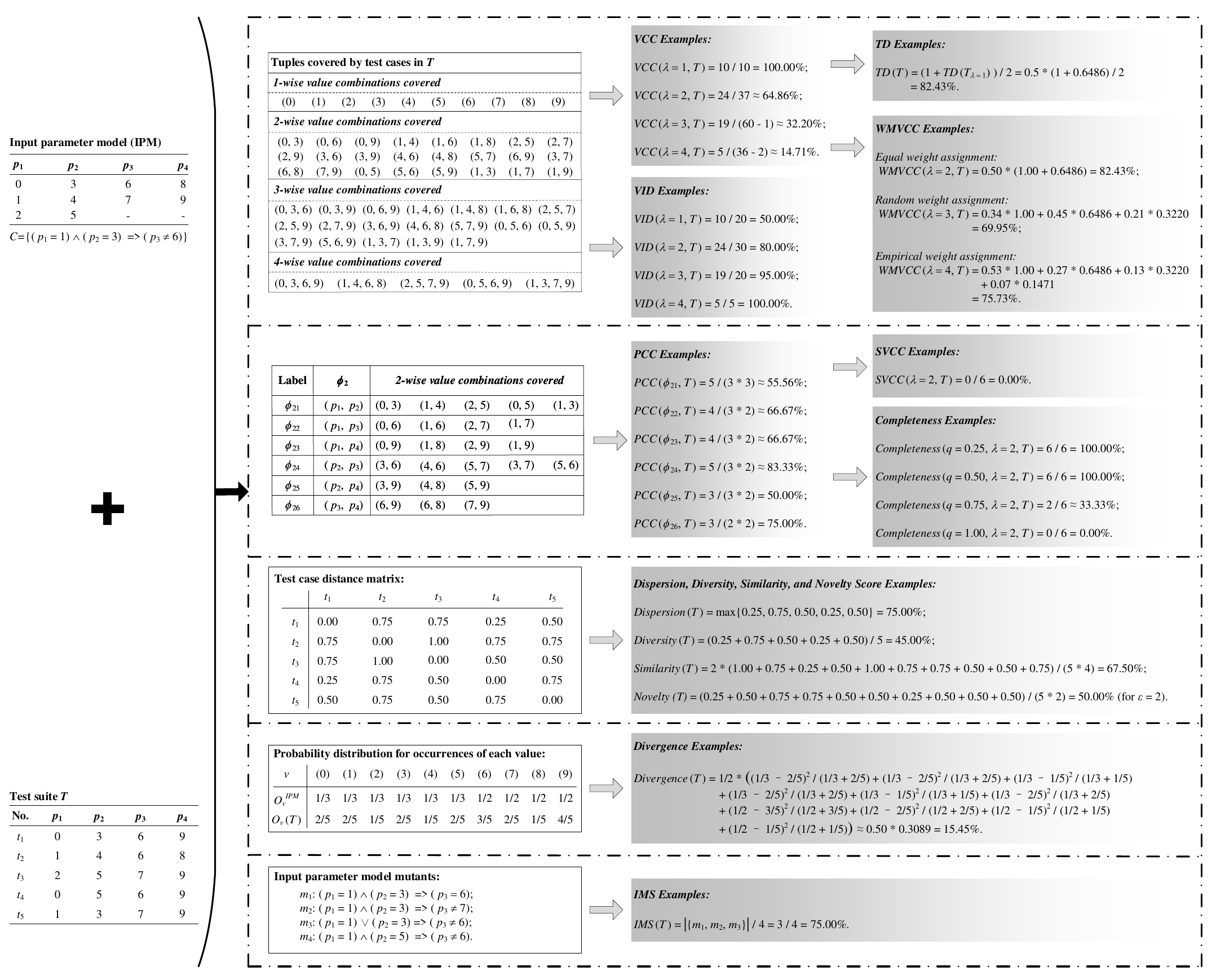}
\caption{An illustrative example for different evaluation metrics.}
\label{FIG:VCC}
\end{figure*}

\subsection{Mutation-Based Metrics}

The \textit{mutation-based metric} (MBM) aims to measure test suites based on mutation analysis~\cite{Jia2011}, which checks the ratio of killed mutants to all non-equivalent mutants (i.e., mutants that are semantically equivalent to the original program under test and can never be killed), namely the \textit{mutation score}.

\begin{definition}[\textbf{Mutation Score}]
Given a test suite $T$ and a set of non-equivalent mutants $\mathcal{M}$ obtained from the subject under mutation, the mutation score of $T$ is defined as:
\begin{equation}
MutationScore(T,\mathcal{M})=\frac{\big|\{\mathcal{M}_i \in \mathcal{M} \mid \mathcal{M}_i \text{ is killed by } T\}\big|}{|\mathcal{M}|}.
\label{Eq.MutationScore}
\end{equation}
\end{definition}

Depending on the subject under mutation, we consider the following two types of mutation score: \textit{program mutation score}, based on program mutation~\cite{Jia2011}; and \textit{input-parameter-model mutation score}, based on IPM mutation~\cite{Papadakis2014}.

\subsubsection{Program Mutation Score}

The \textit{program mutation score} (PMS) aims to measure the percentage of program code faults detected by the test set relative to all injected faults for a particular program~\cite{Jia2011}. For a specific fault $f$ injected into the program $P$, a mutated program $P'$ is produced. Given a test set $T$, if there exists a test case $t \in T$ such that the output of $P'$ differs from the output of $P$ when executing $t$, then fault $f$ is said to be detected by $T$. The PMS is also referred to as the \textit{fault-detection effectiveness} (FDE)~\cite{Wong1995}.

\begin{definition}[\textbf{Program Mutation Score}]
Given a test suite $T$ and a set of program faults $\mathcal{F}$ generated by mutating the SUT, the program mutation score (PMS) of $T$ is defined as:
\begin{equation}
PMS(T,\mathcal{F})=\frac{\big|\{f_i \in \mathcal{F} \mid f_i \text{ is detected by } T\}\big|}{|\mathcal{F}|}.
\label{Eq.PMS}
\end{equation}
\end{definition}

\subsubsection{Input-parameter-model Mutation Score}
The \textit{input-parameter-model mutation}~\cite{Papadakis2014} applies the concept of mutation analysis~\cite{Jia2011} to the IPM, generating defective models (i.e., mutants) by modifying the constraints based on a flattened representation of the IPM. The flattened model is a Boolean version of the original IPM. When a constraint $c$ in the IPM $I$ is replaced by a specific faulty constraint $c'$, a mutated flattened model $I'$ is obtained. Given a test suite $T$, if there exists a test case $t \in T$ such that $t$ violates the mutated constraint $c'$ in $I'$, then $T$ is said to trigger the faulty constraint $c'$.

\begin{definition}[\textbf{Input-parameter-model Mutation Score}]
Given a test suite $T$ and a set of faulty constraints $\mathcal{R}'$ obtained by seeding the constraints $\mathcal{R}$ of the IPM, the input-parameter-model mutation score (IMS) of $T$ is defined as:
\begin{equation}
IMS(T,\mathcal{R}')=\frac{\big|\{r_i \in \mathcal{R}' \mid r_i \text{ is triggered by } T\}\big|}{|\mathcal{R}'|}.
\label{Eq.IMS}
\end{equation}
\end{definition}

Consider an IPM, for example, $IPM(\mathcal{P}=\{p_1,p_2,p_3\}, \mathcal{V}=\{\{c, d\},\{x, y\},\{z, w\}\}, \mathcal{R}=\{(p_1=c) \Rightarrow (p_2 \neq y)\})$. The flattened model involves six Boolean variables instead of three, corresponding to the values of the parameters: $c_{p_1}$, $d_{p_1}$, $x_{p_2}$, $y_{p_2}$, $z_{p_3}$, and $w_{p_3}$. Each of these variables is Boolean, i.e., it can take the values $true$ (labeled $1$) or $false$ (labeled $0$). Additionally, since the parameters $c_{p_1}$ and $d_{p_1}$ cannot both be assigned $1$, the following constraints exist: $c_{p_1} \Rightarrow \neg d_{p_1}$, and $d_{p_1} \Rightarrow \neg c_{p_1}$. As a consequence, the Boolean flattened model $IPM_b$, which is equivalent to $IPM(\mathcal{P},\mathcal{V},\mathcal{R})$, can be defined as:
$IPM_b(\mathcal{P}=\{c_{p_1},d_{p_1},x_{p_2},y_{p_2},z_{p_3},w_{p_3}\}, \mathcal{V}=\{\{1, 0\},\{1, 0\},\{1, 0\},\{1,0\},\{1, 0\},\{1, 0\}\}, \mathcal{R}=\{
(c_{p_1} \Rightarrow \neg d_{p_1}), (d_{p_1} \Rightarrow \neg c_{p_1}),
(x_{p_2} \Rightarrow \neg y_{p_2}), (y_{p_2} \Rightarrow \neg x_{p_2}),
(z_{p_3} \Rightarrow \neg w_{p_3}), (w_{p_3} \Rightarrow \neg z_{p_3}),
(c_{p_1} \Rightarrow \neg y_{p_2})\})$.
After that, the mutated models (i.e., mutants) can be constructed by altering one constraint; for example, the constraint $(x_{p_2} \Rightarrow \neg y_{p_2})$ can be mutated to $(x_{p_2} \Rightarrow y_{p_2})$, or the constraint $(c_{p_1} \Rightarrow \neg y_{p_2})$ can be mutated to $(c_{p_1} \Rightarrow y_{p_2})$.

\begin{example}[\textbf{Evaluation Metrics Examples}]
Given the IPM shown in Table~\ref{Table1} and a test suite $T_1=\{t_1,t_2,t_3,t_4,t_5\}$, where $t_i~(i=1,2,3,4,5)$ is a test case listed in Table~\ref{Table2}, Figure~\ref{FIG:VCC} provides a detailed calculation process for each evaluation metric (except for PMS, as it requires detailed program information).
\end{example}

In effect, only PMS can be considered a \textit{dynamic} evaluation metric, as it requires the SUT to be executed in real time. In contrast, the remaining evaluation metrics are \textit{static}, i.e., they do not require execution of the SUT.

\subsection{Complexity Analysis for Calculating Static Evaluation Metrics}

In this section, we provide a brief analysis of the complexity of calculating static evaluation metrics, including both space and time complexity. The primary reason for focusing solely on static evaluation metrics is that they can be computed directly from test suites and IPMs, without considering the SUT. Before the analysis, we first introduce the basic data structure for storing and searching $\lambda$-wise value combinations used in the evaluation metrics.

Consider the $IPM(\mathcal{P},\mathcal{V},\mathcal{R})$, where $\mathcal{P}=\{p_1,p_2,\cdots,p_k\}$ and $\mathcal{V}=\{\mathcal{V}_1,\mathcal{V}_2,\cdots,\mathcal{V}_k\}$. We assume that $\delta=\max_{i=1}^k{|\mathcal{V}_i|}$, a test suite under evaluation $T=\{t_1,t_2,\cdots,t_n\}$ (where $T_\lambda$ represents a $\lambda$-wise CA), and the strength $\lambda$.

\subsubsection{Storing and Searching $\lambda$-wise Value Combinations}

Following previous studies~\cite{Lei2008}, we employ a similar data structure—a two-layer hierarchy—to store $\lambda$-wise value combinations. More specifically, the first layer is an array of pointers, denoted as $\mathcal{H}$, each of which corresponds to one $\lambda$-wise parameter combination and points to another integer array at the second layer. When searching for a $\lambda$-wise value combination in the data structure, it is necessary to search the first layer to locate the corresponding $\lambda$-wise parameter combination, and then search the second layer to locate the specific $\lambda$-wise value combination.

The index of each $\lambda$-wise parameter combination is determined by lexicographical order in $\mathcal{H}$. For example, consider a parameter list $\mathcal{P}=\{p_1,p_2,p_3\}$ and $\lambda=2$. There are three 2-wise parameter combinations: $(p_1, p_2)$, $(p_1, p_3)$, and $(p_2, p_3)$. According to lexicographical order, $(p_1, p_2)$ ranks first, followed by $(p_1, p_3)$ and $(p_2, p_3)$. Therefore, the index of $(p_1, p_2)$ is $0$ in $\mathcal{H}$, while the indices of $(p_1, p_3)$ and $(p_2, p_3)$ are $1$ and $2$, respectively. The index of each $\lambda$-wise parameter combination in $\mathcal{H}$ can be calculated directly.

At the second layer, each $\lambda$-wise parameter combination $\phi_\lambda$ points to an integer array $\mathcal{H}_{\phi_\lambda}$, where each entry represents the number of times a corresponding $\lambda$-wise value combination derived from $\phi_\lambda$ has been covered. For example, a value of $0$ indicates that the corresponding $\lambda$-wise value combination has not yet been covered, while a positive integer $x~(x > 0)$ indicates that it has been covered $x$ times. Similarly, the indices of all $\lambda$-wise value combinations derived from $\phi_\lambda$ are determined by lexicographical order, which means their positions in the integer array can also be calculated directly. In other words, once the pointer to a $\lambda$-wise parameter combination is known, the position of the corresponding $\lambda$-wise value combination can be quickly located.

As we know, different test cases cover the same $\lambda$-wise parameter combinations, since each parameter must be assigned a specific value. For each $\lambda$-wise parameter combination, a unique $\lambda$-wise value combination is covered by any test case $t$. In other words, when storing and searching $\lambda$-wise value combinations, each test case requires checking all $\lambda$-wise parameter combinations at the first layer $\mathcal{H}$, and then checking the corresponding $\lambda$-wise value combinations at the second layer one by one.

\subsubsection{Space Complexity Analysis}
The space complexity of each evaluation metric is determined by the following two factors: 1) the number of test cases, $n$; and 2) the number of $\lambda$-wise value combinations derived from the IPM. For factor 1), since each test case covers $\lambda$-wise value combinations of size $\mathrm{C}(k,\lambda)$, the space complexity is $O_1 = n \times k + n \times \mathrm{C}(k,\lambda)$. For factor 2), the space complexity is determined by the data structure $\mathcal{H}$. As discussed earlier, since $\mathcal{H}$ contains two layers, the space complexity for factor 2) consists of two parts. More specifically, the first layer of $\mathcal{H}$ requires space complexity $O_2 = O(\mathrm{C}(k,\lambda))$ to store $\lambda$-wise parameter combinations of size $\mathrm{C}(k,\lambda)$; while the second layer of $\mathcal{H}$ must account for the space complexity of all $\lambda$-wise value combinations, i.e.,
\begin{align}
\label{EQ:Q3}
    O_3 =~&O\Bigg(\sum_{1\leq i_1 < i_2 < \cdots < i_\lambda \leq k}(|\mathcal{V}_{i_1}||\mathcal{V}_{i_2}|\cdots|\mathcal{V}_{i_\lambda}|)\Bigg) \nonumber \\
    <~&O\Bigg( \sum_{1\leq i_1 < i_2 < \cdots < i_\lambda \leq k} \big(\delta^\lambda\big)\Bigg) \nonumber \\
    =~&O\Big(\mathrm{C}(k,\lambda)\times\delta^\lambda\Big).
\end{align}
As a result, the space complexity for storing $n$ test cases and the data structure $\mathcal{H}$ is
\begin{align}
\label{EQ:Ospace}
    O_{space}^\lambda =&~O_1 + O_2 + O_3 \nonumber \\
    =&~O\Big(n \times k + n\times \mathrm{C}(k,\lambda)\Big) + O\Big(\mathrm{C}(k, \lambda)\Big) + O\Big(\mathrm{C}(k,\lambda)\times\delta^\lambda\Big) \nonumber \\
    =&~O\Big(\mathrm{C}(k,\lambda) \times \big( n + 1 + \delta^\lambda \big) + n \times k\Big) \nonumber \\
    =&~O\Big(\mathrm{C}(k,\lambda) \times \big(n + \delta^\lambda\big)\Big).
\end{align}
Since different metrics may utilize different information to guide their evaluations, it is necessary to discuss their space complexities individually.

\begin{enumerate}
    \item ICBM Category: Different $\lambda$ values are used in the evaluation metrics of the ICBM category, so we discuss them respectively.
    \begin{itemize}
          \item Since $VCC(\lambda,T)$, $VID(\lambda,T)$, $BVCC(b,T_\lambda)$, and $TD(T_\lambda)$ only need to check $\lambda$-wise value combinations covered by $T$ or $T_\lambda$, their space complexities are equal to $O_{space}^\lambda$, as shown in Eq.~(\ref{EQ:Ospace}).
          
          \item Since $WMVCC(\lambda,T)$ requires consideration of all strength values ranging from $1$ to $\lambda$, its space complexity is equal to $O_{space}^1 + O_{space}^2 + \cdots + O_{space}^\lambda = O(\sum_{i=1}^\lambda(\mathrm{C}(\lambda,i)\times (n+\delta^i)))$.
          
          \item Since $PCC(\phi_\lambda,T)$ only focuses on the $\lambda$-wise parameter combination $\phi_\lambda$, its space complexity can be described as $O(n \times k + n + \delta^\lambda) = O(n \times k + \delta^\lambda)$.
          
          \item For $SVCC(\lambda, T)$ and $Completeness(q,\lambda, T)$, they are similar to $VCC(\lambda, T)$ in that they require storing all $\lambda$-wise value combinations. In other words, their space complexities are equal to $O_{space}^\lambda$.
      \end{itemize}

    \item DBM Category: Since all evaluation metrics in the DBM category are based on test case distribution, they do not need to check whether $\lambda$-wise value combinations are covered or not, which means that it is not necessary to construct the data structure $\mathcal{H}$. However, they use different information to guide the evaluation; therefore, we discuss them separately here.
      \begin{itemize}
          \item According to the definition of $Discrepancy(T)$, a random subdomain $\mathcal{D}_i$ must be constructed from the entire test space $\mathcal{D}$, where $i = 1,2,\cdots, m$ ($m$ is a pre-defined constant). Apart from the test cases in $T$, the elements in $\mathcal{D}_i$ also need to be stored, resulting in space complexity $O(s \times |\mathcal{D}_i|)$, where $s$ is the length of each element. As we know, the maximum length of $s$ is equal to $k$, while the maximum size of $|\mathcal{D}_i|$ is equal to $|\mathcal{D}|$. Therefore, the space complexity of $Discrepancy(T)$ can be described as:
          \begin{align}
              &O(k \times n) + O(s \times |\mathcal{D}_i|) \nonumber \\
              <~& O(k\times n + k \times |\mathcal{D}_i|) \nonumber\\
              <~& O\Big(k \times n + k \times \delta^k\Big) \nonumber\\
              =~&O\Big(k \times \big(n + \delta^k\big)\Big).
          \end{align}

          \item Since $Dispersion(T)$, $Diversity(T)$, and $Similarity(T)$ only need to store the $n$ test cases of $T$, their space complexities can be written as $O(k \times n)$.

          \item $Novelty(T)$ not only needs to store $n$ test cases, but also needs to store the $\varepsilon$ nearest neighbors. Therefore, its space complexity is $O(k \times n + \varepsilon)$.

          \item $Divergence(T)$ differs from the above four metrics of the DBM category, as it needs to collect the number of times each $\lambda$-wise value combination is covered by $T$. However, it only emphasizes strength $\lambda=1$. As a result, its space complexity is equal to $O_{space}^{\lambda=1} = O(k \times (n + \delta))$.
      \end{itemize}

    \item MBM Category: Here, we only focus on the IMS from the MBM category. In addition to storing $n$ test cases, a set of faulty constraints $\mathcal{R}'$ also needs to be stored. As we know, the maximum number of parameters involved in each faulty constraint in $\mathcal{R}'$ is equal to $k$, so its space complexity is less than or equal to $O(n \times k + |\mathcal{R}'| \times k) = O(k \times (n + |\mathcal{R}'|))$.
\end{enumerate}

\subsubsection{Time Complexity Analysis}
\label{SECTION:Time}

Consider a specific strength value $\lambda$. The computational time of each evaluation metric can be divided into the following two parts: (1) the time to construct the two-layer data structure $\mathcal{H}$; and (2) the time to search and update the covered times of each $\lambda$-wise value combination by checking test cases in $T$ one by one. For part (1), the corresponding time complexity is $O'_1 = O(\mathrm{C}(k,\lambda) \times \delta^\lambda)$, according to Eq.~(\ref{EQ:Q3}).
For part (2), each test case in $T$ requires checking and storing a set of its covered $\lambda$-wise value combinations of size $\mathrm{C}(k,\lambda)$, and this check takes $O(1)$ time for each $\lambda$-wise value combination. Therefore, the time complexity for part (2) is $O'_2 = O(n \times \mathrm{C}(k,\lambda))$. By combining parts (1) and (2), the time complexity can be described as:
\begin{align}
\label{EQ:Otime}
    O_{time}^\lambda = &~O'_1 + O'_2 \nonumber\\
    =&~O\Big(\mathrm{C}(k,\lambda) \times \delta^\lambda\Big) + O\Big(n \times \mathrm{C}(k,\lambda)\Big) \nonumber\\
    =&~O\Big(\mathrm{C}(k,\lambda) \times \big(n + \delta^\lambda \big)\Big).
\end{align}
As with space complexity, since different $\lambda$ values are used for different evaluation metrics, we discuss the time complexity of each metric within its respective category.

\begin{enumerate}
    \item ICBM Category:
      \begin{itemize}
          \item As discussed above, $VCC(\lambda,T)$, $VID(\lambda,T)$, $BVCC(b,T_\lambda)$, and $TD(T_\lambda)$ only focus on strength $\lambda$, so their time complexities are the same, i.e., $O_{time}^\lambda$, as shown in Eq.~(\ref{EQ:Otime}).
          
          \item $WMVCC(\lambda, T)$ considers different strength values ranging from $1$ to $\lambda$, so its time complexity can be described as $\sum_{i=1}^\lambda O_{time}^i$.

          \item $PCC(\phi_\lambda,T)$ only checks all $\lambda$-wise value combinations derived from the $\lambda$-wise parameter combination $\phi_\lambda$; its time complexity is $O'_1 + O(n) = O(n + \mathrm{C}(k,\lambda) \times \delta^\lambda)$.

          \item Since $SVCC(\lambda, T)$ and $Completeness(q,\lambda, T)$ require checking all $\lambda$-wise parameter combinations, their time complexities are the same as that of $VCC(\lambda,T)$, i.e., $O_{time}^\lambda$.
      \end{itemize}

    \item DBM Category: As with space complexity, we discuss different DBM metrics across the following subcategories.
      \begin{itemize}
          \item $Discrepancy(T)$ needs to spend time constructing random subdomains $\mathcal{D}_i~(i=1,2,\cdots,m)$ and searching for elements of $T$ that are located in $\mathcal{D}_i$. The subdomain construction process requires time complexity $O(k \times \delta^k)$ at most, while the element search process requires time complexity $O(n \times k \times \delta^k)$ at most. Therefore, the time complexity of $Discrepancy(T)$ is $O(m \times (k \times \delta^k + n \times k \times \delta^k)) = O(m \times n \times k \times \delta^k)$ at most.
          
          \item Since $Dispersion(T)$, $Diversity(T)$, and $Similarity(T)$ need to calculate the Hamming distance between any two test cases in $T$, their time complexities can be calculated as:
          \begin{align}
               &O\Bigg(\sum_{i=1}^{n-1}\big(k \times (n-i) \big)\Bigg) \nonumber \\
              =~&O\Bigg(k \times \sum_{i=1}^{n-1}(n-i)\Bigg) \nonumber \\ 
              =~&O\bigg(k \times \frac{n \times (n-1)}{2}\bigg) \nonumber \\
              =~&O\big(k \times n^2 \big).
          \end{align}

          \item $Novelty(T)$ not only needs to calculate the distance between any two test cases in $T$, but also needs to find the top-$\varepsilon$ neighbors of each test case from $T$. For each test case $t \in T$, the time complexity of searching the $\varepsilon$-nearest neighbors is $O(\log \varepsilon)$ when using a heap data structure of size $\varepsilon$. Therefore, the time complexity of $Novelty(T)$ can be described as:
          \begin{align}
               &O\Bigg(\sum_{i=1}^{n-1}\big(k \times (n-i) + \log \varepsilon \big)\Bigg) \nonumber \\
              =~&O\Bigg(k \times \sum_{i=1}^{n-1}(n-i)\Bigg) + O\Bigg(\sum_{i=1}^{n-1}\log \varepsilon\Bigg) \nonumber \\ 
              =~&O\bigg(k \times \frac{n \times (n-1)}{2}\bigg) + O\big((n-1) \times \log\varepsilon \big) \nonumber \\
              =~&O\big(k \times n^2 + n \times \log\varepsilon \big).
          \end{align}
         
          \item Since $Divergence(T)$ needs to collect the coverage counts of each $1$-wise value combination, its time complexity includes $O_{time}^{\lambda=1}$ (as shown in Eq.~(\ref{EQ:Otime})). In addition, the time complexity also includes the calculation of $Divergence(T)$, i.e., $O(\sum_{i=1}^k{|\mathcal{V}_i|})$. Therefore, the time complexity of $Divergence(T)$ is:
          \begin{align}
              &O_{time}^{\lambda=1} + O\Bigg(\sum_{i=1}^k{|\mathcal{V}_i|}\Bigg) \nonumber \\
              =~&O(k\times (n + \delta)) + O\Bigg(\sum_{i=1}^k{|\mathcal{V}_i|}\Bigg) \nonumber \\
              <~&O(k\times (n + \delta) + k \times \delta) \nonumber\\
              =~&O(k \times (n + \delta)).
          \end{align}
      \end{itemize}

    \item MBM Category: The metric $IMS(T,\mathcal{R}')$ needs to check whether each faulty constraint in $\mathcal{R}'$ can be triggered by $T$ or not. Each faulty constraint $r \in \mathcal{R}'$ is checked by $1$ to $n$ test cases in $T$.
    Therefore, the time complexity of $IMS(T,\mathcal{R}')$ can be described as:
         \begin{align}
              &O\Bigg(\sum_{r\in \mathcal{R}'} \frac{1+2+\cdots+n}{n}\Bigg) \nonumber \\
              =~&O\Bigg(\frac{n \times (n+1)}{2n} \times \sum_{r\in \mathcal{R}'}|r|\Bigg) \nonumber \\
              <~&O\Bigg(\frac{n^2+n}{2n} \times \sum_{r\in \mathcal{R}'}k\Bigg) \nonumber \\
              =~&O\Bigg(\frac{n^2+n}{2n} \times |\mathcal{R}'| \times k\Bigg) \nonumber \\
              =~&O(n \times |\mathcal{R}'| \times k).
          \end{align}
\end{enumerate}

Table~\ref{TABLE:Complexity} summarizes the space and time complexity for each evaluation metric, where $n$ is the number of test cases in $T$, while $k$, $\lambda$, $\delta$, $b$, $m$, $\varepsilon$, and $|\mathcal{R}'|$ are constant integers.

\begin{table}[!b]
\centering
\footnotesize
\caption{Summary of Space and Time Complexity for Each Evaluation Metric}
\label{TABLE:Complexity}
\setlength{\tabcolsep}{3.7mm}{\begin{tabular}{c|l|l|l} \hline

\textbf{Category} &\textbf{Evaluation metric}  &\textbf{Space complexity}      & \textbf{Time Complexity} \\ \hline
\multirow{8}*{ICBM}
&$VCC(\lambda, T)$ &$O(C(k,\lambda) \times (n + \delta^\lambda))$ &$O(C(k,\lambda) \times (n + \delta^\lambda))$ \\
&$VID(\lambda, T)$ &$O(C(k,\lambda) \times (n + \delta^\lambda))$ &$O(C(k,\lambda) \times (n + \delta^\lambda))$ \\
&$BVCC(b, T_\lambda)$ &$O(C(k,\lambda) \times (n + \delta^\lambda))$ &$O(C(k,\lambda) \times (n + \delta^\lambda))$ \\
&$TD(T_\lambda)$ &$O(C(k,\lambda) \times (n + \delta^\lambda))$ &$O(C(k,\lambda) \times (n + \delta^\lambda))$ \\
&$WMVCC(\lambda, T)$ &$O(\sum_{i=1}^\lambda(C(\lambda,i)\times (n+\delta^i)))$ &$O(\sum_{i=1}^\lambda(C(\lambda,i)\times (n+\delta^i)))$ \\
&$PCC(\phi_\lambda,T)$ &$O(n \times k + \delta^\lambda)$ &$O(n+C(k,\lambda) \times \delta^\lambda)$ \\
&$SVCC(\lambda, T)$ &$O(C(k,\lambda) \times (n + \delta^\lambda))$ &$O(C(k,\lambda) \times (n + \delta^\lambda))$ \\
&$Completeness(q,\lambda, T)$ &$O(C(k,\lambda) \times (n + \delta^\lambda))$ &$O(C(k,\lambda) \times (n + \delta^\lambda))$ \\ \hline

\multirow{6}*{DBM} 
&$Discrepancy(T)$ &$(k \times \Big(n+\delta^k))$ &$O(m \times n\times k \times \delta^k)$ \\
&$Dispersion(T)$ &$O(k\times n)$ &$O(k \times n^2 )$ \\
&$Diversity(T)$ &$O(k\times n)$ &$O(k \times n^2 )$ \\
&$Similarity(T)$ &$O(k\times n)$ &$O(k \times n^2 )$ \\ 
&$Novelty(T)$ &$O(k\times n + \varepsilon)$ &$O(k \times n^2 + n \times \log\varepsilon)$\\
&$Divergence(T)$ &$O(k\times (n + \delta))$ &$O(k \times (n + \delta))$ \\ \hline

MBM
&$IMS(T,\mathcal{R}')$ &$O(k \times (n+|\mathcal{R}'|))$ &$O(n \times |\mathcal{R}'| \times k)$ \\\hline
\end{tabular}
}
\end{table}

\section{Experimental Methodology}
\label{SECTION:Design}
In this section, we present the basic experimental methodology, including the research questions, the overall framework, subject programs, software faults, test scenarios, test suites, selected evaluation metrics, and correlation analysis. 

\subsection{Research Questions}
For each test suite $T$, its quality primarily depends on its fault-detection capability; i.e., the better the fault detection, the higher the quality. However, the fault-detection capability of $T$ requires executing $T$ on a given program, which generally entails additional testing resources and effort. As we know, static evaluation metrics do not require executing $T$, which means they are time- and resource-saving. If $T$ with higher values measured by a static evaluation metric $\mathcal{M}$ can produce better fault detection capabilities, $S$ will be very useful as an evaluation metric. Therefore, it is interesting to investigate the correlation between each static evaluation metric and fault detection capability (i.e., PMS). In addition to testing effectiveness, it is important to study testing efficiency, i.e., the time required per evaluation metric. Based on the above analysis, we design the following five research questions (RQs):
\begin{itemize}
    \item \textbf{RQ1:} \textit{Among ICBMs, which one is more correlated to fault-detection effectiveness when evaluating test suites?}
    
    \begin{itemize}
        \item \textbf{RQ1.1:} \textit{Which strength is most suitable for single-strength ICBMs?}
        \item \textbf{RQ1.2:} \textit{Do multiple-strength ICBMs perform better than single-strength ICBMs?}
    \end{itemize}
    
    \textbf{RQ1} is to find the best one among the ICBM category. More specifically, \textbf{RQ1.1} aims at investigating the appropriate strength value for single-strength ICBMs, as different strengths may result in various performances. \textbf{RQ1.2} aims at comparing multiple-strength and single-strength ICBMs, because it is generally believed that the multiple-strength mechanism would be better than the single-strength mechanism.
    
    \item \textbf{RQ2:} \textit{Among DBMs, which one can better reflect the correlation to fault-detection effectiveness for test suite evaluation?}
    
    \textbf{RQ2} aims to identify the best one from the DBM category. As is well known, different test case distribution measures capture distinct perspectives on test set evaluation. This research question investigates the relationship between test case distribution and evaluation.
    
    \item \textbf{RQ3:} \textit{Do higher IMS results indicate higher values of fault-detection effectiveness?}

\textbf{RQ3} aims to identify the correlation between IMS and fault-detection effectiveness.
   
    Both IMS and PMS belong to the MBM category; however, IMS modifies the IPM, whereas PMS modifies the source code. They apply the mutation mechanism to different subjects, so it is interesting to examine whether their performance is consistent across subjects.
    
    \item \textbf{RQ4:} \textit{For the best from each category, which one is the final best?}
    
    \textbf{RQ4} aims to identify the best metric among the investigated metrics. Although evaluation metrics may belong to different categories, testers only consider which to select in practice. So, it's necessary to select the best final metric from the set of evaluation metrics.

    \item \textbf{RQ5:} \textit{How efficient does each metric perform?}

    \textbf{RQ5} aims at checking the computational cost of each evaluation metric. As we know, testing effectiveness (i.e., fault-detection effectiveness) is an important aspect that reflects the performance of each metric; however, testing efficiency (i.e., computational time) is also crucial, especially in resource-constrained scenarios. Although the time complexity has been analyzed in Section \ref{SECTION:Time}, it is also interesting to examine the actual computational time required for each evaluation metric in practice.
    
\end{itemize}

\begin{figure*}[!t]
\graphicspath{{Figures/}}
\centering
\includegraphics[width=0.95\textwidth]{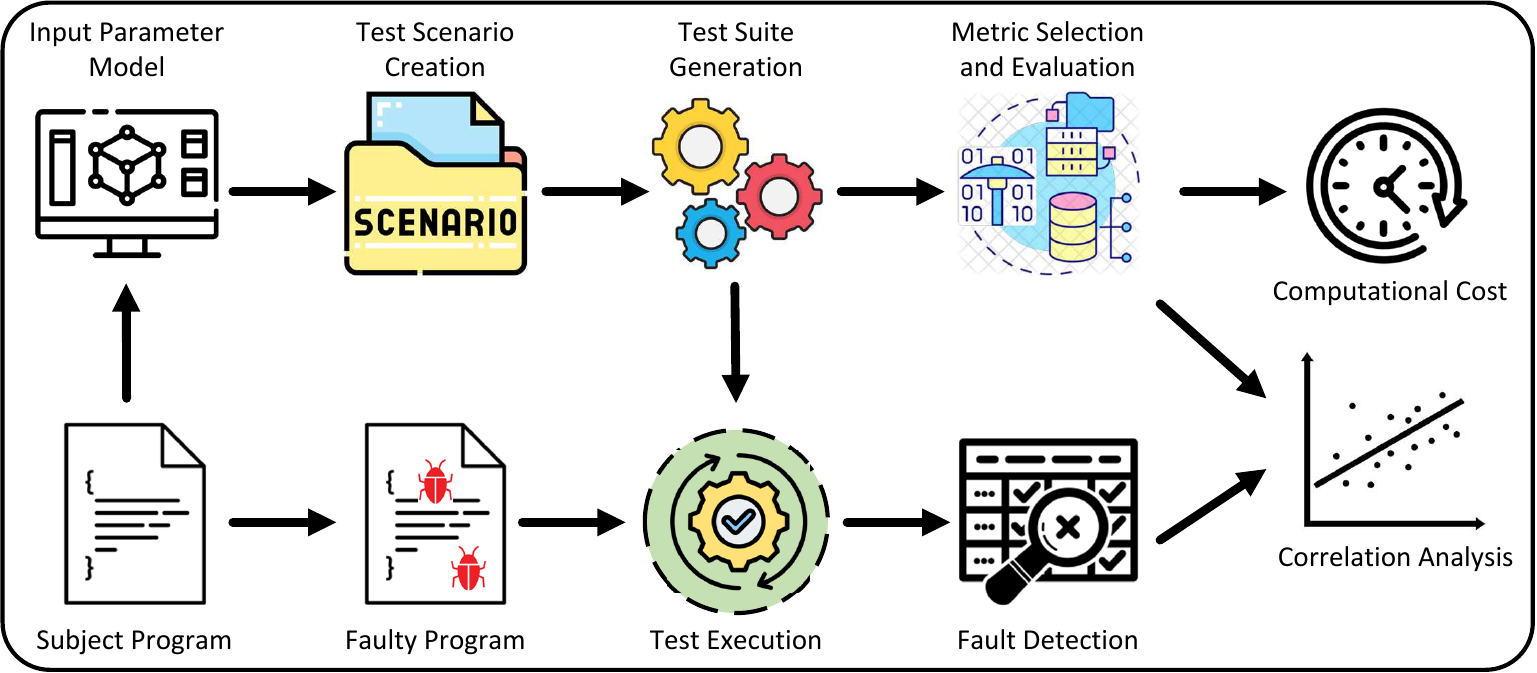}
\caption{The basic framework of the experiments.}
\label{FIG:EM}
\end{figure*}

\subsection{Basic Framework}
\label{SECTION:Basic Framework}
To address the research questions above, each evaluation metric is applied to different test suites. Subsequently, a correlation analysis is conducted between the evaluation metrics to assess their degree of association.

\begin{table}[!b]
\centering
\scriptsize
\caption{Subject Programs}
\label{TABLE:Programs}
\setlength{\tabcolsep}{0.6mm}
{\begin{tabular}{lllrrr} \hline
\textbf{Program} &\textbf{Brief Description}  & \textbf{IPM} &\textbf{LOCs}~\cite{Henard2016,Wu2020}      &\textbf{Faults}  &\textbf{Test Suites}\\ \hline
FLEX &Fast lexical analyzer generator &$IPM(2^63^25^1,\mathcal{R}),|\mathcal{R}|=12$ &  67,654 &149 &32,400\\
GREP &Text searching and processing utility   &$IPM(2^13^34^25^16^18^1,\mathcal{R}),|\mathcal{R}|=83$ &124,833 &285 &30,200\\
GZIP &File compression and decompression utility &$IPM(2^{13}3^1,\mathcal{R}),|\mathcal{R}|=61$ &29,812 &37 &33,600\\
MAKE &Build automation tool for software projects &$IPM(2^{10},\mathcal{R}),|\mathcal{R}|=1$    &122,680  &151 &31,200\\
SED &Stream editor for text parser and transformation &$IPM(2^73^14^16^110^1,\mathcal{R}),|\mathcal{R}|=50$ &101,024 &93 &25,424\\

BUSYBOX &A collection of Unix utilities &$IPM(2^{68},\mathcal{R}),|\mathcal{R}|=16$ &189,722 &9 &39,200\\
DRUPAL &Web content management system &$IPM(2^{47},\mathcal{R}),|\mathcal{R}|=45$ &336,025 &160 &39,200\\
LINUX KERNEL &Component of the Linux operating system &$IPM(2^{104},\mathcal{R}),|\mathcal{R}|=83$ &12,594,584  &28 &39,200\\\hline
\multicolumn{3}{c}{\textbf{\textit{Total}}} &13,566,384	&912	&270,424\\\hline

\end{tabular}
}
\end{table}
 
Figure~\ref{FIG:EM} shows the basic framework of the experiments. In detail, for a given subject program under test: 
(1) the IPM is extracted; 
(2) different test scenarios are created from the IPM; 
(3) various construction approaches generate test suites; 
(4) a static metric is selected and applied to each test suite for collecting the metric value and computational cost 
 within 24 hours (a constraint imposed to ensure computational efficiency); 
(5) each test suite is also executed to detect software faults; and 
(6) a correlation analysis is finally conducted between fault detection and metric evaluation. In the following subsections, we provide further details on the steps outlined above.

\subsection{Subject Programs}

Following previous studies~\cite{Henard2016,Wu2020}, we utilized eight subject programs in this study: FLEX, GREP, GZIP, MAKE, SED, BUSYBOX, DRUPAL, and LINUX KERNEL. These subject programs have also been popularly used in recent CIT studies~\cite{Luo2024a,Luo2025}. Table~\ref{TABLE:Programs} provides the details of these subject programs.

The first five programs, collected from the GNU FTP server~\cite{gnuftp}, are real-world, open-source command-line utilities. Each of them has six versions. Specifically, FLEX is a fast lexical analyzer generator used for generating lexical analyzers; GREP is a popular utility for searching plain-text data for lines that match a regular expression; GZIP is a command-line utility used for file compression and decompression; MAKE is a tool for managing and controlling the compilation and build process of programs; and SED is a stream editor that performs text transformations on input streams.

The last three programs are real-world, open-source, more complex systems due to their high configurability. BUSYBOX is a software package that provides a collection of UNIX utilities within a single executable file~\cite{Sanchez2017}; DRUPAL is an open-source content management system used for building and managing websites and web applications~\cite{Medeiros2016}; and LINUX KERNEL is the core component of the Linux operating system, responsible for managing hardware resources and providing essential services to other software~\cite{Medeiros2016}. Compared with the SIR programs, these three programs are larger in scale and have more complex IPMs (with more parameters).

For each subject program, IPMs can be constructed using different strategies. Consider the subject program GREP, for example. Qu et al.~\cite{Qu2013} used the $IPM(2^23^24^212^1,\mathcal{R}=\emptyset)$ with no constraints, i.e., $|\mathcal{R}|=0$. However, Petke et al.~\cite{Petke2015} constructed the $IPM(2^13^34^25^16^18^1,\mathcal{R})$, where $|\mathcal{R}|=83$; while Ghandehari et al.~\cite{Ghandehari2018} extracted the $IPM(2^74^15^16^38^19^113^1,\mathcal{R})$, where $|\mathcal{R}|=1$. To avoid potential bias, this study used the same IPMs for the nine subject programs as in previous studies~\cite{Wu2020}. Table~\ref{TABLE:Programs} presents the program name, brief description, IPM, and lines of code (LOCs) in the first four columns, respectively.

\subsection{Faults Used}
{
To evaluate the fault-detection effectiveness of each test suite, we utilize subject programs with simulated faults or real faults, allowing us to measure the number of faults detected by the suites. 
The fifth column of Table \ref{TABLE:Programs} provides the number of faults for each program. 
The default faults provided by the SIR repository exhibit a high average detection rate of 60\%, suggesting that they are too trivial to rigorously challenge the test suites. To overcome this limitation, we utilized the mutation faults verified in a prior study ~\cite{henard2016comparing} for the five SIR subjects. We then aggregated these mutants for each version $V_i$ ($1 \le i \le 5$) to form a robust fault evaluation set for each program.
The mutation operators considered in this study consist of relational, logical, arithmetic, bitwise logical, and constant replacement operators, statement deletion, and unary insertion operators.
To mitigate potential threats to validity, we used the Trivial Compiler Equivalence tool~\cite{papadakis2015trivial} to eliminate equivalent mutants~\cite{papadakis2016threats} and duplicated mutants~\cite{papadakis2015trivial}, and removed subsumed mutants~\cite{jia2009higher} by abstract test cases.

For the last three subject programs (BUSYBOX, DRUPAL, and LINUX KERNEL), we adopted an existing corpus of real faults from previous studies~\cite{Sanchez2017,Medeiros2016,Wu2020}, where each fault corresponds to certain value combinations, and a fault can be detected by test cases that cover at least one of those value combinations. As discussed in~\cite{Sanchez2017,Medeiros2016}, these faults were collected from bug-tracking systems, and each had been previously reported, confirmed, and fixed by the original developers. Compared with simulated faults in the SIR programs, the faults used in these three programs are real-world.

\subsection{Test Scenarios}
\label{subsection:Test Scenarios}

To collect sufficient data, we consider various test scenarios for each subject program, adjusting different factors to create diverse IPMs and evaluate multiple metrics. Each test scenario can be regarded as an individual subject program.

Given a subject program, test suite construction is based on the IPM, which involves parameters, values, and constraints. Following previous studies~\cite{Wu2020}, we controlled parameter selection to simulate scenarios in which not all parameters are collected. For each IPM with $k$ parameters, we defined the proportion of parameters as $para$, i.e., we created a series of new IPMs by selecting the first $\lceil para \times k \rceil$ parameters. In this study, $para$ was set to 0.4, 0.6, 0.8, and 1.0. As a result, there are 36 test scenarios across nine subject programs. When $para < 1.0$, indicating that not all parameters were selected, we used default parameter values to construct complete test cases for subsequent evaluation.

As previously discussed, some parameters were omitted for each IPM with $para < 1.0$, and default values were used to fill in the missing parameters. Since non-default values of missing parameters may cause faults, such faults may not be detectable under such an IPM, meaning that not every test case in the test space can detect all faults. For each IPM, we extracted the detectable faults from the set of all faults, yielding the fault set $\mathcal{F}$ for evaluation.

Consequently, each test scenario can be described by a combination $\langle para, \mathcal{F}\rangle$ for each subject program, resulting in four test scenarios per program. For ease of description, we use the term ``\textit{Program Name}-TS$i$'' (where $i = 1, 2, 3, 4$) to label these four test scenarios for each subject program. For example, ``FLEX-TS1'' denotes the first test scenario with $para=0.4$ for the program FLEX, while ``GREP-TS4'' denotes the last test scenario with $para=1.0$ for the program GREP.

\subsection{Test Suite Construction}

The focus of this study is to evaluate the performance of various metrics for assessing CIT test suites, so that more diverse test suites would provide more sufficient data for comparison. To achieve diversity in the test suites, this study focuses on two distinct approaches for test suite construction: \textit{random construction} (RC) and \textit{adaptive random construction} (ARC). In line with prior work~\cite{Hassan2013,Papadakis2014,Gligoric2015}, the upper bound on the number of test suites is set to 100 per approach, providing a sufficient sample size for statistical analysis while keeping construction time manageable.

\subsubsection{Random Construction (RC)}

The random construction generates test cases of a specified size. We randomly generate 100 test suites of varying sizes (2 to 50), yielding approximately 4,900 test suites per test scenario. The minimum size of 2 test cases per test suite was chosen because some metrics in the DBM category (such as dispersion, diversity, and similarity) require at least a pair of test cases to calculate distances. Specifically, for each test scenario $\langle para, \mathcal{F}\rangle$, a random test case is constructed by assigning a random value to each parameter while satisfying the constraints. If the total number of valid test cases is less than 50, the construction process halts once all valid test cases have been constructed.

\subsubsection{Adaptive Random Construction (ARC)}

The adaptive random construction follows the basic mechanism of adaptive random testing (ART)~\cite{Huang2021}, which is considered an enhancement of random testing (RT) by achieving greater diversity among random test cases.

As the representative of ART, the FSCS-ART algorithm~\cite{Chen2010a} was employed in this study, which iteratively selects the best candidate from a set of randomly generated options as the next test case, ensuring it is farthest from previously selected test cases. Following previous studies~\cite{Wu2020}, the adaptive random construction follows these steps: 1) 30 candidates are randomly generated to satisfy the constraints; 2) for each candidate, the Hamming distance to all previously generated test cases is calculated, and the minimum distance is assigned as the fitness value for that candidate; and 3) the candidate with the largest fitness value is selected as the next test case. Similar to random construction, adaptive random construction generates 100 test suites for each size from 2 to 50, yielding up to 4,900 test suites per test scenario.

The last column of Table~\ref{TABLE:Programs} provides the number of test suites used for each subject program, resulting in 295,624 test suites in total.

\subsection{Selected Evaluation Metrics under Comparison}

As illustrated in Section~\ref{SECTION:Metrics}, various metrics have been proposed to evaluate test suites for CIT. However, some evaluation metrics are effectively equivalent, while others require additional parameters to be specified in advance. In this section, we briefly discuss the selection process of the evaluation metrics used in the experiments. Table~\ref{TABLE:comparedMetrics} presents an overview of the investigated evaluation metrics, including category, metric name, parameter, and parameter assignment.

{\sloppy 
For the ICBM category, we select $VCC(\lambda,T)$, $VID(\lambda,T)$, $WMVCC(\lambda,T)$, and $Completeness(q,\lambda,T)$, while discarding $BVCC(b,T_\lambda)$, $TD(T_\lambda)$, $PCC(\phi_\lambda,T)$, and $SVCC(\lambda,T)$ for the following reasons:
\par}
\begin{itemize}
    \item[1)] $BVCC(b,T_\lambda)$ and $TD(T_\lambda)$ are only designed for $\lambda$-wise CAs $T_\lambda$, meaning they cannot be applied to non-CAs. Additionally, these two metrics share the same underlying mechanism as $VCC(\lambda,T)$.
    \item[2)] $PCC(\phi_\lambda,T)$ only focuses on a specific $\lambda$-wise parameter combination $\phi_\lambda$, while ignoring other parameter combinations when evaluating $T$. In other words, different parameter combinations may yield different evaluation results for the same $T$. Therefore, it is not suitable for assessing $T$.
\end{itemize}

{\sloppy 
As for parameter assignments, the strength values $\lambda = 1, 2, 3, 4, 5, 6$ are selected for $VCC(\lambda,T)$, $VID(\lambda,T)$, $SVCC(\lambda,T)$, and $Completeness(q,\lambda,T)$. However, $\lambda = 2, 3, 4, 5, 6$ are chosen for $WMVCC(\lambda,T)$, because $\lambda$ should be at least 2 to encompass multiple strengths. In addition, $q = 0.25, 0.50, 0.75$ are chosen for $Completeness(q,\lambda,T)$; $q = 1.0$ is not considered since it corresponds to $SVCC(\lambda,T)$.
\par}

\begin{table}[!t]
\centering
\footnotesize
\caption{An Overview of Evaluation Metrics and Their Parameter Assignments Investigated in the Experiments}
\label{TABLE:comparedMetrics}
\setlength{\tabcolsep}{2.5mm}{\begin{tabular}{clll} \hline

\textbf{Category} &\textbf{Evaluation metric}  &\textbf{Parameter}  &\textbf{Paramenter Assignment}        \\ \hline
\multirow{5}*{ICBM} 
&$VCC(\lambda, T)$ &Strength, $\lambda$ &$\lambda=1$, $2$, $3$, $4$, $5$, and $6$ \\\cline{2-4}
&$VID(\lambda, T)$ &Strength, $\lambda$ &$\lambda=1$, $2$, $3$, $4$, $5$, and $6$ \\\cline{2-4}
&$WMVCC(\lambda, T)$ &Strength, $\lambda$ &$\lambda=2$, $3$, $4$, $5$, and $6$  \\\cline{2-4}
&$SVCC(\lambda, T)$ &Strength, $\lambda$ &$\lambda=1$, $2$, $3$, $4$, $5$, and $6$  \\\cline{2-4}
&\multirow{2}*{$Completeness(q,\lambda, T)$} &Strength, $\lambda$ &$\lambda=1$, $2$, $3$, $4$, $5$, and $6$ \\\cline{3-4}
& &A float constant, $q$ &$q=0.25$, $0.50$, and $0.75$ \\\hline

\multirow{6}*{DBM} 
&$Discrepancy(T)$ &An integer constant, $m$ &$m=1,000$ \cite{Chen2009b} \\
&$Dispersion(T)$ &No parameters &-- \\
&$Diversity(T)$ &No parameters &--\\
&$Similarity(T)$ &No parameters &-- \\ 
&$Novelty(T)$  &An integer constant, &$\varepsilon=2$, $5$, $15$, $|T|/4$, $|T|/2$, and $3|T|/4$ \cite{Xiang2022} \\
&$Divergence(T)$ &No parameters &-- \\ \hline

MBM
&$IMS(T,\mathcal{R}')$ &A set of faulty constraints $\mathcal{R}'$ &$\mathcal{R}'$ is pre-defined\\\hline
\end{tabular}
}
\end{table}

\begin{table}[!b]
\centering
\footnotesize
\caption{Two Mutant Operators Used to Construct Faulty Constraints for IMS}
\label{TABLE:mutantOperator}
\setlength{\tabcolsep}{2mm}{\begin{tabular}{cll} \hline

\textbf{Name} &\textbf{Input Constraints}  &\textbf{Mutated Constraints}     \\ \hline
Variable negation &$c=(var_1,var_2,\cdots,var_j)$ &$c'=(var_1,var_2,\cdots,\neg var_i, var_{i+1},\cdots, var_j)$\\

Constraint splitting &$c=(var_1,var_2,\cdots,var_j)$ &$c'=(var_1,var_2,\cdots,var_i)$ and $c''=(var_{i+1},var_{i+2},\cdots,var_j)$ \\\hline
\end{tabular}
}
\end{table}

{\sloppy 
For the DBM category, all six evaluation metrics are considered, i.e., $Discrepancy(T)$, $Dispersion(T)$, $Diversity(T)$, $Similarity(T)$, $Novelty(T)$, and $Divergence(T)$. Among them, only two metrics require parameter assignments: 1) $Discrepancy(T)$ needs to assign an integer constant $m$, which is set to $1,000$ following previous study~\cite{Chen2009b}; and 2) $Novelty(T)$ also has an integer constant $\varepsilon$, for which six different assignments are selected following~\cite{Xiang2022}, i.e., $\varepsilon = 2, 5, 15, |T|/4, |T|/2, 3|T|/4$.
\par}

For the MBM category, only $IMS(T,\mathcal{R}')$ is considered, which requires a set of faulty constraints $\mathcal{R}'$ as the parameter. Following previous studies~\cite{Henard2013,Papadakis2014}, we consider two mutation operators to construct faulty constraints, as shown in Table~\ref{TABLE:mutantOperator}. More specifically, the first operator is called \textit{variable negation}, which modifies an existing constraint by selecting a random variable within it and inverting its value (since each variable is Boolean). For example, suppose a constraint $c$ involves some variables $var_1, var_2, \cdots, var_j$, each of which is true or false, i.e., $c = (var_1, var_2, \cdots, var_j)$. The variable negation randomly chooses a variable $var_i~(1 \leq i \leq j)$ and constructs a mutated constraint $c' = (var_1, var_2, \cdots, \neg var_i, \cdots, var_j)$ by negating $var_i$.
The second operator is called \textit{constraint splitting}, which decomposes a single constraint into two distinct constraints. Again, consider the constraint $c = (var_1, var_2, \cdots, var_j)$. Constraint splitting divides this constraint into two constraints at a random position $i~(1 < i < j)$, i.e., $c' = (var_1, var_2, \cdots, var_i)$ and $c'' = (var_{i+1}, var_{i+2}, \cdots, var_j)$.

\subsection{Correlation Analysis}

To assess the relationship between each evaluation metric and fault-detection effectiveness, we computed two correlation measures following previous studies \cite{Gligoric2015,szekely2007measuring}:
For $n$ independent runs, let 
$X=\{x_1,x_2,\ldots,x_n\}$ and 
$Y=\{y_1,y_2,\ldots,y_n\}$ denote the observed values of an evaluation metric and fault-detection effectiveness, respectively. 
This yields $n$ paired observations $(x_i, y_i)$, where 
$x_i$ and $y_i$ are the corresponding values of $X$ and $Y$ in the $i$-th run.
The two correlation measures are defined as follows.

\subsubsection{Coefficient of Determination ($R^2$)}
The coefficient of determination ($R^2$), commonly referred to as \textit{R-Squared}, is a statistical measure used in regression models~\cite{Anderson1958}. 
$R^2$ indicates the proportion of variance in the dependent variable $x$ explained by the independent variable $y$. 
In other words, $R^2$ indicates how well the data fit the regression model, reflecting the model's goodness of fit. Its definition is given as follows:
\begin{equation}
    R^2 = \frac{\Big(n\sum_{i=1}^nx_iy_i-\sum_{i=1}^nx_i\sum_{i=1}^ny_i\Big)^2}
    {\Big(n\sum_{i=1}^nx_i^2-(\sum_{i=1}^nx_i)^2\Big){\Big(n\sum_{i=1}^ny_i^2-(\sum_{i=1}^ny_i)^2\Big)}}.
\end{equation}
$R^2$ ranges from $0.0$ to $1.0$.
A value of 
$R^2=0.0$ indicates that $x$ cannot be predicted from $y$; whereas 
$R^2=1.0$ indicates perfect predictability of $x$ from $y$ without error~\cite{Anderson1958}.

\subsubsection{Distance Correlation}

Distance correlation $\mathcal{R}$ \cite{szekely2007measuring} is a measure of statistical dependence between random vectors. 
It generalizes in two respects: 
(1) $\mathcal{R}$ is defined for random vectors $X$ and $Y$ in arbitrary dimensions; and 
(2) $\mathcal{R}$ satisfies $\mathcal{R} = 0$ if and only if $X$ and $Y$ are independent.
The empirical distance correlation is defined as follows:

\begin{equation}
\label{EQ:R}
\mathcal{R}_n^2(X, Y) = 
\begin{cases} 
\dfrac{\mathcal{V}_n^2(X, Y)}{\sqrt{\mathcal{V}_n^2(X) \, \mathcal{V}_n^2(Y)}}, & \mathcal{V}_n^2(X) \, \mathcal{V}_n^2(Y) > 0, \\[10pt]
0, & \mathcal{V}_n^2(X) \, \mathcal{V}_n^2(Y) = 0,
\end{cases}
\end{equation}

To compute $\mathcal{V}_n^2(X,Y)$, we first define the pairwise Euclidean distance between samples $x_i$ and $x_j$ as
\[
a_{ij} = |x_i - x_j|, \quad i,j = 1,\ldots,n.
\]
Then, the row mean, column mean, and grand mean of the distance matrix are computed as
\[
\bar{a}_{i*} = \frac{1}{n}\sum_{j=1}^n a_{ij}, \quad 
\bar{a}_{*j} = \frac{1}{n}\sum_{i=1}^n a_{ij}, \quad
\bar{a}_{**} = \frac{1}{n^2}\sum_{i,j=1}^n a_{ij}.
\]
The centered distance matrix for $X$ is defined as
\[
A_{ij} = a_{ij} - \bar{a}_{i*} - \bar{a}_{*j} + \bar{a}_{**}.
\]

Similarly, for samples $y_i$ and $y_j$, we define
\[
b_{ij} = |y_i - y_j|, \quad i,j = 1,\ldots,n,
\]
with
\[
\bar{b}_{i*} = \frac{1}{n}\sum_{j=1}^n b_{ij}, \quad 
\bar{b}_{*j} = \frac{1}{n}\sum_{i=1}^n b_{ij}, \quad
\bar{b}_{**} = \frac{1}{n^2}\sum_{i,j=1}^n b_{ij}.
\]
The centered distance matrix for $Y$ is then defined as
\[
B_{ij} = b_{ij} - \bar{b}_{i*} - \bar{b}_{*j} + \bar{b}_{**}.
\]

Then, the empirical distance covariance $\mathcal{V}_n^2(X,Y)$ is calculated as follows:
\begin{equation}
\label{eq:vn2xy}
    \mathcal{V}_n^2(X, Y)=\frac{1}{n^2}\sum_{i, j=1}^n A_{ij}B_{ij}.
\end{equation}

Based on the above equations, the distance correlation $\mathcal{R}$ can be computed, and its value lies in the range $[0, 1]$:
$\mathcal{R}=0.0$ indicates that $X$ and $Y$ are independent, and 
higher values imply stronger dependence between $X$ and $Y$.

\subsection{Running Environments}
To collect the experimental data, our experiments were performed on a machine with the following devices:
a 16-core Intel Xeon Silver 4314 processor at 2.40GHz and 125GB RAM, running Ubuntu 22.04.4 LTS, and Java OpenJDK 64-Bit Server VM, version 21.0.11. We run each metric individually to collect metric values and execution overheads.

 \begin{table*}[h]
\centering
\scriptsize
\caption{$R^2$ values for ICBM metrics under RC and ARC (best)}
\label{TABLE:icbm3}
\setlength{\tabcolsep}{2mm}{
\begin{tabular}{c|ccccc|ccccc}  \hline
\multirow{2}*{\textbf{Subject}} & \multicolumn{5}{c|}{\textbf{RC}} & \multicolumn{5}{c}{\textbf{ARC}} \\
\cline{2-11}
& \textit{VCC} & \textit{VID}   & \textit{WMVCC}  & \textit{SVCC}   & \textit{Completeness}   & \textit{VCC} & \textit{VID}   & \textit{WMVCC}  & \textit{SVCC}   & \textit{Completeness}\\
   
\hline
{FLEX-TS1}	&	0.744 	&	0.705 	&	\cellcolor{green!30}0.750	&	\cellcolor{magenta!20}0.425	&	0.613 	&	\cellcolor{yellow!50}0.641	&	0.622 	&	0.637 	&	\cellcolor{blue!20}0.436	&	0.533 	\\	\hline
{FLEX-TS2}	&	0.802 	&	0.780 	&	\cellcolor{green!30}0.818	&	0.689 	&	\cellcolor{magenta!20}0.669	&	0.661 	&	\cellcolor{yellow!50}0.717	&	0.694 	&	0.644 	&	\cellcolor{blue!20}0.543	\\	\hline
{FLEX-TS3}	&	\cellcolor{green!30}0.890	&	0.805 	&	0.889 	&	\cellcolor{magenta!20}0.776	&	0.838 	&	0.856 	&	\cellcolor{yellow!50}0.870	&	0.864 	&	0.782 	&	\cellcolor{blue!20}0.755	\\	\hline
{FLEX-TS4}	&	\cellcolor{green!30}0.888	&	0.801 	&	0.876 	&	\cellcolor{magenta!20}0.757	&	0.855 	&	\cellcolor{yellow!50}0.918	&	0.903 	&	0.909 	&	0.868 	&	\cellcolor{blue!20}0.854	\\	\hline
{GREP-TS1}	&	0.924 	&	0.883 	&	\cellcolor{green!30}0.947	&	\cellcolor{magenta!20}0.000	&	\cellcolor{magenta!20}0.000	&	0.899 	&	0.890 	&	\cellcolor{yellow!50}0.929	&	0.000 	&	\cellcolor{blue!20}0.000	\\	\hline
{GREP-TS2}	&	0.865 	&	\cellcolor{green!30}0.915	&	0.893 	&	0.230 	&	\cellcolor{magenta!20}0.000	&	0.903 	&	0.911 	&	\cellcolor{yellow!50}0.929	&	0.274 	&	\cellcolor{blue!20}0.000	\\	\hline
{GREP-TS3}	&	\cellcolor{green!30}0.946	&	0.801 	&	0.942 	&	0.701 	&	\cellcolor{magenta!20}0.558	&	\cellcolor{yellow!50}0.948	&	0.938 	&	0.938 	&	0.755 	&	\cellcolor{blue!20}0.543	\\	\hline
{GREP-TS4}	&	0.385 	&	0.345 	&	0.381 	&	\cellcolor{magenta!20}0.000	&	\cellcolor{green!30}0.855	&	0.504 	&	0.478 	&	0.504 	&	\cellcolor{blue!20}0.000	&	\cellcolor{yellow!50}0.868	\\	\hline
{GZIP-TS1}	&	0.096 	&	\cellcolor{magenta!20}0.072	&	\cellcolor{green!30}0.105	&	0.075 	&	0.100 	&	\cellcolor{yellow!50}0.018	&	\cellcolor{blue!20}0.014	&	\cellcolor{yellow!50}0.018	&	0.015 	&	\cellcolor{yellow!50}0.018	\\	\hline
{GZIP-TS2}	&	0.040 	&	0.041 	&	\cellcolor{green!30}0.042	&	\cellcolor{magenta!20}0.026	&	0.042 	&	\cellcolor{yellow!50}0.007	&	0.005 	&	\cellcolor{yellow!50}0.007	&	\cellcolor{blue!20}0.004	&	\cellcolor{yellow!50}0.007	\\	\hline
{GZIP-TS3}	&	0.511 	&	\cellcolor{magenta!20}0.288	&	\cellcolor{green!30}0.517	&	0.429 	&	0.504 	&	0.648 	&	\cellcolor{blue!20}0.500	&	\cellcolor{yellow!50}0.657	&	0.569 	&	0.634 	\\	\hline
{GZIP-TS4}	&	0.305 	&	\cellcolor{magenta!20}0.159	&	\cellcolor{green!30}0.321	&	0.189 	&	0.287 	&	0.537 	&	\cellcolor{blue!20}0.384	&	\cellcolor{yellow!50}0.547	&	0.482 	&	0.528 	\\	\hline
{MAKE-TS1}	&	\cellcolor{green!30}0.259	&	0.235 	&	0.251 	&	0.245 	&	\cellcolor{magenta!20}0.158	&	0.187 	&	0.198 	&	\cellcolor{yellow!50}0.202	&	0.163 	&	\cellcolor{blue!20}0.104	\\	\hline
{MAKE-TS2}	&	0.486 	&	\cellcolor{green!30}0.553	&	0.512 	&	0.388 	&	\cellcolor{magenta!20}0.185	&	0.357 	&	\cellcolor{yellow!50}0.544	&	0.444 	&	0.225 	&	\cellcolor{blue!20}0.127	\\	\hline
{MAKE-TS3}	&	0.339 	&	\cellcolor{green!30}0.362	&	0.352 	&	0.259 	&	\cellcolor{magenta!20}0.126	&	0.376 	&	\cellcolor{yellow!50}0.539	&	0.465 	&	0.223 	&	\cellcolor{blue!20}0.124	\\	\hline
{MAKE-TS4}	&	0.251 	&	\cellcolor{green!30}0.273	&	0.260 	&	0.193 	&	\cellcolor{magenta!20}0.091	&	0.249 	&	\cellcolor{yellow!50}0.352	&	0.309 	&	0.143 	&	\cellcolor{blue!20}0.077	\\	\hline
{SED-TS1}	&	\cellcolor{magenta!20}0.000	&	\cellcolor{magenta!20}0.000	&	\cellcolor{magenta!20}0.000	&	\cellcolor{magenta!20}0.000	&	\cellcolor{magenta!20}0.000	&	\cellcolor{blue!20}0.000	&	\cellcolor{blue!20}0.000	&	\cellcolor{blue!20}0.000	&	\cellcolor{blue!20}0.000	&	\cellcolor{blue!20}0.000	\\	\hline
{SED-TS2}	&	0.696 	&	0.639 	&	\cellcolor{green!30}0.699	&	\cellcolor{magenta!20}0.000	&	\cellcolor{magenta!20}0.000	&	\cellcolor{yellow!50}0.736	&	0.690 	&	0.735 	&	\cellcolor{blue!20}0.000	&	\cellcolor{blue!20}0.000	\\	\hline
{SED-TS3}	&	\cellcolor{green!30}0.839	&	0.777 	&	\cellcolor{green!30}0.839	&	0.658 	&	\cellcolor{magenta!20}0.657	&	\cellcolor{yellow!50}0.859	&	0.834 	&	0.853 	&	0.614 	&	\cellcolor{blue!20}0.603	\\	\hline
{SED-TS4}	&	\cellcolor{green!30}0.889	&	0.810 	&	0.883 	&	\cellcolor{magenta!20}0.670	&	0.801 	&	\cellcolor{yellow!50}0.890	&	0.868 	&	0.887 	&	\cellcolor{blue!20}0.576	&	0.793 	\\	\hline
{BUSYBOX-TS1}	&	0.361 	&	0.360 	&	0.387 	&	\cellcolor{magenta!20}0.353	&	\cellcolor{green!30}0.508	&	0.385 	&	\cellcolor{blue!20}0.349	&	0.388 	&	0.417 	&	\cellcolor{yellow!50}0.457	\\	\hline
{BUSYBOX-TS2}	&	0.413 	&	0.410 	&	\cellcolor{green!30}0.425	&	0.385 	&	\cellcolor{magenta!20}0.361	&	0.347 	&	0.358 	&	\cellcolor{yellow!50}0.361	&	0.304 	&	\cellcolor{blue!20}0.250	\\	\hline
{BUSYBOX-TS3}	&	\cellcolor{magenta!20}0.575	&	0.587 	&	0.587 	&	0.580 	&	\cellcolor{green!30}0.607	&	0.524 	&	\cellcolor{blue!20}0.499	&	0.524 	&	\cellcolor{yellow!50}0.555	&	0.550 	\\	\hline
{BUSYBOX-TS4}	&	0.590 	&	0.584 	&	\cellcolor{green!30}0.599	&	0.582 	&	\cellcolor{magenta!20}0.528	&	0.604 	&	0.598 	&	\cellcolor{yellow!50}0.613	&	0.580 	&	\cellcolor{blue!20}0.490	\\	\hline
{DRUPAL-TS1}	&	0.684 	&	\cellcolor{magenta!20}0.611	&	0.703 	&	0.645 	&	\cellcolor{green!30}0.720	&	0.723 	&	\cellcolor{blue!20}0.650	&	0.713 	&	0.733 	&	\cellcolor{yellow!50}0.755	\\	\hline
{DRUPAL-TS2}	&	0.773 	&	\cellcolor{magenta!20}0.708	&	\cellcolor{green!30}0.784	&	0.736 	&	0.771 	&	0.761 	&	\cellcolor{blue!20}0.702	&	0.752 	&	0.763 	&	\cellcolor{yellow!50}0.787	\\	\hline
{DRUPAL-TS3}	&	0.804 	&	\cellcolor{magenta!20}0.655	&	\cellcolor{green!30}0.815	&	0.767 	&	0.762 	&	0.826 	&	\cellcolor{blue!20}0.784	&	\cellcolor{yellow!50}0.831	&	0.802 	&	0.810 	\\	\hline
{DRUPAL-TS4}	&	0.824 	&	\cellcolor{magenta!20}0.698	&	\cellcolor{green!30}0.837	&	0.786 	&	0.787 	&	0.860 	&	\cellcolor{blue!20}0.832	&	\cellcolor{yellow!50}0.866	&	0.836 	&	0.835 	\\	\hline
{LINUX-TS1}	&	0.533 	&	0.528 	&	\cellcolor{green!30}0.537	&	0.497 	&	\cellcolor{magenta!20}0.481	&	0.486 	&	\cellcolor{yellow!50}0.521	&	0.504 	&	0.414 	&	\cellcolor{blue!20}0.372	\\	\hline
{LINUX-TS2}	&	0.653 	&	0.634 	&	\cellcolor{green!30}0.661	&	0.615 	&	\cellcolor{magenta!20}0.606	&	0.627 	&	0.635 	&	\cellcolor{yellow!50}0.637	&	0.575 	&	\cellcolor{blue!20}0.537	\\	\hline
{LINUX-TS3}	&	0.716 	&	0.711 	&	\cellcolor{green!30}0.727	&	\cellcolor{magenta!20}0.684	&	0.694 	&	0.762 	&	0.759 	&	\cellcolor{yellow!50}0.769	&	0.729 	&	\cellcolor{blue!20}0.695	\\	\hline
{LINUX-TS4}	&	0.809 	&	0.795 	&	\cellcolor{green!30}0.817	&	0.782 	&	\cellcolor{magenta!20}0.707	&	0.800 	&	0.807 	&	\cellcolor{yellow!50}0.808	&	0.759 	&	\cellcolor{blue!20}0.643	\\	\hline

{Standard deviation}	&	0.271 	&	 \cellcolor{green!30}0.257 	&	0.271 	&	0.273 	&	\cellcolor{magenta!20}0.297 	&	0.278 	&	\cellcolor{yellow!50}0.266 	&	0.272 	&	0.289 	&	\cellcolor{blue!20}0.303 	\\	
{Geometric mean}	&	0.398 	&	0.366 	&	 \cellcolor{green!30}0.406 	&	\cellcolor{magenta!20}0.146 	&	0.150 	&	0.360 	&	0.355 	&	\cellcolor{yellow!50}0.371 	&	0.131 	&	\cellcolor{blue!20}0.128 	\\	
{Arithmetic mean}	&	0.590 	&	0.548 	&	 \cellcolor{green!30}0.599 	&	\cellcolor{magenta!20}0.441 	&	0.465 	&	0.591 	&	0.586 	&	\cellcolor{yellow!50}0.603 	&	\cellcolor{blue!20}0.445 	&	0.447 	\\	

{The best results}
&6 & 4 & \cellcolor{green!30}18  & \cellcolor{magenta!20}0 & 5 & 8 & 6 & \cellcolor{yellow!50}14 & \cellcolor{blue!20}1 & 6

\\
{The worst results}
&2 & 8 & \cellcolor{green!30}1 & 11 & \cellcolor{magenta!20}16 & \cellcolor{yellow!50}1 &10 &\cellcolor{yellow!50}1 & 7 & \cellcolor{blue!20}19 
\\
\hline 
\end{tabular}
}
\end{table*}

\section{Results}
\label{SECTION:Results}

In this section, we present the experimental results and answer the research questions. 
In the result tables, background colors are used to highlight the best and worst performance values:
the \greenbg{green} and \redbg{red} backgrounds indicate the best and worst results achieved by RC-constructed test suites, respectively, whereas
the \yellowbg{yellow} and \purplebg{purple} backgrounds indicate the best and worst results achieved by ARC-constructed test suites, respectively.
The notation ``TO'' (Time-Out) denotes an evaluation task that failed to complete within the 24-hour execution threshold (as defined in Section~\ref{SECTION:Basic Framework}). 
We use a ranking mechanism to facilitate correlation analysis:
This mechanism ranks the results for selecting the best-performing strength for the ICBM metric or determining the most effective metric for the DBM category.
The ranking results range from zero to $k$ (where $k$ corresponds to the number of metrics):
A zero result signifies that a program subject was not evaluated; while 
a lower non-zero ranking result indicates a higher correlation. Due to the page limit, the detailed experimental results are reported in the following link: \url{https://github.com/Q-Lisa/More-Details-for-CIT-Metric-Evaluations.git}.

\subsection{Answer to RQ1: Effectiveness Comparisons of ICBM Metrics}
\label{SECTION:ResRQ1}
This section discusses the correlation results between fault-detection effectiveness and ICBM metrics, and analyzes the performance of different interaction strengths, systematically comparing the validity of single-strength and multiple-strength paradigms.

\begin{table*}[!t]
\centering
\scriptsize
\caption{$\mathcal{R}$ values for ICBM metrics under RC and ARC (best)}
\label{TABLE:icbm4}
\setlength{\tabcolsep}{2mm}{
\begin{tabular}{c|ccccc|ccccc}  \hline
\multirow{2}*{\textbf{Subject}} & \multicolumn{5}{c|}{\textbf{RC}} & \multicolumn{5}{c}{\textbf{ARC}} \\
\cline{2-11}
& \textit{VCC} & \textit{VID}   & \textit{WMVCC}  & \textit{SVCC}   & \textit{Completeness}   & \textit{VCC} & \textit{VID}   & \textit{WMVCC}  & \textit{SVCC}   & \textit{Completeness}\\
   
\hline
{FLEX-TS1}	&	\cellcolor{green!30}0.840	&	0.814 	&	0.750 	&	\cellcolor{magenta!20}0.650	&	0.777 	&	\cellcolor{yellow!50}0.856	&	0.846 	&	\cellcolor{blue!20}0.637	&	0.702 	&	0.789 	\\	\hline
{FLEX-TS2}	&	\cellcolor{green!30}0.871	&	0.866 	&	0.818 	&	0.797 	&	\cellcolor{magenta!20}0.795	&	0.820 	&	\cellcolor{yellow!50}0.848	&	\cellcolor{blue!20}0.694	&	0.789 	&	0.741 	\\	\hline
{FLEX-TS3}	&	\cellcolor{green!30}0.924	&	0.891 	&	0.889 	&	\cellcolor{magenta!20}0.857	&	0.903 	&	0.908 	&	\cellcolor{yellow!50}0.911	&	0.864 	&	0.855 	&	\cellcolor{blue!20}0.852	\\	\hline
{FLEX-TS4}	&	\cellcolor{green!30}0.933	&	0.906 	&	0.876 	&	\cellcolor{magenta!20}0.850	&	0.918 	&	\cellcolor{yellow!50}0.950	&	0.947 	&	\cellcolor{blue!20}0.909	&	0.918 	&	0.926 	\\	\hline
{GREP-TS1}	&	\cellcolor{green!30}0.957	&	0.938 	&	0.947 	&	\cellcolor{magenta!20}0.000	&	\cellcolor{magenta!20}0.000	&	\cellcolor{yellow!50}0.938	&	0.937 	&	0.929 	&	\cellcolor{blue!20}0.000	&	\cellcolor{blue!20}0.000	\\	\hline
{GREP-TS2}	&	0.928 	&	\cellcolor{green!30}0.947	&	0.893 	&	0.561 	&	\cellcolor{magenta!20}0.000	&	0.942 	&	\cellcolor{yellow!50}0.945	&	0.929 	&	0.607 	&	\cellcolor{blue!20}0.000	\\	\hline
{GREP-TS3}	&	\cellcolor{green!30}0.966	&	0.908 	&	0.942 	&	0.851 	&	\cellcolor{magenta!20}0.808	&	\cellcolor{yellow!50}0.967	&	0.963 	&	0.938 	&	0.879 	&	\cellcolor{blue!20}0.797	\\	\hline
{GREP-TS4}	&	0.645 	&	0.603 	&	0.381 	&	\cellcolor{magenta!20}0.000	&	\cellcolor{green!30}0.929	&	0.699 	&	0.668 	&	0.504 	&	\cellcolor{blue!20}0.000	&	\cellcolor{yellow!50}0.941	\\	\hline
{GZIP-TS1}	&	0.358 	&	0.316 	&	\cellcolor{magenta!20}0.105	&	0.333 	&	\cellcolor{green!30}0.380	&	0.152 	&	0.139 	&	\cellcolor{blue!20}0.018	&	0.154 	&	\cellcolor{yellow!50}0.164	\\	\hline
{GZIP-TS2}	&	0.240 	&	0.240 	&	\cellcolor{magenta!20}0.042	&	0.207 	&	\cellcolor{green!30}0.260	&	0.095 	&	0.086 	&	\cellcolor{blue!20}0.007	&	0.081 	&	\cellcolor{yellow!50}0.108	\\	\hline
{GZIP-TS3}	&	0.752 	&	0.588 	&	\cellcolor{magenta!20}0.517	&	0.668 	&	\cellcolor{green!30}0.754	&	0.828 	&	0.727 	&	\cellcolor{blue!20}0.657	&	0.753 	&	\cellcolor{yellow!50}0.835	\\	\hline
{GZIP-TS4}	&	\cellcolor{green!30}0.577	&	0.421 	&	\cellcolor{magenta!20}0.321	&	0.417 	&	0.553 	&	\cellcolor{yellow!50}0.735	&	0.645 	&	\cellcolor{blue!20}0.547	&	0.682 	&	0.732 	\\	\hline
{MAKE-TS1}	&	\cellcolor{green!30}0.503	&	0.485 	&	\cellcolor{magenta!20}0.251	&	0.492 	&	0.398 	&	0.424 	&	\cellcolor{yellow!50}0.435	&	\cellcolor{blue!20}0.202	&	0.402 	&	0.322 	\\	\hline
{MAKE-TS2}	&	0.719 	&	\cellcolor{green!30}0.766	&	0.512 	&	0.634 	&	\cellcolor{magenta!20}0.434	&	0.631 	&	\cellcolor{yellow!50}0.762	&	0.444 	&	0.490 	&	\cellcolor{blue!20}0.359	\\	\hline
{MAKE-TS3}	&	0.639 	&	\cellcolor{green!30}0.663	&	\cellcolor{magenta!20}0.352	&	0.560 	&	0.398 	&	0.670 	&	\cellcolor{yellow!50}0.777	&	0.465 	&	0.504 	&	\cellcolor{blue!20}0.367	\\	\hline
{MAKE-TS4}	&	0.559 	&	\cellcolor{green!30}0.578	&	\cellcolor{magenta!20}0.26	&	0.499 	&	0.352 	&	0.563 	&	\cellcolor{yellow!50}0.644	&	0.309 	&	0.423 	&	\cellcolor{blue!20}0.304	\\	\hline
{SED-TS1}	&	\cellcolor{magenta!20}0.000	&	\cellcolor{magenta!20}0.000	&	\cellcolor{magenta!20}0.000	&	\cellcolor{magenta!20}0.000	&	\cellcolor{magenta!20}0.000	&	0.000 	&	0.000 	&	0.000 	&	0.000 	&	\cellcolor{blue!20}0.000	\\	\hline
{SED-TS2}	&	\cellcolor{green!30}0.862	&	0.837 	&	0.699 	&	\cellcolor{magenta!20}0.000	&	\cellcolor{magenta!20}0.000	&	\cellcolor{yellow!50}0.873	&	0.856 	&	0.735 	&	0.000 	&	\cellcolor{blue!20}0.000	\\	\hline
{SED-TS3}	&	\cellcolor{green!30}0.904	&	0.895 	&	0.839 	&	0.784 	&	\cellcolor{magenta!20}0.782	&	\cellcolor{yellow!50}0.903	&	0.902 	&	0.853 	&	0.731 	&	\cellcolor{blue!20}0.729	\\	\hline
{SED-TS4}	&	\cellcolor{green!30}0.929	&	0.908 	&	0.883 	&	\cellcolor{magenta!20}0.801	&	0.880 	&	\cellcolor{yellow!50}0.927	&	0.922 	&	0.887 	&	\cellcolor{blue!20}0.754	&	0.883 	\\	\hline
{BUSYBOX-TS1}	&	0.577 	&	0.575 	&	\cellcolor{magenta!20}0.387	&	0.601 	&	\cellcolor{green!30}0.715	&	0.586 	&	0.560 	&	\cellcolor{blue!20}0.388	&	0.642 	&	\cellcolor{yellow!50}0.667	\\	\hline
{BUSYBOX-TS2}	&	\cellcolor{green!30}0.519	&	\cellcolor{green!30}0.519	&	\cellcolor{magenta!20}0.425	&	0.506 	&	0.473 	&	0.488 	&	\cellcolor{yellow!50}0.499	&	\cellcolor{blue!20}0.361	&	0.450 	&	0.393 	\\	\hline
{BUSYBOX-TS3}	&	0.727 	&	0.730 	&	\cellcolor{magenta!20}0.587	&	\cellcolor{green!30}0.750	&	0.742 	&	0.701 	&	0.684 	&	\cellcolor{blue!20}0.524	&	\cellcolor{yellow!50}0.742	&	0.713 	\\	\hline
{BUSYBOX-TS4}	&	0.677 	&	0.672 	&	\cellcolor{magenta!20}0.599	&	\cellcolor{green!30}0.683	&	0.639 	&	\cellcolor{yellow!50}0.677	&	0.675 	&	0.613 	&	0.667 	&	\cellcolor{blue!20}0.591	\\	\hline
{DRUPAL-TS1}	&	0.832 	&	0.797 	&	\cellcolor{magenta!20}0.703	&	0.833 	&	\cellcolor{green!30}0.855	&	0.862 	&	0.820 	&	\cellcolor{blue!20}0.713	&	\cellcolor{yellow!50}0.887	&	0.881 	\\	\hline
{DRUPAL-TS2}	&	0.879 	&	0.845 	&	\cellcolor{magenta!20}0.784	&	0.879 	&	\cellcolor{green!30}0.880	&	0.885 	&	0.850 	&	\cellcolor{blue!20}0.752	&	\cellcolor{yellow!50}0.906	&	0.902 	\\	\hline
{DRUPAL-TS3}	&	\cellcolor{green!30}0.878	&	\cellcolor{magenta!20}0.79	&	0.815 	&	0.871 	&	0.864 	&	\cellcolor{yellow!50}0.919	&	0.899 	&	\cellcolor{blue!20}0.831	&	\cellcolor{yellow!50}0.919	&	0.900 	\\	\hline
{DRUPAL-TS4}	&	\cellcolor{green!30}0.888	&	\cellcolor{magenta!20}0.81	&	0.837 	&	0.878 	&	0.875 	&	\cellcolor{yellow!50}0.936	&	0.920 	&	\cellcolor{blue!20}0.866	&	0.935 	&	0.910 	\\	\hline
{LINUX-TS1}	&	\cellcolor{green!30}0.657	&	\cellcolor{green!30}0.657	&	\cellcolor{magenta!20}0.537	&	0.639 	&	0.629 	&	0.647 	&	\cellcolor{yellow!50}0.667	&	\cellcolor{blue!20}0.504	&	0.587 	&	0.558 	\\	\hline
{LINUX-TS2}	&	\cellcolor{green!30}0.732	&	0.720 	&	\cellcolor{magenta!20}0.661	&	0.720 	&	0.709 	&	0.708 	&	\cellcolor{yellow!50}0.713	&	\cellcolor{blue!20}0.637	&	0.680 	&	0.652 	\\	\hline
{LINUX-TS3}	&	\cellcolor{green!30}0.787	&	0.785 	&	\cellcolor{magenta!20}0.727	&	0.784 	&	0.776 	&	\cellcolor{yellow!50}0.808	&	0.806 	&	0.769 	&	0.799 	&	\cellcolor{blue!20}0.768	\\	\hline
{LINUX-TS4}	&	\cellcolor{green!30}0.865	&	0.856 	&	0.817 	&	0.862 	&	\cellcolor{magenta!20}0.814	&	0.855 	&	\cellcolor{yellow!50}0.858	&	0.808 	&	0.840 	&	\cellcolor{blue!20}0.758	\\	\hline

{Standard deviation}	&	0.220 	&	\cellcolor{green!30}0.219 	&	0.271 	&	0.278 	&	\cellcolor{magenta!20}0.293 	&	0.249 	&	\cellcolor{yellow!50}0.246 	&	0.272 	&	0.301 	&	\cellcolor{blue!20}0.314 	\\
{Geometric mean}	&	\cellcolor{green!30}0.543 	&	0.523 	&	0.406 	&	\cellcolor{magenta!20}0.216 	&	0.218 	&	\cellcolor{yellow!50}0.517 	&	0.514 	&	0.371 	&	0.205 	&	\cellcolor{blue!20}0.201 	\\
{Arithmetic mean}	&	\cellcolor{green!30}0.723 	&	0.698 	&	0.599 	&	\cellcolor{magenta!20}0.593 	&	0.603 	&	\cellcolor{yellow!50}0.717 	&	0.716 	&	0.603 	&	0.587 	&	\cellcolor{blue!20}0.579 	\\

{The best results}
&\cellcolor{green!30}18 & 6 & \cellcolor{magenta!20}0 & 2 & 7 & \cellcolor{yellow!50}12 & 11 & \cellcolor{blue!20}0 & 4 & 5

\\
{The worst results}
&\cellcolor{green!30}1 & 3 & \cellcolor{magenta!20}17 & 8 & 9 & \cellcolor{yellow!50}1 & \cellcolor{yellow!50}1& \cellcolor{blue!20}18 & 5 & 13
\\
\hline 

\end{tabular}
}
\end{table*}

\begin{table}[t!]
\centering
\scriptsize
\caption{Ranking scores of different assignments for ICBM metrics}
\label{TABLE:rank0}
\setlength{\tabcolsep}{2mm}{
\begin{tabular}{lcccccc}  \hline
\multicolumn{7}{c}{\textbf{Parameter assignments}}   \\ 
{} & \textbf{$\lambda=1$} & \textbf{$\lambda=2$}   & \textbf{$\lambda=3$}  & \textbf{$\lambda=4$}   & \textbf{$\lambda=5$}   & \textbf{$\lambda=6$} \\
\hline
\multicolumn{7}{c}{\textit{VCC}} \\ \hline
$R^2$ values under RC &	3.88 	&	2.25 	&	2.22 	&	2.81 	&	3.84 	&	4.85 

\\ 

$R^2$ values under ARC 	&	3.97 	&	2.53 	&	2.03 	&	2.63 	&	3.52	&	4.70 	
 \\ 

$\mathcal{R}$ values under RC 
&	3.84 	&	2.25 	&	2.06 	&	2.84 	&	3.56 	&	4.66 	
 \\ 

$\mathcal{R}$ values under ARC &	4.25 	&	2.53 	&	2.03 	&	2.66 	&	3.28 	&	4.70
  \\ 
\hline \textit{Average}	&	3.99 	&	2.39 	&	2.09 	&	2.74 	&	3.55 	&	4.73 
\\

\hline
\multicolumn{7}{c}{\textit{VID}} \\ \hline
$R^2$ values under RC 	&	2.94 	&	2.41 	&	2.69 	&	3.31 	&	3.64 	&	5.01 	\\
$R^2$ values under ARC	&	2.72 	&	2.41 	&	2.66 	&	3.25 	&	3.52 	&	4.94 	\\
$\mathcal{R}$ values under RC 	&	2.13 	&	2.22 	&	2.69 	&	3.34 	&	3.80 	&	5.10 	\\
$\mathcal{R}$ values under ARC	&	2.59 	&	2.31 	&	2.56 	&	3.19 	&	3.60 	&	5.10 	\\
\hline \textit{Average}	&	2.60 	&	2.34 	&	2.65 	&	3.27 	&	3.64 	&	5.04 	\\
\hline
\multicolumn{7}{c}{\textit{WMVCC}} \\ \hline
$R^2$ values under RC 	&	0.00 	&	3.03 	&	2.34 	&	2.41 	&	2.68 	&	3.25 \\	
$R^2$ values under ARC	&	0.00	&	2.94	&	2.31	&	2.31	&	2.52	&	3.15	\\
$\mathcal{R}$ values under RC 	&	0.00 	&	2.75 	&	2.22 	&	2.25 	&	2.52 	&	3.20 \\	
$\mathcal{R}$ values under ARC	&	0.00	&	2.91	&	2.28	&	2.13	&	2.44	&	3.25	\\
\hline \textit{Average}	&	0.00 	&	2.91 	&	2.29 	&	2.28 	&	2.54 	&	3.21 	\\
\hline
\multicolumn{7}{c}{\textit{SVCC}} \\ \hline
$R^2$ values under RC 	&	1.91 	&	1.66 	&	2.69 	&	3.63 	&	4.12 	&	4.66 	\\
$R^2$ values under ARC	&	2.16 	&	1.63 	&	2.69 	&	3.50 	&	4.00 	&	4.66 	\\
$\mathcal{R}$ values under RC 	&	2.00 	&	1.59 	&	2.63 	&	3.66 	&	4.12 	&	4.45 	\\
$\mathcal{R}$ values under ARC	&	2.28 	&	1.69 	&	2.50 	&	3.41 	&	4.00 	&	4.45 	\\
\hline \textit{Average}	&	2.09 	&	1.64 	&	2.63 	&	3.55 	&	4.06 	&	4.56 	\\

\hline
\multicolumn{7}{c}{\textbf{Completeness(q=0.25)}} \\ \hline
$R^2$ values under RC 	&	4.25 	&	3.75 	&	2.44 	&	1.88 	&	2.24 	&	3.74 	\\
$R^2$ values under ARC	&	4.16 	&	3.66 	&	2.56 	&	1.88 	&	2.12 	&	3.34 	\\
$\mathcal{R}$ values under RC 	&	4.25 	&	3.72 	&	2.38 	&	2.06 	&	2.16 	&	3.70 	\\
$\mathcal{R}$ values under ARC	&	4.13 	&	3.59 	&	2.63 	&	2.00 	&	2.00 	&	3.34 	\\
\hline \textit{Average}	&	4.20 	&	3.68 	&	2.50 	&	1.96 	&	2.13 	&	3.53 	\\

\hline

\multicolumn{7}{c}{\textbf{Completeness(q=0.50)}} \\ \hline
$R^2$ values under RC 	&	4.44 	&	2.53 	&	1.84 	&	2.69 	&	3.37 	&	4.45 	\\
$R^2$ values under ARC	&	4.44 	&	2.59 	&	1.91 	&	2.66 	&	3.24 	&	4.40 	\\
$\mathcal{R}$ values under RC 	&	4.44 	&	2.56 	&	1.97 	&	2.56 	&	3.32 	&	4.45 	\\
$\mathcal{R}$ values under ARC	&	4.41 	&	2.69 	&	2.09 	&	2.47 	&	3.12 	&	4.40 	\\
\hline \textit{Average}	&	4.43 	&	2.59 	&	1.95 	&	2.60 	&	3.26 	&	4.43 	\\

\hline

\multicolumn{7}{c}{\textbf{Completeness(q=0.75)}} \\ \hline
$R^2$ values under RC 	&	2.34 	&	1.72 	&	2.59 	&	3.59 	&	3.92 	&	5.01 	\\
$R^2$ values under ARC	&	2.81 	&	1.94 	&	2.28 	&	3.41 	&	3.88 	&	4.94 	\\
$\mathcal{R}$ values under RC 	&	2.56 	&	1.84 	&	2.31 	&	3.53 	&	3.88 	&	5.01 	\\
$\mathcal{R}$ values under ARC	&	2.72 	&	2.16 	&	2.03 	&	3.06 	&	3.88 	&	4.80 	\\
\hline \textit{Average}	&	2.61 	&	1.92 	&	2.30 	&	3.40 	&	3.89 	&	4.94 	\\

\hline

\end{tabular}
}
\end{table}

\begin{table}[t!]
\centering
\scriptsize
\caption{Ranking scores of different assignments for Completeness under best strength }
\label{TABLE:rankcomp}
\setlength{\tabcolsep}{2mm}{
\begin{tabular}{lcccc}  \hline

{\textbf{Parameter}} & $q=0.25$ & $q=0.50$   & $q=0.75$     \\
\hline
$R^2$ values under RC 	&	2.47	&	1.63	&	1.59	\\
$R^2$ values under ARC	&	2.25	&	1.50	&	1.91	\\
$\mathcal{R}$ values under RC 	&	2.44	&	1.59	&	1.63	\\
$\mathcal{R}$ values under ARC	&	2.31	&	1.56	&	1.75	\\
\hline \textit{Average}	&	2.37 	&	1.57 	&	1.72 	
\\
\hline
\end{tabular}
}
\end{table}

\subsubsection{Answer to RQ1.1: }
\label{SECTION:subsubrq1.1}
Tables~\ref{TABLE:icbm3} and~\ref{TABLE:icbm4} present the exact correlation values for $R^2$ and $\mathcal{R}$.
Table~\ref{TABLE:rank0} provides the detailed ranking scores of their various parameter assignments. 
Based on the results, we have the following observations:

\begin{itemize}
    \item 
    Regarding the $R^2$ values examined on RC-generated test suites, VCC delivered its optimal performance at $\lambda=3$ (with an arithmetic mean value of $0.590$):
    This secured the lowest average ranking score of $2.22$, whereas VID minimized its ranking score to $2.41$ at $\lambda=2$. 
    At $\lambda=3$, WMVCC achieved a ranking score of $2.34$. 
    Meanwhile, the SVCC metric performed best at $\lambda=2$. 

    \item
    A similar trend is observed for the ARC-generated test suites.
    For VCC, a peak arithmetic mean of $0.591$ was recorded at $\lambda=3$. 
    VID reached its lowest ranking score of $2.41$ under the $\lambda=2$. 
    WMVCC attained an identical ranking score of $2.31$ at $\lambda=3$ and $\lambda=4$.
    SVCC accomplished its top capability at $\lambda=2$.
    
    \item 
    Regarding the $\mathcal{R}$ values, the VCC metric achieved its worst ranking scores at $\lambda=3$ under both RC-generated and ARC-generated test suites. 
    VID performed optimally at $\lambda=1$ (with an arithmetic mean of $0.698$) for RC-generated suites; and at $\lambda=2$ (the arithmetic mean is $0.716$) for ARC-generated ones. 
    WMVCC achieved its lowest ranking scores at $\lambda=3$ for RC-generated test suites and $\lambda=4$ for ARC-generated test suites, respectively. 
    SVCC excelled consistently at $\lambda=2$ across both RC and ARC test-suite generation approaches.

    \item 
    The Completeness metric depends heavily on the choice of $q$. 
    For $q=0.25$, the metric peaked at $\lambda=4$ under $R^2$, but shifts to $\lambda=5$ under $\mathcal{R}$. 
    For $q=0.50$, $\lambda=3$ consistently yielded the lowest ranking score across all evaluations.
    For $q=0.75$, the performance universally optimized at $\lambda=2$, except under the $\mathcal{R}$-ARC configuration which requires $\lambda=3$. 

    \item 
    Overall, $\lambda=3$ yielded the most competitive results for VCC, achieving the lowest average score of $2.09$. 
    VID and SVCC both optimized their performance at $\lambda=2$; while 
    WMVCC demonstrated the strongest correlation at $\lambda=4$. 
    For completeness, the group of $q=0.50$ and $\lambda=3$ is the optimal combination (details in Table~\ref{TABLE:rankcomp}).

\end{itemize}

We find that the metrics usually perform well at strength $\lambda=2$ or $\lambda=3$.
To better understand this finding, we make the following analysis: 
\begin{itemize}
    \item 
    VCC considers the number of all possible valid $\lambda$-wise value combinations covered by a given test suite. 
    When $\lambda\leq2$, the $\lambda$-wise VCC is prone to reaching $1.0$, failing to reflect the differentiation in fault-detection effectiveness for high-wise test scenarios;
    When $\lambda\geq4$, the size of the set of all possible valid $\lambda$-wise value combinations derived from the IPM experiences an exponential growth, which significantly diminishes the sensitivity of VCC. 
    $\lambda=3$ performed best, with a peak efficiency attributable to its ability to effectively reconcile the dual limitations of low-wise saturation and high-wise sparsity.
  
    \item 
    According to the definition of VID in Equation~\ref{Equation10}, 
    when $\lambda\geq3$, the denominator of VID grows too rapidly, rendering the metric ineffective in distinguishing the ``avoidance of redundancy'' among test suites. 
    However, when $\lambda=2$, the diversity of pairwise interactions can be most accurately measured.
    
    \item 
    A strict limitation of SVCC is that it requires a PCC value of $1.0$ to perform the simplification. 
    While it is easy to achieve 100\% coverage at $\lambda=2$, the SVCC score frequently drops to 0.0 when $\lambda\geq3$, making it difficult to accurately reflect the quality of the test suite.
    
    \item 
    The $\lambda$-wise WMVCC aggregated information from the $(\lambda-1)$-wise VCC and applies a weighting mechanism for analysis. 
    This mitigates the combinatorial explosion problem in high-strength scenarios, except for cases where $\lambda=5$ or $\lambda=6$, where the combination space becomes prohibitively large.
    As a result, $\lambda=4$ emerged as the optimal strength, representing a critical trade-off between computational efficiency and fault-detection effectiveness.

    \item 
    Based on the preceding combination coverage analysis, $\lambda=3$ represented a well-balanced strength threshold between interaction coverage and computational feasibility.
    Furthermore, setting the threshold at $q=0.50$ establishes a robust benchmark that effectively balances metric sensitivity, avoiding both premature saturation and unachievable strictness. Consequently, the configuration $(q=0.50, \lambda=3)$ emerges as the optimal parameter combination for WMVCC.
\end{itemize}

\finding{1.1}{
The relative effectiveness of the metrics varies substantially with the choice of strength assignments, and most of them perform well within the range of $\lambda=2$ and $\lambda=3$. 
This phenomenon is consistent with the findings of prior studies~\cite{Kuhn2002, Kuhn2004, Kuhn2006}, which report that more than 70\% of faults can be identified by $2$-wise testing, and approximately 90\% by $3$-wise testing.}

\subsubsection{Answer to RQ1.2: }
Table~\ref{TABLE:rank0} shows the ranking scores of various parameter assignments for the ICBM metrics.
Based on the results, we have the following observations:
\begin{itemize}
     \item 
     Regarding the VCC metric, the single-strength configuration outperformed the multiple-strength counterpart:
     the single-strength at $\lambda = 3$ achieved an average score of $2.09$
     ---
     approximately 14\% to 126\% lower than those of other strengths. 
     This substantial reduction clearly demonstrated that a single strength of $\lambda = 3$ is sufficient to reflect the relative effectiveness for VCC. 
     A similar trend is observed for the VID, SVCC, and the completeness metric, where single-strength consistently proved more suitable than multiple-strength.

     \item
     For the WMVCC, the multiple-strength was preferable due to its stable and comparable average scores across five valid strengths
     ---
     the ranking scores for $\lambda = 2, 3, 4, 5,$ and $6$ are $2.91$, $2.29$, $2.28$, $2.54$, and $3.21$, respectively.
     This stability implied that single-strength may fail to adequately evaluate relative effectiveness.
   
\end{itemize}

Under single-strength settings, VCC, VID, SVCC, and the completeness metric performed robustly because they are directly driven by specific combinatorial coverage states; shifting the strength ($\lambda$) inherently introduces distinct, step-like performance gaps due to the non-linear expansion of the combination space.
Since WMVCC employs an equal weight assignment, each strength level can uniformly leverage information from lower strengths to evaluate the test suite. 
Consequently, this multi-strength integration narrows the performance variance across different strengths. 
Therefore, regarding the WMVCC metric, the multiple-strength mechanism performs better.

\finding{1.2}{In summary, single-strength configurations frequently outperform their multiple-strength counterparts.}

\begin{table*}[t!]
\centering
\scriptsize
\caption{$R^2$ values for DBM metrics under RC and ARC}
\label{TABLE:dbm3}
\setlength{\tabcolsep}{2.2mm}{
\begin{tabular}{c|cccccc|cccccc}  \hline
\multirow{2}*{\textbf{Subject}} & \multicolumn{6}{c|}{\textbf{RC}} & \multicolumn{6}{c}{\textbf{ARC}} \\
\cline{2-13}
& \textit{Disc.} & \textit{Disp.}   & \textit{Dive.}  & \textit{Simi.}   & \textit{Nove.}   & \textit{Divg.}  & \textit{Disc.} & \textit{Disp.}   & \textit{Dive.}  & \textit{Simi.}   & \textit{Nove.}   & \textit{Divg.} \\
   
\hline
{FLEX-TS1}	&	\cellcolor{green!30}0.721 	&	0.303 	&	0.306 	&	\cellcolor{magenta!20}0.005 	&	0.504 	&	0.504 	&	0.605 	&	0.502 	&	0.525 	&	0.495 	&	\cellcolor{yellow!50}0.614 	&	\cellcolor{blue!20}0.264 	\\	\hline
{FLEX-TS2}	&	\cellcolor{green!30}0.821 	&	0.331 	&	0.298 	&	\cellcolor{magenta!20}0.001 	&	0.602 	&	0.449 	&	\cellcolor{yellow!50}0.732 	&	0.597 	&	0.596 	&	0.542 	&	0.700 	&	\cellcolor{blue!20}0.071 	\\	\hline
{FLEX-TS3}	&	\cellcolor{green!30}0.840 	&	0.325 	&	0.318 	&	\cellcolor{magenta!20}0.032 	&	0.600 	&	0.587 	&	0.840 	&	0.714 	&	0.799 	&	0.707 	&	\cellcolor{yellow!50}0.849 	&	\cellcolor{blue!20}0.239 	\\	\hline
{FLEX-TS4}	&	\cellcolor{green!30}0.804 	&	0.318 	&	0.398 	&	\cellcolor{magenta!20}0.017 	&	0.624 	&	0.556 	&	\cellcolor{yellow!50}0.866 	&	0.744 	&	0.766 	&	0.612 	&	0.850 	&	\cellcolor{blue!20}0.378 	\\	\hline
{GREP-TS1}	&	\cellcolor{green!30}0.399 	&	0.144 	&	0.189 	&	\cellcolor{magenta!20}0.019 	&	0.183 	&	0.273 	&	\cellcolor{yellow!50}0.504 	&	\cellcolor{blue!20}0.293 	&	0.368 	&	0.380 	&	0.502 	&	\cellcolor{blue!20}0.492 	\\	\hline
{GREP-TS2}	&	\cellcolor{green!30}0.918 	&	0.400 	&	0.350 	&	\cellcolor{magenta!20}0.032 	&	0.725 	&	0.709 	&	0.893 	&	0.768 	&	0.854 	&	0.807 	&	\cellcolor{yellow!50}0.916 	&	\cellcolor{blue!20}0.498 	\\	\hline
{GREP-TS3}	&	\cellcolor{green!30}0.821 	&	0.388 	&	0.481 	&	\cellcolor{magenta!20}0.031 	&	0.780 	&	0.806 	&	0.877 	&	0.736 	&	0.877 	&	0.771 	&	\cellcolor{yellow!50}0.921 	&	\cellcolor{blue!20}0.588 	\\	\hline
{GREP-TS4}	&	\cellcolor{green!30}0.840 	&	0.264 	&	0.321 	&	\cellcolor{magenta!20}0.126 	&	0.572 	&	0.729 	&	0.839 	&	0.741 	&	0.828 	&	0.652 	&	\cellcolor{yellow!50}0.876 	&	\cellcolor{blue!20}0.482 	\\	\hline
{GZIP-TS1}	&	0.069 	&	\cellcolor{magenta!20}0.028 	&	0.041 	&	0.035 	&	0.034 	&	\cellcolor{green!30}0.140 	&	\cellcolor{blue!20}0.010 	&	0.016 	&	0.015 	&	0.013 	&	0.013 	&	\cellcolor{yellow!50}0.026 	\\	\hline
{GZIP-TS2}	&	0.031 	&	0.023 	&	0.026 	&	\cellcolor{magenta!20}0.005 	&	0.023 	&	\cellcolor{green!30}0.054 	&	0.004 	&	0.004 	&	0.004 	&	\cellcolor{blue!20}0.002 	&	0.004 	&	\cellcolor{yellow!50}0.019 	\\	\hline
{GZIP-TS3}	&	\cellcolor{green!30}0.385 	&	0.105 	&	\cellcolor{magenta!20}0.075 	&	0.121 	&	0.178 	&	0.369 	&	\cellcolor{yellow!50}0.563 	&	0.441 	&	0.457 	&	\cellcolor{blue!20}0.269 	&	0.521 	&	0.511 	\\	\hline
{GZIP-TS4}	&	0.201 	&	\cellcolor{magenta!20}0.033 	&	0.056 	&	0.064 	&	0.102 	&	\cellcolor{green!30}0.222 	&	\cellcolor{yellow!50}0.450 	&	0.373 	&	0.402 	&	\cellcolor{blue!20}0.178 	&	0.401 	&	0.445 	\\	\hline
{MAKE-TS1}	&	\cellcolor{green!30}0.257 	&	0.067 	&	0.069 	&	\cellcolor{magenta!20}0.002 	&	0.131 	&	0.144 	&	\cellcolor{yellow!50}0.209 	&	0.140 	&	0.129 	&	0.093 	&	0.169 	&	\cellcolor{blue!20}0.062 	\\	\hline
{MAKE-TS2}	&	\cellcolor{green!30}0.652 	&	0.191 	&	0.140 	&	\cellcolor{magenta!20}0.001 	&	0.404 	&	0.278 	&	\cellcolor{yellow!50}0.659 	&	0.443 	&	0.376 	&	0.265 	&	0.507 	&	\cellcolor{blue!20}0.131 	\\	\hline
{MAKE-TS3}	&	\cellcolor{green!30}0.390 	&	0.163 	&	0.129 	&	\cellcolor{magenta!20}0.000 	&	0.284 	&	0.179 	&	\cellcolor{yellow!50}0.587 	&	0.406 	&	0.464 	&	0.272 	&	0.481 	&	\cellcolor{blue!20}0.182 	\\	\hline
{MAKE-TS4}	&	\cellcolor{green!30}0.282 	&	0.103 	&	0.167 	&	\cellcolor{magenta!20}0.001 	&	0.233 	&	0.123 	&	\cellcolor{yellow!50}0.385 	&	0.312 	&	0.354 	&	0.178 	&	0.346 	&	\cellcolor{blue!20}0.134 	\\	\hline
{SED-TS1}	&	\cellcolor{magenta!20}0.000 	&	\cellcolor{magenta!20}0.000 	&	\cellcolor{magenta!20}0.000 	&	\cellcolor{magenta!20}0.000 	&	\cellcolor{magenta!20}0.000 	&	\cellcolor{magenta!20}0.000 	&	\cellcolor{blue!20}0.000 	&	\cellcolor{blue!20}0.000 	&	\cellcolor{blue!20}0.000 	&	\cellcolor{blue!20}0.000 	&	\cellcolor{blue!20}0.000 	&	\cellcolor{blue!20}0.000 	\\	\hline
{SED-TS2}	&	\cellcolor{green!30}0.684 	&	0.128 	&	0.137 	&	\cellcolor{magenta!20}0.003 	&	0.360 	&	0.541 	&	\cellcolor{yellow!50}0.702 	&	0.411 	&	\cellcolor{blue!20}0.338 	&	0.601 	&	0.531 	&	0.391 	\\	\hline
{SED-TS3}	&	\cellcolor{green!30}0.777 	&	0.345 	&	0.266 	&	\cellcolor{magenta!20}0.011 	&	0.608 	&	0.566 	&	\cellcolor{yellow!50}0.814 	&	0.669 	&	0.665 	&	0.618 	&	0.794 	&	\cellcolor{blue!20}0.510 	\\	\hline
{SED-TS4}	&	\cellcolor{green!30}0.795 	&	0.280 	&	0.291 	&	\cellcolor{magenta!20}0.059 	&	0.581 	&	0.506 	&	\cellcolor{yellow!50}0.819 	&	0.629 	&	0.712 	&	0.713 	&	0.758 	&	\cellcolor{blue!20}0.187 	\\	\hline

{BUSYBOX-TS1}
&	TO&	0.110 	&	0.248 	& \cellcolor{magenta!20}0.022 	&	0.234 	&\cellcolor{green!30}0.508 	&
	TO&	\cellcolor{blue!20}0.346 	&	0.386 	&	0.446 	&	0.379 	&	\cellcolor{yellow!50}0.510 

 \\
\hline
{BUSYBOX-TS2}
& TO	&	0.120 	&	0.267 	& 	\cellcolor{magenta!20}0.001 	&	0.307 	&	\cellcolor{green!30}0.416 	&
	TO&	0.338 	&	0.362 	&	0.326 	&\cellcolor{yellow!50}0.364 	&\cellcolor{blue!20}0.295

 \\
\hline
{BUSYBOX-TS3}
& TO	&	0.242 	&	0.411 	& \cellcolor{magenta!20}0.004 	&	0.434 	&	\cellcolor{green!30}0.616 	&
TO	&\cellcolor{blue!20}0.471 	&	0.515 	&	0.542 	&	0.514 	&	\cellcolor{yellow!50}0.583 

 \\
\hline
{BUSYBOX-TS4}
&	TO&	0.181 	&	0.379 	& \cellcolor{magenta!20}0.011 	&	0.425 	&\cellcolor{green!30}0.591 	&
TO	&	0.541 	&	0.585 	&	\cellcolor{blue!20}0.539 	&\cellcolor{yellow!50}0.590 	&	0.571 

  \\
\hline

{DRUPAL-TS1}
&TO	&	0.195 	&	0.404 	& \cellcolor{magenta!20}0.050 	&	0.395 	&	\cellcolor{green!30}0.716 	&
TO	&	\cellcolor{blue!20}0.571 	&	0.638 	&	0.685 	&	0.643 	&	\cellcolor{yellow!50}0.702 

  \\
\hline
{DRUPAL-TS2}
&TO	&	0.198 	&	0.448 	&\cellcolor{magenta!20}0.032 	&	0.449 	&	\cellcolor{green!30}0.770 	&
TO	&\cellcolor{blue!20}0.623 	&	0.679 	&	0.700 	&	0.676 	&\cellcolor{yellow!50}0.751 

  \\
\hline
{DRUPAL-TS3}
&TO	&	0.149 	&	0.426 	& \cellcolor{magenta!20}0.081 	&	0.362 	&\cellcolor{green!30}0.764 	&
	TO&	0.670 	&	0.729 	&	\cellcolor{blue!20}0.640 	&	0.727 	&	\cellcolor{yellow!50}0.813 

 \\
\hline
{DRUPAL-TS4}
&	TO	&	0.134 	&0.437 	& \cellcolor{magenta!20}0.076 	&	0.399 	&\cellcolor{green!30}0.768 	&
TO	&	0.697 	&	0.764 	&	\cellcolor{blue!20}0.620 	&	0.764 	&\cellcolor{yellow!50}0.834 

 \\
\hline 

{LINUX-TS1}
&TO	&	0.171 	&	0.378 	& \cellcolor{magenta!20}0.005 	&	0.406 	&\cellcolor{green!30}0.458 	&
TO	&	0.468 	& 	0.500 	&	0.406 	&\cellcolor{yellow!50}0.511 	&\cellcolor{blue!20}0.350 

  \\
\hline
{LINUX-TS2}
&TO	&	0.181 	&	0.377 	& 	\cellcolor{magenta!20}0.019 	&	0.419 	&\cellcolor{green!30}0.600 	&
TO	&	0.541 	& 	0.615 	&	0.545 	&	\cellcolor{yellow!50}0.625 	&\cellcolor{blue!20}0.532 

  \\
\hline
{LINUX-TS3}
& TO	&	0.217 	&	0.484 	& \cellcolor{magenta!20}0.000 	&	0.501 	&	\cellcolor{green!30}0.712 	&
TO	&	\cellcolor{blue!20}0.654 	& 	0.720 	&	0.655 	&	\cellcolor{yellow!50}0.732 	&	0.708 

 \\
\hline
{LINUX-TS4}
&TO	&	0.192 	&	0.494 	& \cellcolor{magenta!20}0.023 	&	0.557 	&	\cellcolor{green!30}0.742 	&
TO	&	0.732 	& 	0.777 	&	\cellcolor{blue!20}0.677 	&	\cellcolor{yellow!50}0.790 	&	0.697 

 \\
\hline 

{Standard deviation}	&	\cellcolor{magenta!20}0.303 	&0.106&0.149&  \cellcolor{green!30}0.033	&0.202 &	0.234 &\cellcolor{blue!20}0.296 &	\cellcolor{yellow!50}0.219 &	0.244 &	0.236 &0.254	&0.244

	\\	
{Geometric mean}	&	0.286 	&	0.127 	&	0.180 	&	\cellcolor{magenta!20}0.009 	&	0.252 	&	\cellcolor{green!30}0.326 	&	0.257 	&	0.299 	&	0.317 	&	0.269 	&	\cellcolor{yellow!50}0.340 	&	\cellcolor{blue!20}0.241 	\\	
{Arithmetic mean}	&	\cellcolor{green!30}0.534 	&	0.188 	&	0.275 	&	\cellcolor{magenta!20}0.028 	&	0.388 	&	0.481 	&	\cellcolor{yellow!50}0.568 	&	0.487 	&	0.525 	&	0.467 	&	0.565 	&	\cellcolor{blue!20}0.405 	\\

{The best results}
&\cellcolor{green!30}16	&\cellcolor{magenta!20}0	&\cellcolor{magenta!20}0	&\cellcolor{magenta!20}0&	\cellcolor{magenta!20}0	&15&
\cellcolor{yellow!50}12&	\cellcolor{blue!20}0	&\cellcolor{blue!20}0	&\cellcolor{blue!20}0&	11	&8

\\
{The worst results}
& \cellcolor{green!30}1	&3	&2&	\cellcolor{magenta!20}29&	\cellcolor{green!30}1&	\cellcolor{green!30}1&
2	&7	&2	&8&	\cellcolor{yellow!50}1&	\cellcolor{blue!20}17
\\
\hline 

\end{tabular}
}
\end{table*}

\begin{table*}[t!]
\centering
\scriptsize
\cellcolor{magenta!20}\caption{$\mathcal{R}$ values for DBM metrics under RC and ARC}
\label{TABLE:dbm4}
\setlength{\tabcolsep}{2.2mm}{
\begin{tabular}{c|cccccc|cccccc}  \hline
\multirow{2}*{\textbf{Subject}} & \multicolumn{6}{c|}{\textbf{RC}} & \multicolumn{6}{c}{\textbf{ARC}} \\
\cline{2-13}
& \textit{Disc.} & \textit{Disp.}   & \textit{Dive.}  & \textit{Simi.}   & \textit{Nove.}   & \textit{Divg.}  & \textit{Disc.} & \textit{Disp.}   & \textit{Dive.}  & \textit{Simi.}   & \textit{Nove.}   & \textit{Divg.} \\
   
\hline
{FLEX-TS1}	&	\cellcolor{green!30}0.831 	&	0.516 	&	0.546 	&	\cellcolor{magenta!20}0.462 	&	0.718 	&	0.693 	&	0.835 	&	0.731 	&	0.775 	&	0.794 	&	\cellcolor{yellow!50}0.839 	&	\cellcolor{blue!20}0.588 	\\	\hline
{FLEX-TS2}	&	\cellcolor{green!30}0.876 	&	0.551 	&	0.574 	&	\cellcolor{magenta!20}0.495 	&	0.816 	&	0.698 	&	\cellcolor{yellow!50}0.846 	&	0.748 	&	0.775 	&	0.807 	&	0.843 	&	\cellcolor{blue!20}0.497 	\\	\hline
{FLEX-TS3}	&	\cellcolor{green!30}0.894 	&	0.540 	&	0.593 	&	\cellcolor{magenta!20}0.451 	&	0.808 	&	0.760 	&	0.888 	&	0.769 	&	0.858 	&	0.865 	&	\cellcolor{yellow!50}0.898 	&	\cellcolor{blue!20}0.550 	\\	\hline
{FLEX-TS4}	&	\cellcolor{green!30}0.894 	&	0.513 	&	0.679 	&	\cellcolor{magenta!20}0.341 	&	0.823 	&	0.752 	&	0.920 	&	0.832 	&	0.892 	&	0.846 	&	\cellcolor{yellow!50}0.935 	&	\cellcolor{blue!20}0.634 	\\	\hline
{GREP-TS1}	&	\cellcolor{green!30}0.651 	&	\cellcolor{magenta!20}0.373 	&	0.421 	&	0.400 	&	0.522 	&	0.586 	&	0.699 	&	\cellcolor{blue!20}0.529 	&	0.579 	&	0.595 	&	\cellcolor{yellow!50}0.700 	&	0.688 	\\	\hline
{GREP-TS2}	&	\cellcolor{green!30}0.950 	&	0.608 	&	0.629 	&	\cellcolor{magenta!20}0.541 	&	0.890 	&	0.875 	&	0.924 	&	0.827 	&	0.910 	&	0.916 	&	\cellcolor{yellow!50}0.945 	&	\cellcolor{blue!20}0.727 	\\	\hline
{GREP-TS3}	&	0.894 	&	0.541 	&	0.715 	&	\cellcolor{magenta!20}0.463 	&	\cellcolor{green!30}0.905 	&	0.888 	&	0.919 	&	0.798 	&	0.906 	&	0.894 	&	\cellcolor{yellow!50}0.948 	&	\cellcolor{blue!20}0.770 	\\	\hline
{GREP-TS4}	&	0.898 	&	\cellcolor{magenta!20}0.464 	&	0.633 	&	0.547 	&	0.816 	&	0.889 	&	0.908 	&	0.833 	&	0.925 	&	0.874 	&	\cellcolor{yellow!50}0.949 	&	\cellcolor{blue!20}0.731 	\\	\hline
{GZIP-TS1}	&	0.311 	&	\cellcolor{magenta!20}0.218 	&	0.272 	&	0.300 	&	0.251 	&	\cellcolor{green!30}0.423 	&	\cellcolor{blue!20}0.120 	&	0.131 	&	0.147 	&	0.156 	&	0.139 	&	\cellcolor{yellow!50}0.173 	\\	\hline
{GZIP-TS2}	&	0.213 	&	0.166 	&	0.207 	&	\cellcolor{magenta!20}0.152 	&	0.192 	&	\cellcolor{green!30}0.273 	&	0.072 	&	\cellcolor{blue!20}0.068 	&	0.082 	&	0.088 	&	0.075 	&	\cellcolor{yellow!50}0.133 	\\	\hline
{GZIP-TS3}	&	0.649 	&	0.346 	&	\cellcolor{magenta!20}0.310 	&	0.512 	&	0.490 	&	\cellcolor{green!30}0.684 	&	0.749 	&	0.651 	&	0.684 	&	\cellcolor{blue!20}0.567 	&	0.729 	&	\cellcolor{yellow!50}0.758 	\\	\hline
{GZIP-TS4}	&	0.459 	&	\cellcolor{magenta!20}0.177 	&	0.268 	&	0.354 	&	0.375 	&	\cellcolor{green!30}0.500 	&	0.669 	&	0.583 	&	0.635 	&	\cellcolor{blue!20}0.473 	&	0.656 	&	\cellcolor{yellow!50}0.704 	\\	\hline
{MAKE-TS1}	&	\cellcolor{green!30}0.498 	&	0.284 	&	0.288 	&	\cellcolor{magenta!20}0.273 	&	0.388 	&	0.405 	&	\cellcolor{yellow!50}0.444 	&	0.392 	&	0.377 	&	0.334 	&	0.426 	&	\cellcolor{blue!20}0.300 	\\	\hline
{MAKE-TS2}	&	\cellcolor{green!30}0.799 	&	0.450 	&	0.421 	&	\cellcolor{magenta!20}0.396 	&	0.712 	&	0.614 	&	\cellcolor{yellow!50}0.807 	&	0.695 	&	0.678 	&	0.641 	&	0.774 	&	\cellcolor{blue!20}0.424 	\\	\hline
{MAKE-TS3}	&	\cellcolor{green!30}0.663 	&	0.414 	&	0.445 	&	\cellcolor{magenta!20}0.311 	&	0.623 	&	0.522 	&	\cellcolor{yellow!50}0.790 	&	0.678 	&	0.704 	&	0.672 	&	0.772 	&	\cellcolor{blue!20}0.523 	\\	\hline
{MAKE-TS4}	&	\cellcolor{green!30}0.571 	&	0.313 	&	0.484 	&	\cellcolor{magenta!20}0.276 	&	0.545 	&	0.441 	&	0.653 	&	0.584 	&	0.657 	&	0.547 	&	\cellcolor{yellow!50}0.660 	&	\cellcolor{blue!20}0.486 	\\	\hline
{SED-TS1}	&	\cellcolor{magenta!20}0.000 	&	\cellcolor{magenta!20}0.000 	&	\cellcolor{magenta!20}0.000 	&	\cellcolor{magenta!20}0.000 	&	\cellcolor{magenta!20}0.000 	&	\cellcolor{magenta!20}0.000 	&	\cellcolor{blue!20}0.000 	&	\cellcolor{blue!20}0.000 	&	\cellcolor{blue!20}0.000 	&	\cellcolor{blue!20}0.000 	&	\cellcolor{blue!20}0.000 	&	\cellcolor{blue!20}0.000 	\\	\hline
{SED-TS2}	&	\cellcolor{green!30}0.856 	&	\cellcolor{magenta!20}0.376 	&	0.399 	&	0.411 	&	0.666 	&	0.757 	&	\cellcolor{yellow!50}0.858 	&	0.660 	&	\cellcolor{blue!20}0.613 	&	0.811 	&	0.769 	&	0.679 	\\	\hline
{SED-TS3}	&	\cellcolor{green!30}0.881 	&	0.585 	&	0.564 	&	\cellcolor{magenta!20}0.519 	&	0.841 	&	0.790 	&	0.886 	&	0.782 	&	0.825 	&	0.832 	&	\cellcolor{yellow!50}0.894 	&	\cellcolor{blue!20}0.724 	\\	\hline
{SED-TS4}	&	\cellcolor{green!30}0.885 	&	\cellcolor{magenta!20}0.476 	&	0.590 	&	0.579 	&	0.824 	&	0.751 	&	0.891 	&	0.753 	&	0.861 	&	0.894 	&	\cellcolor{yellow!50}0.900 	&	\cellcolor{blue!20}0.545 	\\	\hline

{BUSYBOX-TS1}
&	TO	&	\cellcolor{magenta!20}0.314 	&	0.469 	&	0.320 	&	0.473 	&\cellcolor{green!30}0.648 	&	TO	&	\cellcolor{blue!20}0.518 	&	0.544 	&	0.615 	&	0.556 	&	\cellcolor{yellow!50}0.652 

 \\
\hline
{BUSYBOX-TS2}
&TO	&	0.257 	&	0.413 	&\cellcolor{magenta!20}0.249 	&	0.474 	&	\cellcolor{green!30}0.507 	&	TO	&	0.469 	&	0.498 	&	0.492 	&\cellcolor{yellow!50}0.505 	&\cellcolor{blue!20}0.447 

 \\
\hline
{BUSYBOX-TS3}
&	TO	&	0.455 	&	0.602 	&	\cellcolor{magenta!20}0.343 	&	0.639 	&\cellcolor{green!30}0.737 	&	TO	&\cellcolor{blue!20}0.631 	&	0.668 	&	0.704 	&	0.672 	&	\cellcolor{yellow!50}0.731 

 \\
\hline
{BUSYBOX-TS4}
&TO	&	0.371 	&	0.542 	&\cellcolor{magenta!20}0.298 	&	0.583 	&	\cellcolor{green!30}0.680 	&	TO	&\cellcolor{blue!20}0.625 	&	0.657 	&	0.644 	&	0.665 	&\cellcolor{yellow!50}0.668 

  \\
\hline 

{DRUPAL-TS1}
&TO	&	0.432 	&	0.654 	&\cellcolor{magenta!20}0.369 	&	0.678 	&\cellcolor{green!30}0.814 	&	TO	&	\cellcolor{blue!20}0.718 	&	0.763 	&	\cellcolor{yellow!50}0.831 	&	0.793 	&	0.820 

  \\
\hline
{DRUPAL-TS2}
&	TO	&	0.439 	&	0.681 	&\cellcolor{magenta!20}0.421 	&	0.711 	&\cellcolor{green!30}0.863 	&	TO	&	\cellcolor{blue!20}0.762 	&	0.807 	&	0.840 	&	0.816 	&	\cellcolor{yellow!50}0.863 

  \\
\hline
{DRUPAL-TS3}
&	TO	&	\cellcolor{magenta!20}0.350 	&	0.650 	&	0.428 	&	0.624 	&	\cellcolor{green!30}0.856 	&	TO	&\cellcolor{blue!20}0.811 	&	0.852 	&	0.815 	&	0.865 	&	\cellcolor{yellow!50}0.889 

 \\
\hline
{DRUPAL-TS4}
&	TO	&\cellcolor{magenta!20}0.347 	&	0.640 	&	0.370 	&	0.635 	&\cellcolor{green!30}0.847 	&	TO	&	0.824 	&	0.875 	&	\cellcolor{blue!20}0.807 	&	0.886 	&\cellcolor{yellow!50}0.900 

 \\
\hline

{LINUX-TS1}
&	TO	&	0.343 	&	0.550 	&	\cellcolor{magenta!20}0.314 	&	0.589 	&\cellcolor{green!30}0.623 	&	TO	&	0.618 	&	0.664 	&	0.627 	&	\cellcolor{yellow!50}0.671 	&	\cellcolor{blue!20}0.575 

  \\
\hline
{LINUX-TS2}
&	TO	&	0.373 	&	0.541 	&	\cellcolor{magenta!20}0.351 	&	0.602 	&	\cellcolor{green!30}0.710 	&	TO	&	\cellcolor{blue!20}0.642 	&	0.700 	&	0.676 	&	\cellcolor{yellow!50}0.711 	&	0.660 

  \\
\hline
{LINUX-TS3}
&TO	&	0.410 	&	0.651 	&	\cellcolor{magenta!20}0.333 	&	0.693 	&	\cellcolor{green!30}0.775 	&	TO	&	\cellcolor{blue!20}0.728 	&	0.776 	&	0.763 	&	\cellcolor{yellow!50}0.789 	&	0.785

 \\
\hline
{LINUX-TS4}
&TO	&	0.405 	&	0.664 	&\cellcolor{magenta!20}0.349 	&	0.731 	&	\cellcolor{green!30}0.842 	&	TO	&	0.808 	&	0.840 	&	\cellcolor{blue!20}0.796 	&\cellcolor{yellow!50}0.851 	&	0.808

 \\
\hline 
{Standard deviation}	&	
\cellcolor{magenta!20}0.261& 	0.130& 	0.164& 	\cellcolor{green!30}0.117 &0.204	&0.196& \cellcolor{blue!20}0.289& 	\cellcolor{yellow!50}0.213 &	0.231& 0.242 &	0.241 &	0.214 

\\	
{Geometric mean}	&	0.434 	&	0.296 	&	0.380 	&	\cellcolor{magenta!20}0.287 	&	\cellcolor{green!30}0.459 	&	0.503 	&	\cellcolor{blue!20}0.409 	&	0.455 	&	0.485 	&	0.478 	&	\cellcolor{yellow!50}0.507 	&	0.446 	\\
{Arithmetic mean}	&	\cellcolor{green!30}0.684 	&	0.388 	&	0.503 	&	\cellcolor{magenta!20}0.373 	&	0.614 	&	0.663 	&	0.694 	&	0.631 	&	0.673 	&	0.663 	&	\cellcolor{yellow!50}0.707 	&	\cellcolor{blue!20}0.607 	\\	

{The best results}
&14	&\cellcolor{magenta!20}0	&\cellcolor{magenta!20}0	&\cellcolor{magenta!20}0	&1&	\cellcolor{green!30}16&
5	&\cellcolor{blue!20}0	&\cellcolor{blue!20}0	&1	&\cellcolor{yellow!50}15&	10

\\
{The worst results}
&\cellcolor{green!30}1&	10&	2	&\cellcolor{magenta!20}22&	\cellcolor{green!30}1&	\cellcolor{green!30}1&
2	&11&	2&	5&	\cellcolor{yellow!50}1	&\cellcolor{blue!20}16
\\
\hline 

\end{tabular}
}
\end{table*}

\begin{table}[!t]
\centering
\scriptsize
\caption{Ranking scores of DBM metrics}
\label{TABLE:rank3}
\setlength{\tabcolsep}{2mm}{
\begin{tabular}{lccccccc}  \hline

\textbf{Metric} & \textit{Discrepancy} & \textit{Dispersion}   & \textit{Diversity}  & \textit{Similarity}   & \textit{Novelty}  & \textit{Divergence}  \\
\hline
$R^2$ values under RC 	&	1.15 	&	4.31 	&	3.53 	&	5.22 	&	2.50 	&	1.81 	\\
$R^2$ values under ARC	&	1.55 	&	3.88 	&	2.97 	&	4.00 	&	2.06 	&	3.88 	\\
$\mathcal{R}$ values under RC 	&	1.25 	&	4.72 	&	3.59 	&	4.97 	&	2.38 	&	1.84 	\\
$\mathcal{R}$ values under ARC	&	2.20 	&	4.59 	&	3.34 	&	3.41 	&	1.84 	&	3.66 	\\
\hline \textit{Average}	&	1.54 	&	4.38 	&	3.36 	&	4.40 	&	2.20 	&	2.80 	\\

\hline
\end{tabular}
}
\end{table}

\begin{table*}[t!]
\centering
\scriptsize
\caption{$R^2$ values for MBM metrics under RC and ARC}
\label{TABLE:mbm3}
\setlength{\tabcolsep}{1.6pt}{
\begin{tabular}{c|c|c}  \hline
\multirow{2}*{\textbf{Subject}} & \multicolumn{1}{c|}{\textbf{RC}} & \multicolumn{1}{c}{\textbf{ARC}} \\
\cline{2-3}
& \textit{Input-parameter-model Mutation Score (IMS)} & \textit{Input-parameter-model Mutation Score (IMS)} \\
   
\hline
{FLEX-TS1}	&	0.565 	&	0.429 	\\	\hline
{FLEX-TS2}	&	0.593 	&	0.422 	\\	\hline
{FLEX-TS3}	&	0.695 	&	0.653 	\\	\hline
{FLEX-TS4}	&	0.629 	&	0.530 	\\	\hline
{GREP-TS1}	&	0.342 	&	0.451 	\\	\hline
{GREP-TS2}	&	0.887 	&	0.851 	\\	\hline
{GREP-TS3}	&	0.891 	&	0.862 	\\	\hline
{GREP-TS4}	&	0.846 	&	0.764 	\\	\hline
{GZIP-TS1}	&	0.104 	&	0.003 	\\	\hline
{GZIP-TS2}	&	0.051 	&	0.011 	\\	\hline
{GZIP-TS3}	&	0.510 	&	0.631 	\\	\hline
{GZIP-TS4}	&	0.371 	&	0.565 	\\	\hline
{MAKE-TS1}	&	0.098 	&	0.029 	\\	\hline
{MAKE-TS2}	&	0.124 	&	0.030 	\\	\hline
{MAKE-TS3}	&	0.081 	&	0.032 	\\	\hline
{MAKE-TS4}	&	0.069 	&	0.027 	\\	\hline
{SED-TS1}	&	0.000 	&	0.000 	\\	\hline
{SED-TS2}	&	0.659 	&	0.646 	\\	\hline
{SED-TS3}	&	0.730 	&	0.727 	\\	\hline
{SED-TS4}	&	0.747 	&	0.794 	\\	\hline

{BUSYBOX-TS1}
&  0.520 	&	 0.441 

 \\
\hline
{BUSYBOX-TS2}
& 0.355 	&	 0.196

 \\
\hline
{BUSYBOX-TS3}
& 0.580 	&	 0.484 

 \\
\hline
{BUSYBOX-TS4}
& 0.532 	&	 0.453 

  \\
\hline 

{DRUPAL-TS1}
& 0.558 	&	 0.494 

  \\
\hline
{DRUPAL-TS2}
& 0.637 	&	 0.540 

  \\
\hline
{DRUPAL-TS3}
& 0.724 	&	 0.638 

 \\
\hline
{DRUPAL-TS4}
& 0.757 	&	 0.693 

 \\
\hline 

{LINUX-TS1}
&  0.369 	&	 0.186 

  \\
\hline
{LINUX-TS2}
& 0.511 	&	 0.396 

  \\
\hline
{LINUX-TS3}
& 0.613 	&	 0.557 

 \\
\hline
{LINUX-TS4}
& 0.716 	&	 0.675 

 \\
\hline 
{Standard deviation}	&	0.261 	&	0.272	\\	
{Geometric mean}	&	0.313 	&	0.213 	\\	
{Arithmetic mean}	&	0.496 	&	0.444 	\\	

\hline 

\end{tabular}
}
\end{table*}

\begin{table*}[t!]
\centering
\scriptsize
\caption{$\mathcal{R}$ values for MBM metrics under RC and ARC}
\label{TABLE:mbm4}
\setlength{\tabcolsep}{1.8pt}{
\begin{tabular}{c|c|c}  \hline
\multirow{2}*{\textbf{Subject}} & \multicolumn{1}{c|}{\textbf{RC}} & \multicolumn{1}{c}{\textbf{ARC}} \\
\cline{2-3}
& \textit{Input-parameter-model Mutation Score (IMS)} & \textit{Input-parameter-model Mutation Score (IMS)} \\
   
\hline
{FLEX-TS1}	&	0.756 	&	0.699 	\\	\hline
{FLEX-TS2}	&	0.746 	&	0.652 	\\	\hline
{FLEX-TS3}	&	0.833 	&	0.798 	\\	\hline
{FLEX-TS4}	&	0.804 	&	0.756 	\\	\hline
{GREP-TS1}	&	0.619 	&	0.680 	\\	\hline
{GREP-TS2}	&	0.955 	&	0.936 	\\	\hline
{GREP-TS3}	&	0.928 	&	0.917 	\\	\hline
{GREP-TS4}	&	0.939 	&	0.895 	\\	\hline
{GZIP-TS1}	&	0.344 	&	0.139 	\\	\hline
{GZIP-TS2}	&	0.273 	&	0.119 	\\	\hline
{GZIP-TS3}	&	0.763 	&	0.841 	\\	\hline
{GZIP-TS4}	&	0.660 	&	0.765 	\\	\hline
{MAKE-TS1}	&	0.320 	&	0.170 	\\	\hline
{MAKE-TS2}	&	0.371 	&	0.180 	\\	\hline
{MAKE-TS3}	&	0.327 	&	0.202 	\\	\hline
{MAKE-TS4}	&	0.318 	&	0.186 	\\	\hline
{SED-TS1}	&	0.000 	&	0.000 	\\	\hline
{SED-TS2}	&	0.841 	&	0.842 	\\	\hline
{SED-TS3}	&	0.843 	&	0.851 	\\	\hline
{SED-TS4}	&	0.851 	&	0.884 	\\	\hline

{BUSYBOX-TS1}
&0.689 	&	0.000 

 \\
\hline
{BUSYBOX-TS2}
&0.464 	&	0.344 

 \\
\hline
{BUSYBOX-TS3}
&0.721 	&	0.669 

 \\
\hline
{BUSYBOX-TS4}
&0.647 	&	0.580 

  \\
\hline 

{DRUPAL-TS1}
&0.773 	&	0.692 

  \\
\hline
{DRUPAL-TS2}
&0.803 	&	0.727 

  \\
\hline
{DRUPAL-TS3}
&0.860 	&	0.772 

 \\
\hline
{DRUPAL-TS4}
&0.883 	&	0.814 

 \\
\hline 

{LINUX-TS1}
&0.547 	&	0.378 

  \\
\hline
{LINUX-TS2}
&0.631 	&	0.553 

  \\
\hline
{LINUX-TS3}
&0.705 	&	0.681 

 \\
\hline
{LINUX-TS4}
&0.809 	&	0.794 
 \\
\hline 

{Standard deviation}	&	0.233 	&	0.295 	\\	
{Geometric mean}	&	0.486 	&	0.310 	\\	
{Arithmetic mean}	&	0.657 	&	0.579 	\\	
\hline 

\end{tabular}
}
\end{table*}

\subsection{Answer to RQ2: Effectiveness Comparisons of DBM Metrics}
\label{subsection:Answer to RQ2}
Tables~\ref{TABLE:dbm3} and~\ref{TABLE:dbm4} provide the detailed exact correlation values for $R^2$ and $\mathcal{R}$ under RC and ARC test suite generation approaches.
Table~\ref{TABLE:rank3} reports the ranking score among all DBM metrics. 
Based on the results, we have the following results:

\begin{itemize}
    \item 
    Regarding the $R^2$ values for DBM metrics under RC and ARC, the discrepancy metric consistently outperformed its counterparts:
    It achieved a top-ranked average score of $1.54$, yielding arithmetic mean values of $0.534$ and $0.568$ for RC- and ARC-generated test suites, respectively.
    
    \item 
    The divergence metric excelled on RC-generated test suites, whereas the novelty metric showed a distinct advantage on ARC-generated test suites. 
    The exact $\mathcal{R}$ values further supported this distinction, where discrepancy is validated as a robust predictor of fault-detection effectiveness for RC-generated test suites, while novelty yielded superior predictive performance for ARC-generated test suites.

\end{itemize}

Discrepancy measures the density degree of sample test cases located in different subdomains of a domain. 
A higher discrepancy value indicates a more evenly distributed set of test cases.
This characteristic aligns with the core philosophy of both test suite construction approaches, which posits that spreading test cases widely across the input space maximizes the probability of intersecting clustered failure regions.
The technique-specific preference stems from the underlying mechanisms of the test generation techniques: 
(1) Traditional random testing generates test cases via a uniform random process, lacking awareness of specific value coverage. 
The divergence metric addresses this by evaluating individual parameter value distributions against an ideal state under a complementary RT assumption. 
(2) Adaptive random testing explicitly optimizes the spatial diversity of the test suite by maximizing pairwise test case distances. 
Consequently, the novelty metric, which assesses test suite sparseness, inherently captures the fundamental philosophy of ART.

\finding{2}{
Across all test scenarios and test generation approaches, the discrepancy metric exhibited the highest correlation, closely followed by the novelty metric. 
These results suggest that the choice of evaluation metrics should be closely aligned with the inherent properties of the test-suite generation approaches, thereby ensuring a more accurate and reliable test-suite evaluation.}

\subsection{Answer to RQ3: Effectiveness Comparisons of MBM Metrics}
Tables~\ref{TABLE:mbm3} and \ref{TABLE:mbm4} show $R^2$ and $\mathcal{R}$ values across all test scenarios and test-suite-generated approaches.
Based on the results, we have the following observations: 

\begin{itemize}
    \item 
    The mean values exhibited a moderate level, with the arithmetic means nearly $0.5$. 
    However, the geometric means tend toward $0.3$, and the correlation for a certain test scenario, SED-TS1, is exactly $0.0$. 
   
    \item 
    The correlation values within the test scenarios tend to increase as $para$ becomes larger
    ---
    For instance, GREP-TS4, SED-TS4, and DRUPAL-TS4 exhibited higher correlations 
    compared to scenarios with fewer parameters, i.e., GREP-TS1, SED-TS1, and DRUPAL-TS1. 
    Notably, this dependency on parameter size is also observed in the ARC-generated test suites.

\end{itemize}

To better understand these observations, we make a theoretical analysis.
As the first test scenario for the SED program ($para=0.4$), SED-TS1 contains only six test suites:
This limited size makes it difficult to establish a correlation between IMS and fault-detection effectiveness, resulting in a correlation value of $0.0$ for this scenario.
As $para$ scales up toward TS4 scenarios, the flattened boolean space experiences an exponential explosion of mutually exclusive constraints. 
This rigorous space-probing capability inherently aligns with the prerequisites for uncovering complex interactive bugs.
We also observe that $\mathcal{R}$ value is higher than $R^2$ value. This is because $R^2$ is intended for exploring linear relationships, and its validity for relative metrics may be compromised by non-linear relationships.

\finding{3}{
We recommend utilizing \textit{Input-parameter-model Mutation Score} (IMS) to evaluate complex test suites:
They consistently demonstrate strong predictive capability for test scenarios with a large number of parameters, suggesting that the IMS serves as a viable alternative for predicting the quality of test suites with many parameters.}

\begin{table}[t!]
\centering
\scriptsize
\caption{Ranking scores of investigated metrics (best)}
\label{TABLE:rankf}
\setlength{\tabcolsep}{2mm}{
\begin{tabular}{lccc}  \hline

\textbf{Metric} & \textbf{\textit{VCC}} & \textbf{\textit{Discrepancy}}   & \textbf{\textit{IMS}}  \\
\hline
$R^2$ values under RC &  
1.41 &	1.95 &	2.16 

 \\ 

$R^2$ values under ARC & 1.31 	&1.80 	&2.34 
 
\\ 

$\mathcal{R}$ values under RC & 1.38 &	2.00 &	2.16 

 \\ 

$\mathcal{R}$ values under ARC &1.31 &	1.90 &	2.28

 \\ 

\hline \textit{Average} & 1.35 & 1.91 &2.24
 \\
\hline
\end{tabular}
}
\end{table}

\begin{table}[h]
\centering
\scriptsize
\caption{Ranking scores of VCC metric under different strengths against the discrepancy metric}
\label{TABLE:rankstrength}
\setlength{\tabcolsep}{2mm}{
\begin{tabular}{lcc}  \hline

\textbf{VCC} & \textbf{\textit{VCC}} & \textbf{\textit{Discrepancy}}     \\
\hline
VCC at strength $\lambda=1$ & 1.42 &	1.29

 \\ \hline

VCC at strength $\lambda=2$ & 1.34 &	1.59 
\\  \hline

VCC at strength $\lambda=3$  & 1.19 &	1.65
 \\ \hline

VCC at strength $\lambda=4$ &1.14 &	1.74
 \\ \hline

VCC at strength $\lambda=5$ &1.20 &1.58
 \\\hline

VCC at strength $\lambda=6$ &1.30 &1.50
 \\
\hline
\end{tabular}
}
\end{table}

\subsection{Answer to RQ4: Effectiveness Comparisons of All Metrics}
Table~\ref{TABLE:rankf} shows the ranking scores of the top-performing metric from each category under both the RC and ARC test suite generation approaches. 

Based on the results, we have the following observations:
\begin{itemize}
    \item 
    VCC emerged as the top-performing metric among all investigated metrics, achieving the lowest ranking score $1.35$, 
    whereas the average ranking score of the best DBM metric is $1.91$. The MBM category achieves the lowest correlation.
    
    \item 
    VCC performed best when limited to $3$-wise testing, while the discrepancy metric applied to all test suites regardless of strength. 
    According to details in Appendix C, the discrepancy outperformed VCC exclusively at $\lambda=1$. 
    In all other scenarios, VCC consistently exhibited superior performance, confirming it as the most effective metric for suite evaluation.
    As shown in Table~\ref{TABLE:rankstrength}, for $\lambda=2,3,4,5,6$, VCC achieved ranking scores of $1.34$, $1.19$, $1.14$, $1.20$, and $1.30$, which are all lower than the corresponding scores of discrepancy ($1.59$, $1.65$, $1.74$, $1.58$, and $1.50$), respectively.

\end{itemize}

To better understand why VCC performed better than the discrepancy or novelty metrics, we conduct an in-depth analysis.
\begin{itemize}
      \item 
      VCC is an interaction coverage-based metric that strictly computes the absolute coverage of individual value combinations. 
      Discrepancy relies on randomly selected subdomains, which inherently measures macroscopic geometric uniformity. 
      Consequently, a test suite with a high discrepancy score might achieve proportional distribution across fined subdomains but still suffer from critical combinatorial gaps at specific parameter-value boundaries. 
      
      \item 
      When $\lambda=1$, the VCC metric suffers from premature metric saturation
      ---
      a sparse test suite containing only two test cases achieved a VCC value exceeding $0.5$
      ---
      VCC fails to accurately reflect the true quality and fault-detection capability of a test suite.
  \end{itemize}

\finding{4}{In summary, the VCC metric is the most effective predictor for comparing combinatorial test suites, owing to its exceptionally high correlation with fault-detection effectiveness.}

\begin{table*}[h]
\centering
\scriptsize
\caption{Total computation time of ICBM metrics}
\label{TABLE:timetotal}
\setlength{\tabcolsep}{2.9mm}{
\begin{tabular}{c|r|r|r|r|r}  \hline

{\textbf{Subject}}
& \textit{VCC} & \textit{VID}   &  \textit{WMVCC}  &  \textit{SVCC}    &\textit{Completeness} 
   \\
\hline
{$\lambda=1$}	&	103.94	&	84.04	&	TO	&	93.17	&	89.61	\\	\hline
{$\lambda=2$}	&	2122.85	&	1950.44	&	2228.47	&	1909.05	&	2105.42	\\	\hline
{$\lambda=3$}	&	58,260.65	&	58,181.86	&	60,489.06	&	56,866.55	&	60,220.15	\\	\hline
{$\lambda=4$}	&	1,373,244.44	&	1,383,555.44	&	1,433,733.35	&	1,354,910.97	&	1,410,320.31	\\	\hline
{$\lambda=5$}	&	557,222.94	&	540,103.29	&	632,691.93	&	571,402.17	&	514,723.99\\	\hline
{$\lambda=6$}		&	633,003.31	&	558,189.30	&	767,949.80	&	496,373.79	&	507,459.75	\\	\hline

\end{tabular}
}
\end{table*}

\begin{table*}[h]
\centering
\scriptsize
\caption{Computation time of DBM metrics}
\label{TABLE:dbmtime1}
\setlength{\tabcolsep}{2.9mm}{
\begin{tabular}{c|r|r|r|r|r|r}  \hline

{\textbf{Subject}}
& \textit{Discrepancy} & \textit{Dispersion}   &  \textit{Diversity}  &  \textit{Similarity}    &\textit{Novelty} & \textit{Divergence} 
   \\
\hline
{FLEX-TS1}	&	552.08 	&	0.11 	&	0.28 	&	0.16 	&	0.18 	&	0.05 	\\	\hline
{FLEX-TS2}	&	1420.11 	&	0.71 	&	1.17 	&	0.86 	&	0.93 	&	0.16 	\\	\hline
{FLEX-TS3}	&	3101.54 	&	1.34 	&	1.28 	&	1.61 	&	1.07 	&	0.19 	\\	\hline
{FLEX-TS4}	&	3101.54 	&	1.34 	&	1.28 	&	1.61 	&	1.05 	&	0.19 	\\	\hline
{GREP-TS1}	&	1240.74 	&	0.04 	&	0.10 	&	0.06 	&	0.12 	&	0.02 	\\	\hline
{GREP-TS2}	&	78,184.83 	&	0.71 	&	1.92 	&	0.95 	&	0.93 	&	0.17 	\\	\hline
{GREP-TS3}	&	70,585.78 	&	0.83 	&	1.18 	&	0.97 	&	1.11 	&	0.17 	\\	\hline
{GREP-TS4}	&	58,499.66 	&	1.00 	&	1.40 	&	1.05 	&	1.13 	&	0.19 	\\	\hline
{GZIP-TS1}	&	1074.50 	&	0.17 	&	0.37 	&	0.27 	&	0.20 	&	0.06 	\\	\hline
{GZIP-TS2}	&	2745.40 	&	0.83 	&	1.17 	&	0.97 	&	0.85 	&	0.15 	\\	\hline
{GZIP-TS3}	&	9991.52 	&	1.92 	&	1.71 	&	1.26 	&	0.93 	&	0.21 	\\	\hline
{GZIP-TS4}	&	58,337.61 	&	1.25 	&	1.49 	&	2.05 	&	1.10 	&	0.26 	\\	\hline
{MAKE-TS1}	&	594.29 	&	0.10 	&	0.19 	&	0.13 	&	0.16 	&	0.03 	\\	\hline
{MAKE-TS2}	&	2188.47	&	0.63 	&	1.03 	&	0.76 	&	0.87 	&	0.11 	\\	\hline
{MAKE-TS3}	&	7156.66 	&	0.82 	&	1.27 	&	0.96 	&	0.99 	&	0.15 	\\	\hline
{MAKE-TS4}	&	23,070.18 	&	1.08 	&	1.54 	&	1.11 	&	1.05 	&	0.20 	\\	\hline
{SED-TS1}	&	91.17 	&	0.17 	&	1.00 	&	0.50 	&	0.33 	&	0.17 	\\	\hline
{SED-TS2}	&	1343.50 	&	0.33 	&	0.71 	&	0.43 	&	0.35 	&	0.11 	\\	\hline
{SED-TS3}	&	9545.19 	&	0.89 	&	1.38 	&	1.02 	&	1.03 	&	0.18 	\\	\hline
{SED-TS4}	&	72,533.55 	&	1.02 	&	1.86 	&	1.15 	&	1.11 	&	0.20 	\\	\hline
{BUSYBOX-TS1}	&	TO	&	1.83 	&	2.76 	&	1.93 	&	1.27 	&	0.31 	\\	\hline
{BUSYBOX-TS2}	&	TO	&	2.81 	&	2.95 	&	2.99 	&	1.62 	&	0.34 	\\	\hline
{BUSYBOX-TS3}	&	TO	&	3.24 	&	5.62 	&	6.40 	&	1.67 	&	0.48 	\\	\hline
{BUSYBOX-TS4}	&	TO	&	7.05 	&	8.21 	&	7.02 	&	2.21 	&	0.68 	\\	\hline
{DRUPAL-TS1}	&	TO	&	1.81 	&	2.48 	&	2.34 	&	1.09 	&	0.33 	\\	\hline
{DRUPAL-TS2}	&	TO	&	2.58 	&	2.45 	&	2.18 	&	1.38 	&	0.42 	\\	\hline
{DRUPAL-TS3}	&	TO	&	5.28 	&	3.59 	&	3.40 	&	1.38 	&	0.56 	\\	\hline
{DRUPAL-TS4}	&	TO	&	4.07 	&	5.19 	&	3.76 	&	1.62 	&	0.48 	\\	\hline
{LINUX-TS1}	&	TO	&	4.37 	&	4.82 	&	3.57 	&	1.84 	&	0.79 	\\	\hline
{LINUX-TS2}	&	TO	&	6.08 	&	8.81 	&	8.25 	&	2.01 	&	0.51 	\\	\hline
{LINUX-TS3}	&	TO	&	5.31 	&	6.95 	&	5.91 	&	2.71 	&	0.56 	\\	\hline
{LINUX-TS4}	&	TO	&	10.78 	&	9.49 	&	9.99 	&	2.94 	&	0.86 	\\	\hline
{Total}	&	405,358.32 	&	70.50 	&	85.65 	&	75.62 	&	37.23 	&	9.29 	\\	\hline

\end{tabular}
}
\end{table*}

\subsection{Answer to RQ5: Efficiency Comparisons}
Table~\ref{TABLE:timetotal} presents the total overhead of measuring ICBM metrics across various strengths. 
Table~\ref{TABLE:dbmtime1} exhibits the computational overhead of measuring the DBM metrics. 
Details are reported in Appendix~\ref{appdendix:C}. 
Based on the results, we have the following observations:
\begin{itemize}
    \item 
    As the coverage strength increases, the computational time rise and peaks at $\lambda=4$, followed by a subsequent decline. 
    This reduction occurred because execution time limits and stringent parameter constraints prevent certain programs from being fully evaluated when the strength was set to $\lambda=5$ or $\lambda=6$.
        
    \item 
    Evaluating test suites using the discrepancy metric incurred a massive overhead compared to other DBM metrics. For instance, while discrepancy required a substantial execution time of 405,358.32\,ms, diversity required merely 85.65\,ms. Consequently, the runtime of discrepancy exceeded that of the divergence metric by a factor of over 43,633\,s. 
    This substantial disparity between the two metrics stems from the fact that computing discrepancy requires 1000 repetitions.
    
    \item
    The total overhead for computing the correlation values between the \textit{Input-parameter-model Mutation Score} (IMS) and fault detection was 433,257.25\,ms (approximately 433.26\,s), which is comparable to that of the discrepancy metric. 
    The computational cost of the ICBM metrics was generally higher than that of the DBM metrics: 
    This elevated cost stems from the necessity of processing a substantial volume of parameter combinations, which inherently scales up the computational complexity.

\end{itemize}

\finding{5}{ 
In practice, the actual computational time required for each evaluation metric aligns closely with its theoretical time complexity; specifically, the overhead of the ICBM metrics is significantly higher than that of the DBM metrics.
}

\subsection{Threats to Validity}
This section examines some potential threats to the validity of our study, including external validity, internal validity, and construct validity.

\subsubsection{External Validity}
Threats to external validity refer to the extent to which research findings can be generalized to other settings. 
In this study, we define the proportion of parameters used to simulate scenarios where certain parameters are missing. In practice, parameter interactions are more complex, and simplifying the IPM may either underestimate or overestimate the effectiveness of the metrics.
To address this potential threat, we used eight programs, and IPMs can be constructed using different strategies involving different parameter selections.
There are various distance measures available for evaluating the difference between two test cases. However, in this study, we exclusively adopt the Hamming distance for all DBM metrics. 
This choice does not pose a significant threat to validity, as our test suites are constructed randomly, where the Hamming distance serves as a suitable and commonly used measure.
A potential threat to validity is the limited number of subject programs and test suites. However, we generated 100 test suites per construction approach for each scenario, each with varying sizes, resulting in approximately 4,900 test suites per test scenario. 
Consequently, the large volume of test suites provides sufficient data for statistical analysis, reducing the impact of the relatively small subject program set.
A new indicator~\cite{Zhang2024} ranges from 0.0 to 1.0 and is therefore more general. 
However, since all indicators are compared on the same benchmarks, whether other indicators reach 1.0 does not affect the fairness of the comparison.

\subsubsection{Internal Validity}
Threats to internal validity relate to potential issues with the execution of our study. One prominent threat is that the evaluation process for highly configurable and complex systems typically requires a substantial amount of time, which may restrict the scalability of the empirical study. To mitigate this threat, we enforced a timeout threshold that stops the examination and designates the corresponding subject as ``TO''.
To ensure the correctness of the experiment, we manually cross-validated our analysis scripts on small-scale examples, thereby reducing the risk of implementation errors in the experimental process.

\subsubsection{Construct Validity}

Construct validity refers to whether we have conducted the studies fairly. 
In this study, we focus on the evaluation metrics for combinatorial test suites and the effectiveness of fault detection.
We used $R^2$ values and $\mathcal{R}$ values to measure correlation, which are commonly used in correlation analysis~\cite{huang2025research,martinez2014distance}.

\section{Practical Guidelines}
\label{SECTION:Guidelines}
Our study provides several practical guidelines for CIT testers on selecting evaluation metrics to assess combinatorial test suites. 
\begin{itemize}
    \item[(1)] 
    \textbf{VCC demonstrates comprehensive dominance as the premier evaluation metric.}
    If computational overhead is ignored, the VCC metric stands out as the most suitable choice for evaluating combinatorial test suites due to its exceptional correlation with fault-detection effectiveness. 
    However, when specifically considering the ICBM category, VCC remains the optimal choice owing to its compelling balance of high effectiveness and minimal overhead.
    
    \item[(2)] 
    \textbf{Single-strength configuration demonstrates general superiority and serves as the foundational baseline for test suite evaluation.}
    Single-strength ICBMs generally outperform their multiple-strength counterparts, leading us to recommend a single-strength metric as a valid predictor for test suite evaluation. 
    Although our experiments indicate that WMVCC necessitates a multiple-strength evaluation, its performance is still fundamentally dependent on the baseline metrics of VCC.
    
    \item[(3)] 
    \textbf{DBM metric selection is intrinsically linked to the test-suite generation approach.}
    Our investigation into DBM metrics demonstrates that test-suite generation approaches significantly influence the subsequent evaluation process. Consequently, we recommend selecting evaluation metrics that align with the specific properties of the test suite to ensure a more reliable and accurate assessment.
    \item[(4)] 
    \textbf{The \textit{Input-parameter-model Mutation Score} (IMS) serves as a premier predictor for assessing high-complexity test suites.}
    Since the IMS consistently demonstrates strong predictive capability for subjects with a large number of parameters, we recommend utilizing this metric to evaluate complex test suites, particularly those characterized by numerous constraints.
    \item[(5)] 
    \textbf{DBM metrics fundamentally outperform ICBMs in execution speed.}
    Considering time efficiency, DBM metrics (excluding Discrepancy) are preferable to ICBM metrics for test suite evaluation due to their significantly lower computational overhead.
\end{itemize}

\section{Conclusions and Future Work}
\label{SECTION:Conclusions}
In this paper, we have conducted an extensive empirical study to evaluate metrics used in black-box testing, which involves eight open-source projects comprising 32 test scenarios and 295,624 combinatorial test suites. 
We classified these metrics into three distinct categories, and within each category, we investigated their correlation with fault-detection effectiveness using two correlation measures across test suites generated by two distinct approaches. 
We evaluated the overall performance via a ranking mechanism to identify the most effective test suite predictor
---
one that delivers strong predictive capability for test suite quality while maintaining a low computational overhead. 
Experimental results indicated that the VCC metric performs best among all evaluated candidates, while DBM metrics generally incur low computational cost. 
Additionally, single-strength metrics prove to be more cost-effective than their multiple-strength counterparts. 
The \textit{Input-parameter-model Mutation Score} (IMS) is highly suitable for subjects characterized by numerous parameters.

Future work should evaluate these and additional metrics across a broader range of subject programs and test suites. 
Since the current study is restricted to ten evaluation metrics, the benchmarks could be further extended to include more intricate and larger-scale programs. 
In addition, we plan to combine multiple evaluation metrics to further explore their synergistic effects, as the comparative performance between individual metrics and aggregated metric groups for predicting test suite quality remains unexamined. 
Furthermore, exploring the correlation between these evaluation metrics and test suites generated via alternative paradigms, such as search-based tools like BenCIGenS~\cite{arcaini2025search} or large language models (LLMs), is also worth investigating.

\section*{Acknowledgment}
This work in this paper was in part supported by the Science and Technology Development Fund of Macau, Macao SAR, under Grant Nos. 0069/2025/RIB2 and 0021/2023/RIA1.

~\\

\bibliographystyle{ACM-Reference-Format}
\bibliography{reference}

\clearpage
\newpage

\appendix
\setcounter{page}{1} 
\pagenumbering{arabic} 
\section*{Appendices} 

\addcontentsline{toc}{section}{Appendices} 

\setcounter{figure}{0}
\renewcommand{\thefigure}{\thesection.\arabic{figure}}

\setcounter{table}{0}
\renewcommand{\thetable}{\thesection.\arabic{table}}

\section{Correlation Values for ICBM}
\label{appendixA}
{
This appendix reports the correlation values for all investigated subjects and metrics under RT and ART, as referenced in Section~\ref{SECTION:ResRQ1}. The notation "NA" indicates that a subject was unevaluated due to strength constraints, while "TO" signifies that the evaluation was unsuccessful because it exceeded the computational time budget (24 hours). In each row, the best and worst values are highlighted in green/yellow and red/purple, respectively. Tables~\ref{TABLE:icbm3} and~\ref{TABLE:icbm4} show the $R^2$ and $\mathcal{R}$ values for the selected top-performing ICBM metrics under RT and ART, respectively. The optimal metric is defined as the one achieving the lowest correlation across different strengths. For instance, VCC involves six strength assignments and obtains the lowest ranking score when $\lambda=3$. Consequently, the values under the $\lambda=3$ configuration are selected as the final results presented in Tables~\ref{TABLE:icbm3} and~\ref{TABLE:icbm4}.
}

\begin{table*}[h]
\centering
\scriptsize

\caption{$R^2$ values of different strengths for VCC under RT and ART}
\label{TABLE:ap3}
\setlength{\tabcolsep}{2mm}{
\begin{tabular}{c|cccccc|cccccc}  \hline
\multirow{2}*{\textbf{Subject}} & \multicolumn{6}{c|}{\textbf{RT}} & \multicolumn{6}{c}{\textbf{ART}} \\
\cline{2-13}
& \textbf{$\lambda=1$} & \textbf{$\lambda=2$}   & \textbf{$\lambda=3$}  & \textbf{$\lambda=4$}   & \textbf{$\lambda=5$}   & \textbf{$\lambda=6$} & \textbf{$\lambda=1$} & \textbf{$\lambda=2$}   & \textbf{$\lambda=3$}  & \textbf{$\lambda=4$}   & \textbf{$\lambda=5$}   & \textbf{$\lambda=6$} 
   \\
\hline
\multirow{1}{*}{FLEX-TS1}	&	\cellcolor{magenta!20}0.622 	&	\cellcolor{green!30}0.745 	&	0.744 	&	0.678 	&	NA	&	NA	&	\cellcolor{blue!20}0.448 	&	0.616 	&	\cellcolor{yellow!50}0.641 	&	0.568 	&	NA	&	NA	\\	\hline
\multirow{1}{*}{FLEX-TS2}	&	\cellcolor{magenta!20}0.580 	&	0.720 	&	0.802 	&	\cellcolor{green!30}0.824 	&	0.791 	&	0.725 	&	\cellcolor{blue!20}0.433 	&	0.567 	&	0.661 	&	\cellcolor{yellow!50}0.711 	&	0.695 	&	0.630 	\\	\hline
\multirow{1}{*}{FLEX-TS3}	&	\cellcolor{magenta!20}0.716 	&	0.848 	&	\cellcolor{green!30}0.890 	&	0.880 	&	0.843 	&	0.797 	&	\cellcolor{blue!20}0.669 	&	0.776 	&	\cellcolor{yellow!50}0.856 	&	\cellcolor{yellow!50}0.856 	&	0.805 	&	0.739 	\\	\hline
\multirow{1}{*}{FLEX-TS4}	&	\cellcolor{magenta!20}0.649 	&	0.834 	&	0.888 	&	\cellcolor{green!30}0.890 	&	0.873 	&	0.853 	&	\cellcolor{blue!20}0.558 	&	0.818 	&	0.918 	&	\cellcolor{yellow!50}0.919 	&	0.882 	&	0.841 	\\	\hline
\multirow{1}{*}{GREP-TS1}	&	0.877 	&	\cellcolor{green!30}0.952 	&	0.924 	&	0.879 	&	0.836 	&	\cellcolor{magenta!20}0.800 	&	0.791 	&	\cellcolor{yellow!50}0.938 	&	0.899 	&	0.840 	&	0.775 	&	\cellcolor{blue!20}0.712 	\\	\hline
\multirow{1}{*}{GREP-TS2}	&	0.881 	&	\cellcolor{green!30}0.899 	&	0.865 	&	0.829 	&	0.797 	&	\cellcolor{magenta!20}0.772 	&	0.801 	&	\cellcolor{yellow!50}0.936 	&	0.903 	&	0.850 	&	0.800 	&	\cellcolor{blue!20}0.760 	\\
\hline 

\end{tabular}
}
\end{table*}

\begin{table*}[h]
\centering
\scriptsize
\caption{$\mathcal{R}$ values of different strengths for VCC under RT and ART}
\label{TABLE:ap4}
\setlength{\tabcolsep}{2mm}{
\begin{tabular}{c|cccccc|cccccc}  \hline
\multirow{2}*{\textbf{Subject}} & \multicolumn{6}{c|}{\textbf{RT}} & \multicolumn{6}{c}{\textbf{ART}} \\
\cline{2-13}
& \textbf{$\lambda=1$} & \textbf{$\lambda=2$}   & \textbf{$\lambda=3$}  & \textbf{$\lambda=4$}   & \textbf{$\lambda=5$}   & \textbf{$\lambda=6$} & \textbf{$\lambda=1$} & \textbf{$\lambda=2$}   & \textbf{$\lambda=3$}  & \textbf{$\lambda=4$}   & \textbf{$\lambda=5$}   & \textbf{$\lambda=6$} 
   \\
\hline

\multirow{1}{*}{FLEX-TS1}	&	\cellcolor{magenta!20}0.762 	&	0.831 	&	\cellcolor{green!30}0.840 	&	0.818 	&	NA	&	NA	&	\cellcolor{blue!20}0.693 	&	0.831 	&	\cellcolor{yellow!50}0.856 	&	0.823 	&	NA	&	NA	\\	\hline
\multirow{1}{*}{FLEX-TS2}	&	\cellcolor{magenta!20}0.749 	&	0.839 	&	0.871 	&	\cellcolor{green!30}0.878 	&	0.868 	&	0.850 	&	\cellcolor{blue!20}0.655 	&	0.775 	&	0.820 	&	\cellcolor{yellow!50}0.843 	&	0.838 	&	0.816 	\\	\hline
\multirow{1}{*}{FLEX-TS3}	&	\cellcolor{magenta!20}0.843 	&	0.911 	&	\cellcolor{green!30}0.924 	&	0.918 	&	0.904 	&	0.887 	&	\cellcolor{blue!20}0.804 	&	0.877 	&	\cellcolor{yellow!50}0.908 	&	0.904 	&	0.883 	&	0.859 	\\	\hline
\multirow{1}{*}{FLEX-TS4}	&	\cellcolor{magenta!20}0.817 	&	0.913 	&	\cellcolor{green!30}0.933 	&	\cellcolor{green!30}0.933 	&	0.927 	&	0.919 	&	\cellcolor{blue!20}0.766 	&	0.917 	&	\cellcolor{yellow!50}0.950 	&	0.948 	&	0.935 	&	0.921 	\\	\hline
\multirow{1}{*}{GREP-TS1}	&	0.945 	&	\cellcolor{green!30}0.970 	&	0.957 	&	0.941 	&	0.925 	&	\cellcolor{magenta!20}0.911 	&	\cellcolor{blue!20}0.699 	&	\cellcolor{yellow!50}0.956 	&	0.938 	&	0.917 	&	0.892 	&	0.867 	\\	\hline
\multirow{1}{*}{GREP-TS2}	&	0.919 	&	\cellcolor{green!30}0.935 	&	0.928 	&	0.919 	&	0.910 	&	\cellcolor{magenta!20}0.903 	&	0.899 	&	\cellcolor{yellow!50}0.953 	&	0.942 	&	0.925 	&	0.909 	&	\cellcolor{blue!20}0.895 	\\
\hline 

\end{tabular}
}
\end{table*}

\begin{table*}[t!]
\centering
\scriptsize
\caption{$R^2$ values of different strengths for VID under RT and ART}
\label{TABLE:ap6}
\setlength{\tabcolsep}{2mm}{
\begin{tabular}{c|cccccc|cccccc}  \hline
\multirow{2}*{\textbf{Subject}} & \multicolumn{6}{c|}{\textbf{RT}} & \multicolumn{6}{c}{\textbf{ART}} \\
\cline{2-13}
& \textbf{$\lambda=1$} & \textbf{$\lambda=2$}   & \textbf{$\lambda=3$}  & \textbf{$\lambda=4$}   & \textbf{$\lambda=5$}   & \textbf{$\lambda=6$} & \textbf{$\lambda=1$} & \textbf{$\lambda=2$}   & \textbf{$\lambda=3$}  & \textbf{$\lambda=4$}   & \textbf{$\lambda=5$}   & \textbf{$\lambda=6$} 
   \\
\hline

\multirow{1}{*}{FLEX-TS1}	&	\cellcolor{green!30}0.730 	&	0.705 	&	0.611 	&	\cellcolor{magenta!20}0.000 	&	NA	&	NA	&	\cellcolor{yellow!50}0.641 	&	0.622 	&	0.481 	&	\cellcolor{blue!20}0.000 	&	NA	&	NA	\\	\hline
\multirow{1}{*}{FLEX-TS2}	&	0.723 	&	0.780 	&	\cellcolor{green!30}0.792 	&	0.760 	&	0.696 	&	\cellcolor{magenta!20}0.000 	&	0.669 	&	\cellcolor{yellow!50}0.717 	&	0.694 	&	0.624 	&	0.539 	&	\cellcolor{blue!20}0.000 	\\	\hline
\multirow{1}{*}{FLEX-TS3}	&	0.764 	&	0.805 	&	\cellcolor{green!30}0.813 	&	0.792 	&	0.753 	&	\cellcolor{magenta!20}0.707 	&	0.834 	&	\cellcolor{yellow!50}0.870 	&	0.832 	&	0.741 	&	0.626 	&	\cellcolor{blue!20}0.496 	\\	\hline
\multirow{1}{*}{FLEX-TS4}	&	0.733 	&	0.801 	&	\cellcolor{green!30}0.823 	&	0.810 	&	0.774 	&	\cellcolor{magenta!20}0.725 	&	0.805 	&	0.903 	&	\cellcolor{yellow!50}0.917 	&	0.862 	&	0.772 	&	\cellcolor{blue!20}0.672 	\\	\hline
\multirow{1}{*}{GREP-TS1}	&	0.851 	&	\cellcolor{green!30}0.883 	&	0.847 	&	0.794 	&	0.738 	&	\cellcolor{magenta!20}0.000 	&	\cellcolor{yellow!50}0.927 	&	0.890 	&	0.768 	&	0.617 	&	0.457 	&	\cellcolor{blue!20}0.000 	\\	\hline
\multirow{1}{*}{GREP-TS2}	&	0.914 	&	\cellcolor{green!30}0.915 	&	0.875 	&	0.819 	&	0.762 	&	\cellcolor{magenta!20}0.705 	&	\cellcolor{yellow!50}0.930 	&	0.911 	&	0.824 	&	0.715 	&	0.600 	&	\cellcolor{blue!20}0.468 	\\
\hline 

\end{tabular}
}
\end{table*}

\begin{table*}[t!]
\centering
\scriptsize
\caption{$\mathcal{R}$ values of different strengths for VID under RT and ART}
\label{TABLE:ap7}
\setlength{\tabcolsep}{2mm}{
\begin{tabular}{c|cccccc|cccccc}  \hline
\multirow{2}*{\textbf{Subject}} & \multicolumn{6}{c|}{\textbf{RT}} & \multicolumn{6}{c}{\textbf{ART}} \\
\cline{2-13}
& \textbf{$\lambda=1$} & \textbf{$\lambda=2$}   & \textbf{$\lambda=3$}  & \textbf{$\lambda=4$}   & \textbf{$\lambda=5$}   & \textbf{$\lambda=6$} & \textbf{$\lambda=1$} & \textbf{$\lambda=2$}   & \textbf{$\lambda=3$}  & \textbf{$\lambda=4$}   & \textbf{$\lambda=5$}   & \textbf{$\lambda=6$} 
   \\
\hline

\multirow{1}{*}{FLEX-TS1}	&	\cellcolor{green!30}0.832 	&	0.814 	&	0.764 	&	\cellcolor{magenta!20}0.000 	&	NA	&	NA	&	\cellcolor{yellow!50}0.859 	&	0.846 	&	0.775 	&	\cellcolor{blue!20}0.000 	&	NA	&	NA	\\	\hline
\multirow{1}{*}{FLEX-TS2}	&	0.855 	&	\cellcolor{green!30}0.866 	&	\cellcolor{green!30}0.866 	&	0.855 	&	0.833 	&	\cellcolor{magenta!20}0.000 	&	0.841 	&	\cellcolor{yellow!50}0.848 	&	0.840 	&	0.816 	&	0.782 	&	\cellcolor{blue!20}0.000 	\\	\hline
\multirow{1}{*}{FLEX-TS3}	&	0.889 	&	\cellcolor{green!30}0.891 	&	0.884 	&	0.869 	&	0.849 	&	\cellcolor{magenta!20}0.828 	&	0.909 	&	\cellcolor{yellow!50}0.911 	&	0.896 	&	0.865 	&	0.820 	&	\cellcolor{blue!20}0.757 	\\	\hline
\multirow{1}{*}{FLEX-TS4}	&	0.893 	&	\cellcolor{green!30}0.906 	&	0.903 	&	0.890 	&	0.868 	&	\cellcolor{magenta!20}0.840 	&	0.925 	&	0.947 	&	\cellcolor{yellow!50}0.948 	&	0.931 	&	0.900 	&	\cellcolor{blue!20}0.861 	\\	\hline
\multirow{1}{*}{GREP-TS1}	&	\cellcolor{green!30}0.942 	&	0.938 	&	0.915 	&	0.886 	&	0.856 	&	\cellcolor{magenta!20}0.000 	&	\cellcolor{yellow!50}0.953 	&	0.937 	&	0.894 	&	0.831 	&	0.743 	&	\cellcolor{blue!20}0.000 	\\	\hline
\multirow{1}{*}{GREP-TS2}	&	\cellcolor{green!30}0.956 	&	0.947 	&	0.928 	&	0.902 	&	0.874 	&	\cellcolor{magenta!20}0.843 	&	\cellcolor{yellow!50}0.953 	&	0.945 	&	0.919 	&	0.884 	&	0.840 	&	\cellcolor{blue!20}0.779 	\\
\hline 

\end{tabular}
}
\end{table*}

\begin{table*}[t!]
\centering
\scriptsize
\caption{$R^2$ values of different strengths for WMVCC under RT and ART}
\label{TABLE:ap9}
\setlength{\tabcolsep}{2mm}{
\begin{tabular}{c|ccccc|ccccc}  \hline
\multirow{2}*{\textbf{Subject}} & \multicolumn{5}{c|}{\textbf{RT}} & \multicolumn{5}{c}{\textbf{ART}} \\
\cline{2-11}
&\textbf{$\lambda=2$}   & \textbf{$\lambda=3$}  & \textbf{$\lambda=4$}   & \textbf{$\lambda=5$}   & \textbf{$\lambda=6$} & \textbf{$\lambda=2$}   & \textbf{$\lambda=3$}  & \textbf{$\lambda=4$}   & \textbf{$\lambda=5$}   & \textbf{$\lambda=6$} 
   \\
\hline

\multirow{1}{*}{FLEX-TS1}	&	\cellcolor{magenta!20}0.719 	&	\cellcolor{green!30}0.752 	&	0.750 	&	NA	&	NA	&	\cellcolor{blue!20}0.588 	&	0.635 	&	\cellcolor{yellow!50}0.637 	&	NA	&	NA	\\	\hline
\multirow{1}{*}{FLEX-TS2}	&	\cellcolor{magenta!20}0.688 	&	0.766 	&	0.818 	&	0.842 	&	\cellcolor{green!30}0.846 	&	\cellcolor{blue!20}0.548 	&	0.629 	&	0.694 	&	0.733 	&	\cellcolor{yellow!50}0.745 	\\	\hline
\multirow{1}{*}{FLEX-TS3}	&	\cellcolor{magenta!20}0.814 	&	0.866 	&	0.889 	&	\cellcolor{green!30}0.894 	&	0.890 	&	\cellcolor{blue!20}0.758 	&	0.827 	&	0.864 	&	\cellcolor{yellow!50}0.872 	&	0.865 	\\	\hline
\multirow{1}{*}{FLEX-TS4}	&	\cellcolor{magenta!20}0.784 	&	0.847 	&	0.876 	&	0.888 	&	\cellcolor{green!30}0.893 	&	\cellcolor{blue!20}0.759 	&	0.864 	&	0.909 	&	0.924 	&	\cellcolor{yellow!50}0.926 	\\	\hline
\multirow{1}{*}{GREP-TS1}	&	0.942 	&	\cellcolor{green!30}0.953 	&	0.947 	&	0.936 	&	\cellcolor{magenta!20}0.924 	&	0.920 	&	\cellcolor{yellow!50}0.938 	&	0.929 	&	0.910 	&	\cellcolor{blue!20}0.890 	\\	\hline
\multirow{1}{*}{GREP-TS2}	&	\cellcolor{green!30}0.906 	&	0.902 	&	0.893 	&	0.882 	&	\cellcolor{magenta!20}0.873 	&	0.920 	&	\cellcolor{yellow!50}0.937 	&	0.929 	&	0.916 	&	\cellcolor{blue!20}0.902 	\\
\hline 

\end{tabular}
}
\end{table*}

\begin{table*}[t!]
\centering
\scriptsize
\caption{$\mathcal{R}$ values of different strengths for WMVCC under RT and ART}
\label{TABLE:ap10}
\setlength{\tabcolsep}{2mm}{
\begin{tabular}{c|ccccc|ccccc}  \hline
\multirow{2}*{\textbf{Subject}} & \multicolumn{5}{c|}{\textbf{RT}} & \multicolumn{5}{c}{\textbf{ART}} \\
\cline{2-11}
&\textbf{$\lambda=2$}   & \textbf{$\lambda=3$}  & \textbf{$\lambda=4$}   & \textbf{$\lambda=5$}   & \textbf{$\lambda=6$} & \textbf{$\lambda=2$}   & \textbf{$\lambda=3$}  & \textbf{$\lambda=4$}   & \textbf{$\lambda=5$}   & \textbf{$\lambda=6$} 
   \\
\hline

\multirow{1}{*}{FLEX-TS1}	&	\cellcolor{magenta!20}0.819 	&	0.839 	&	\cellcolor{green!30}0.842 	&	NA	&	NA	&	\cellcolor{blue!20}0.809 	&	0.846 	&	\cellcolor{yellow!50}0.851 	&	NA	&	NA	\\	\hline
\multirow{1}{*}{FLEX-TS2}	&	\cellcolor{magenta!20}0.824 	&	0.862 	&	0.881 	&	\cellcolor{green!30}0.888 	&	\cellcolor{green!30}0.888 	&	\cellcolor{blue!20}0.761 	&	0.808 	&	0.838 	&	0.853 	&	\cellcolor{yellow!50}0.856 	\\	\hline
\multirow{1}{*}{FLEX-TS3}	&	\cellcolor{magenta!20}0.898 	&	0.919 	&	\cellcolor{green!30}0.925 	&	\cellcolor{green!30}0.925 	&	0.922 	&	\cellcolor{blue!20}0.867 	&	0.900 	&	\cellcolor{yellow!50}0.911 	&	0.910 	&	0.904 	\\	\hline
\multirow{1}{*}{FLEX-TS4}	&	\cellcolor{magenta!20}0.894 	&	0.921 	&	0.931 	&	0.935 	&	\cellcolor{green!30}0.936 	&	\cellcolor{blue!20}0.896 	&	0.937 	&	0.949 	&	\cellcolor{yellow!50}0.952 	&	0.951 	\\	\hline
\multirow{1}{*}{GREP-TS1}	&	0.970 	&	\cellcolor{green!30}0.971 	&	0.967 	&	0.962 	&	\cellcolor{magenta!20}0.957 	&	0.955 	&	\cellcolor{yellow!50}0.956 	&	0.949 	&	0.941 	&	\cellcolor{blue!20}0.932 	\\	\hline
\multirow{1}{*}{GREP-TS2}	&	0.934 	&	\cellcolor{green!30}0.936 	&	0.934 	&	0.932 	&	\cellcolor{magenta!20}0.930 	&	0.947 	&	\cellcolor{yellow!50}0.953 	&	0.949 	&	0.944 	&	\cellcolor{blue!20}0.940 	\\
\hline 

\end{tabular}
}
\end{table*}

\begin{table*}[t!]
\centering
\scriptsize
\caption{$R^2$ values of different strengths for SVCC under RT and ART}
\label{TABLE:ap12}
\setlength{\tabcolsep}{2mm}{
\begin{tabular}{c|cccccc|cccccc}  \hline
\multirow{2}*{\textbf{Subject}} & \multicolumn{6}{c|}{\textbf{RT}} & \multicolumn{6}{c}{\textbf{ART}} \\
\cline{2-13}
& \textbf{$\lambda=1$} & \textbf{$\lambda=2$}   & \textbf{$\lambda=3$}  & \textbf{$\lambda=4$}   & \textbf{$\lambda=5$}   & \textbf{$\lambda=6$} & \textbf{$\lambda=1$} & \textbf{$\lambda=2$}   & \textbf{$\lambda=3$}  & \textbf{$\lambda=4$}   & \textbf{$\lambda=5$}   & \textbf{$\lambda=6$} 
   \\
\hline

\multirow{1}{*}{FLEX-TS1}	&	\cellcolor{green!30}0.612 	&	0.425 	&	\cellcolor{magenta!20}0.000 	&	\cellcolor{magenta!20}0.000 	&	NA	&	NA	&	\cellcolor{yellow!50}0.443 	&	0.436 	&	\cellcolor{blue!20}0.000 	&	\cellcolor{blue!20}0.000 	&	NA	&	NA	\\	\hline
\multirow{1}{*}{FLEX-TS2}	&	0.581 	&	\cellcolor{green!30}0.689 	&	0.638 	&	0.318 	&	\cellcolor{magenta!20}0.000 	&	\cellcolor{magenta!20}0.000 	&	0.438 	&	\cellcolor{yellow!50}0.644 	&	0.518 	&	0.308 	&	\cellcolor{blue!20}0.000 	&	\cellcolor{blue!20}0.000 	\\	\hline
\multirow{1}{*}{FLEX-TS3}	&	0.717 	&	\cellcolor{green!30}0.776 	&	0.454 	&	0.200 	&	\cellcolor{magenta!20}0.000 	&	\cellcolor{magenta!20}0.000 	&	0.666 	&	\cellcolor{yellow!50}0.782 	&	0.607 	&	0.213 	&	\cellcolor{blue!20}0.000 	&	\cellcolor{blue!20}0.000 	\\	\hline
\multirow{1}{*}{FLEX-TS4}	&	0.683 	&	\cellcolor{green!30}0.757 	&	0.473 	&	0.070 	&	\cellcolor{magenta!20}0.000 	&	\cellcolor{magenta!20}0.000 	&	0.606 	&	\cellcolor{yellow!50}0.868 	&	0.770 	&	0.245 	&	\cellcolor{blue!20}0.000 	&	\cellcolor{blue!20}0.000 	\\	\hline
\multirow{1}{*}{GREP-TS1}	&	\cellcolor{green!30}0.885 	&	\cellcolor{magenta!20}0.000 	&	\cellcolor{magenta!20}0.000 	&	\cellcolor{magenta!20}0.000 	&	\cellcolor{magenta!20}0.000 	&	\cellcolor{magenta!20}0.000 	&	\cellcolor{yellow!50}0.846 	&	\cellcolor{blue!20}0.000 	&	\cellcolor{blue!20}0.000 	&	\cellcolor{blue!20}0.000 	&	\cellcolor{blue!20}0.000 	&	\cellcolor{blue!20}0.000 	\\
\hline 

\end{tabular}
}
\end{table*}

\begin{table*}[t!]
\centering
\scriptsize
\caption{$\mathcal{R}$ values of different strengths for SVCC under RT and ART}
\label{TABLE:ap13}
\setlength{\tabcolsep}{2mm}{
\begin{tabular}{c|cccccc|cccccc}  \hline
\multirow{2}*{\textbf{Subject}} & \multicolumn{6}{c|}{\textbf{RT}} & \multicolumn{6}{c}{\textbf{ART}} \\
\cline{2-13}
& \textbf{$\lambda=1$} & \textbf{$\lambda=2$}   & \textbf{$\lambda=3$}  & \textbf{$\lambda=4$}   & \textbf{$\lambda=5$}   & \textbf{$\lambda=6$} & \textbf{$\lambda=1$} & \textbf{$\lambda=2$}   & \textbf{$\lambda=3$}  & \textbf{$\lambda=4$}   & \textbf{$\lambda=5$}   & \textbf{$\lambda=6$} 
   \\
\hline
				
\multirow{1}{*}{FLEX-TS1}	&	\cellcolor{green!30}0.754 	&	0.650 	&	\cellcolor{magenta!20}0.000 	&	\cellcolor{magenta!20}0.000 	&	NA	&	NA	&	0.688 	&	\cellcolor{yellow!50}0.702 	&	\cellcolor{blue!20}0.000 	&	\cellcolor{blue!20}0.000 	&	NA	&	NA	\\	\hline
\multirow{1}{*}{FLEX-TS2}	&	0.744 	&	\cellcolor{green!30}0.797 	&	0.794 	&	0.595 	&	\cellcolor{magenta!20}0.000 	&	\cellcolor{magenta!20}0.000 	&	0.654 	&	\cellcolor{yellow!50}0.789 	&	0.732 	&	0.588 	&	\cellcolor{blue!20}0.000 	&	\cellcolor{blue!20}0.000 	\\	\hline
\multirow{1}{*}{FLEX-TS3}	&	0.837 	&	\cellcolor{green!30}0.857 	&	0.734 	&	0.526 	&	\cellcolor{magenta!20}0.000 	&	\cellcolor{magenta!20}0.000 	&	0.801 	&	\cellcolor{yellow!50}0.855 	&	0.800 	&	0.544 	&	\cellcolor{blue!20}0.000 	&	\cellcolor{blue!20}0.000 	\\	\hline
\multirow{1}{*}{FLEX-TS4}	&	0.818 	&	\cellcolor{green!30}0.850 	&	0.761 	&	0.278 	&	\cellcolor{magenta!20}0.000 	&	\cellcolor{magenta!20}0.000 	&	0.781 	&	\cellcolor{yellow!50}0.918 	&	0.894 	&	0.583 	&	\cellcolor{blue!20}0.000 	&	\cellcolor{blue!20}0.000 	\\	\hline
\multirow{1}{*}{GREP-TS1}	&	\cellcolor{green!30}0.936 	&	\cellcolor{magenta!20}0.000 	&	\cellcolor{magenta!20}0.000 	&	\cellcolor{magenta!20}0.000 	&	\cellcolor{magenta!20}0.000 	&	\cellcolor{magenta!20}0.000 	&	\cellcolor{yellow!50}0.909 	&	\cellcolor{blue!20}0.000 	&	\cellcolor{blue!20}0.000 	&	\cellcolor{blue!20}0.000 	&	\cellcolor{blue!20}0.000 	&	\cellcolor{blue!20}0.000 	\\
\hline 

\end{tabular}
}
\end{table*}

\begin{table*}[t!]
\centering
\scriptsize
\caption{$R^2$ values of different strengths for Completeness(q=0.25) under RT and ART}
\label{TABLE:ap15}
\setlength{\tabcolsep}{2mm}{
\begin{tabular}{c|cccccc|cccccc}  \hline
\multirow{2}*{\textbf{Subject}} & \multicolumn{6}{c|}{\textbf{RT}} & \multicolumn{6}{c}{\textbf{ART}} \\
\cline{2-13}
& \textbf{$\lambda=1$} & \textbf{$\lambda=2$}   & \textbf{$\lambda=3$}  & \textbf{$\lambda=4$}   & \textbf{$\lambda=5$}   & \textbf{$\lambda=6$} & \textbf{$\lambda=1$} & \textbf{$\lambda=2$}   & \textbf{$\lambda=3$}  & \textbf{$\lambda=4$}   & \textbf{$\lambda=5$}   & \textbf{$\lambda=6$} 
   \\
\hline

\multirow{1}{*}{FLEX-TS1}	&	\cellcolor{magenta!20}0.000 	&	0.339 	&	\cellcolor{green!30}0.650 	&	0.515 	&	NA	&	NA	&	\cellcolor{blue!20}0.000 	&	0.301 	&	\cellcolor{yellow!50}0.512 	&	0.403 	&	NA	&	NA	\\	\hline
\multirow{1}{*}{FLEX-TS2}	&	\cellcolor{magenta!20}0.000 	&	0.228 	&	0.413 	&	0.675 	&	\cellcolor{green!30}0.678 	&	0.322 	&	\cellcolor{blue!20}0.000 	&	0.190 	&	0.323 	&	0.549 	&	\cellcolor{yellow!50}0.558 	&	0.294 	\\	\hline
\multirow{1}{*}{FLEX-TS3}	&	\cellcolor{magenta!20}0.000 	&	0.225 	&	0.452 	&	\cellcolor{green!30}0.798 	&	0.789 	&	0.441 	&	\cellcolor{blue!20}0.000 	&	0.205 	&	0.407 	&	\cellcolor{yellow!50}0.739 	&	0.731 	&	0.437 	\\	\hline
\multirow{1}{*}{FLEX-TS4}	&	\cellcolor{magenta!20}0.020 	&	0.249 	&	0.564 	&	\cellcolor{green!30}0.854 	&	0.739 	&	0.377 	&	\cellcolor{blue!20}0.000 	&	0.216 	&	0.552 	&	\cellcolor{yellow!50}0.858 	&	0.794 	&	0.581 	\\	\hline
\multirow{1}{*}{GREP-TS1}	&	\cellcolor{magenta!20}0.000 	&	\cellcolor{green!30}0.224 	&\cellcolor{magenta!20}0.000 	&	\cellcolor{magenta!20}0.000 	&	\cellcolor{magenta!20}0.000 	&	NA	&	\cellcolor{blue!20}0.000 	&	\cellcolor{yellow!50}0.285 	&	\cellcolor{blue!20}0.000 	&	\cellcolor{blue!20}0.000 	&	\cellcolor{blue!20}0.000 	&	NA	\\	\hline
\multirow{1}{*}{GREP-TS2}	&	0.046 	&	\cellcolor{green!30}0.807 	&	0.763 	&	\cellcolor{magenta!20}0.000 	&	\cellcolor{magenta!20}0.000 	&	\cellcolor{magenta!20}0.000 	&	\cellcolor{blue!20}0.000 	&	\cellcolor{yellow!50}0.734 	&	0.663 	&	\cellcolor{blue!20}0.000 	&	\cellcolor{blue!20}0.000 	&	\cellcolor{blue!20}0.000 	\\
\hline 

\end{tabular}
}
\end{table*}

\begin{table*}[t!]
\centering
\scriptsize

\caption{$\mathcal{R}$ values of different strengths for Completeness(q=0.25) under RT and ART}
\label{TABLE:ap16}
\setlength{\tabcolsep}{2mm}{
\begin{tabular}{c|cccccc|cccccc}  \hline
\multirow{2}*{\textbf{Subject}} & \multicolumn{6}{c|}{\textbf{RT}} & \multicolumn{6}{c}{\textbf{ART}} \\
\cline{2-13}
& \textbf{$\lambda=1$} & \textbf{$\lambda=2$}   & \textbf{$\lambda=3$}  & \textbf{$\lambda=4$}   & \textbf{$\lambda=5$}   & \textbf{$\lambda=6$} & \textbf{$\lambda=1$} & \textbf{$\lambda=2$}   & \textbf{$\lambda=3$}  & \textbf{$\lambda=4$}   & \textbf{$\lambda=5$}   & \textbf{$\lambda=6$} 
   \\
\hline

\multirow{1}{*}{FLEX-TS1}	&	\cellcolor{magenta!20}0.000 	&	0.582 	&	\cellcolor{green!30}0.779 	&	0.731 	&	NA	&	NA	&	\cellcolor{blue!20}0.000 	&	0.576 	&	\cellcolor{yellow!50}0.767 	&	0.712 	&	NA	&	NA	\\	\hline
\multirow{1}{*}{FLEX-TS2}	&	\cellcolor{magenta!20}0.000 	&	0.498 	&	0.657 	&	0.799 	&	\cellcolor{green!30}0.821 	&	0.616 	&	\cellcolor{blue!20}0.000 	&	0.427 	&	0.596 	&	0.744 	&	\cellcolor{yellow!50}0.774 	&	0.605 	\\	\hline
\multirow{1}{*}{FLEX-TS3}	&	\cellcolor{magenta!20}0.000 	&	0.504 	&	0.712 	&	0.888 	&	\cellcolor{green!30}0.897 	&	0.713 	&	\cellcolor{blue!20}0.000 	&	0.437 	&	0.662 	&	0.846 	&	\cellcolor{yellow!50}0.867 	&	0.707 	\\	\hline
\multirow{1}{*}{FLEX-TS4}	&	\cellcolor{magenta!20}0.137 	&	0.517 	&	0.789 	&	\cellcolor{green!30}0.924 	&	0.883 	&	0.692 	&	\cellcolor{blue!20}0.000 	&	0.463 	&	0.770 	&	\cellcolor{yellow!50}0.922 	&	0.903 	&	0.810 	\\	\hline
\multirow{1}{*}{GREP-TS1}	&	\cellcolor{magenta!20}0.000 	&	\cellcolor{green!30}0.544 	&	\cellcolor{magenta!20}0.000 	&	\cellcolor{magenta!20}0.000 	&	\cellcolor{magenta!20}0.000 	&	NA	&	\cellcolor{blue!20}0.000 	&	\cellcolor{yellow!50}0.584 	&	\cellcolor{blue!20}0.000 	&	\cellcolor{blue!20}0.000 	&	\cellcolor{blue!20}0.000 	&	NA	\\	\hline
\multirow{1}{*}{GREP-TS2}	&	0.207 	&	\cellcolor{green!30}0.917 	&	0.892 	&	\cellcolor{magenta!20}0.000 	&	\cellcolor{magenta!20}0.000 	&	\cellcolor{magenta!20}0.000 	&	\cellcolor{blue!20}0.000 	&	\cellcolor{yellow!50}0.871 	&	0.839 	&	\cellcolor{blue!20}0.000 	&	\cellcolor{blue!20}0.000 	&	\cellcolor{blue!20}0.000 	\\
\hline 

\end{tabular}
}
\end{table*}

\begin{table*}[t!]
\centering
\scriptsize

\caption{$R^2$ values of different strengths for Completeness(q=0.50) under RT and ART}
\label{TABLE:ap18}
\setlength{\tabcolsep}{2mm}{
\begin{tabular}{c|cccccc|cccccc}  \hline
\multirow{2}*{\textbf{Subject}} & \multicolumn{6}{c|}{\textbf{RT}} & \multicolumn{6}{c}{\textbf{ART}} \\
\cline{2-13}
& \textbf{$\lambda=1$} & \textbf{$\lambda=2$}   & \textbf{$\lambda=3$}  & \textbf{$\lambda=4$}   & \textbf{$\lambda=5$}   & \textbf{$\lambda=6$} & \textbf{$\lambda=1$} & \textbf{$\lambda=2$}   & \textbf{$\lambda=3$}  & \textbf{$\lambda=4$}   & \textbf{$\lambda=5$}   & \textbf{$\lambda=6$} 
   \\
\hline

\multirow{1}{*}{FLEX-TS1}	&	0.253 	&	\cellcolor{green!30}0.626 	&	0.613 	&	\cellcolor{magenta!20}0.000 	&	NA	&	NA	&	0.077 	&	0.478 	&	\cellcolor{yellow!50}0.533 	&	\cellcolor{blue!20}0.000 	&	NA	&	NA	\\	\hline
\multirow{1}{*}{FLEX-TS2}	&	0.233 	&	0.475 	&	0.669 	&	\cellcolor{green!30}0.682 	&	0.462 	&	\cellcolor{magenta!20}0.000 	&	0.070 	&	0.349 	&	\cellcolor{yellow!50}0.543 	&	0.531 	&	0.397 	&	\cellcolor{blue!20}0.000 	\\	\hline
\multirow{1}{*}{FLEX-TS3}	&	0.233 	&	0.544 	&	\cellcolor{green!30}0.838 	&	0.709 	&	0.358 	&	\cellcolor{magenta!20}0.011 	&	0.160 	&	0.467 	&	\cellcolor{yellow!50}0.755 	&	0.712 	&	0.412 	&	\cellcolor{blue!20}0.074 	\\	\hline
\multirow{1}{*}{FLEX-TS4}	&	0.267 	&	0.581 	&	\cellcolor{green!30}0.855 	&	0.684 	&	0.373 	&	\cellcolor{magenta!20}0.000 	&	0.162 	&	0.537 	&	\cellcolor{yellow!50}0.854 	&	0.844 	&	0.529 	&	\cellcolor{blue!20}0.110 	\\	\hline
\multirow{1}{*}{GREP-TS1}	&	\cellcolor{green!30}0.181 	&	\cellcolor{magenta!20}0.000 	&	\cellcolor{magenta!20}0.000 	&	\cellcolor{magenta!20}0.000 	&	\cellcolor{magenta!20}0.000 	&	NA	&	\cellcolor{yellow!50}0.314 	&	\cellcolor{blue!20}0.000 	&	\cellcolor{blue!20}0.000 	&	\cellcolor{blue!20}0.000 	&	\cellcolor{blue!20}0.000 	&	NA	\\
\hline 

\end{tabular}
}
\end{table*}

\begin{table*}[t!]
\centering
\scriptsize

\caption{$\mathcal{R}$ values of different strengths for Completeness(q=0.50) under RT and ART}
\label{TABLE:ap19}
\setlength{\tabcolsep}{2mm}{
\begin{tabular}{c|cccccc|cccccc}  \hline
\multirow{2}*{\textbf{Subject}} & \multicolumn{6}{c|}{\textbf{RT}} & \multicolumn{6}{c}{\textbf{ART}} \\
\cline{2-13}
& \textbf{$\lambda=1$} & \textbf{$\lambda=2$}   & \textbf{$\lambda=3$}  & \textbf{$\lambda=4$}   & \textbf{$\lambda=5$}   & \textbf{$\lambda=6$} & \textbf{$\lambda=1$} & \textbf{$\lambda=2$}   & \textbf{$\lambda=3$}  & \textbf{$\lambda=4$}   & \textbf{$\lambda=5$}   & \textbf{$\lambda=6$} 
   \\
\hline

\multirow{1}{*}{FLEX-TS1}	&	0.474 	&	0.765 	&	\cellcolor{green!30}0.777 	&	\cellcolor{magenta!20}0.000 	&	NA	&	NA	&	0.306 	&	0.745 	&	\cellcolor{yellow!50}0.789 	&	\cellcolor{blue!20}0.000 	&	NA	&	NA	\\	\hline
\multirow{1}{*}{FLEX-TS2}	&	0.456 	&	0.716 	&	0.795 	&	\cellcolor{green!30}0.819 	&	0.712 	&	\cellcolor{magenta!20}0.000 	&	0.255 	&	0.638 	&	0.741 	&	\cellcolor{yellow!50}0.745 	&	0.682 	&	\cellcolor{blue!20}0.000 	\\	\hline
\multirow{1}{*}{FLEX-TS3}	&	0.460 	&	0.776 	&	\cellcolor{green!30}0.903 	&	0.858 	&	0.687 	&	\cellcolor{magenta!20}0.118 	&	0.370 	&	0.725 	&	0.852 	&	\cellcolor{yellow!50}0.854 	&	0.716 	&	\cellcolor{blue!20}0.306 	\\	\hline
\multirow{1}{*}{FLEX-TS4}	&	0.520 	&	0.804 	&	\cellcolor{green!30}0.918 	&	0.853 	&	0.681 	&	\cellcolor{magenta!20}0.000 	&	0.391 	&	0.776 	&	0.926 	&	\cellcolor{yellow!50}0.928 	&	0.807 	&	\cellcolor{blue!20}0.361 	\\	\hline
\multirow{1}{*}{GREP-TS1}	&	\cellcolor{green!30}0.453 	&	\cellcolor{magenta!20}0.000 	&	\cellcolor{magenta!20}0.000 	&	\cellcolor{magenta!20}0.000 	&	\cellcolor{magenta!20}0.000 	&	NA	&	\cellcolor{yellow!50}0.530 	&	\cellcolor{blue!20}0.000 	&	\cellcolor{blue!20}0.000 	&	\cellcolor{blue!20}0.000 	&	\cellcolor{blue!20}0.000 	&	NA	\\
\hline 

\end{tabular}
}
\end{table*}

\begin{table*}[t!]
\centering
\scriptsize

\caption{$R^2$ values of different strengths for Completeness(q=0.75) under RT and ART}
\label{TABLE:ap21}
\setlength{\tabcolsep}{2mm}{
\begin{tabular}{c|cccccc|cccccc}  \hline
\multirow{2}*{\textbf{Subject}} & \multicolumn{6}{c|}{\textbf{RT}} & \multicolumn{6}{c}{\textbf{ART}} \\
\cline{2-13}
& \textbf{$\lambda=1$} & \textbf{$\lambda=2$}   & \textbf{$\lambda=3$}  & \textbf{$\lambda=4$}   & \textbf{$\lambda=5$}   & \textbf{$\lambda=6$} & \textbf{$\lambda=1$} & \textbf{$\lambda=2$}   & \textbf{$\lambda=3$}  & \textbf{$\lambda=4$}   & \textbf{$\lambda=5$}   & \textbf{$\lambda=6$} 
   \\
\hline

\multirow{1}{*}{FLEX-TS1}	&	0.612 	&	\cellcolor{green!30}0.661 	&	0.306 	&	\cellcolor{magenta!20}0.000 	&	NA	&	NA	&	0.443 	&	\cellcolor{yellow!50}0.527 	&	0.296 	&	\cellcolor{blue!20}0.000 	&	NA	&	NA	\\	\hline
\multirow{1}{*}{FLEX-TS2}	&	0.581 	&	\cellcolor{green!30}0.733 	&	\cellcolor{green!30}0.733 	&	0.464 	&	\cellcolor{magenta!20}0.000 	&	\cellcolor{magenta!20}0.000 	&	0.438 	&	0.546 	&	\cellcolor{yellow!50}0.618 	&	0.458 	&	\cellcolor{blue!20}0.000 	&	\cellcolor{blue!20}0.000 	\\	\hline
\multirow{1}{*}{FLEX-TS3}	&	0.717 	&	\cellcolor{green!30}0.812 	&	0.769 	&	0.398 	&	0.065 	&	\cellcolor{magenta!20}0.000 	&	0.666 	&	0.742 	&	\cellcolor{yellow!50}0.803 	&	0.572 	&	0.171 	&	\cellcolor{blue!20}0.000 	\\	\hline
\multirow{1}{*}{FLEX-TS4}	&	0.638 	&	\cellcolor{green!30}0.818 	&	0.753 	&	0.337 	&	\cellcolor{magenta!20}0.000 	&	\cellcolor{magenta!20}0.000 	&	0.542 	&	0.795 	&	\cellcolor{yellow!50}0.882 	&	0.707 	&	0.224 	&	\cellcolor{blue!20}0.000 	\\	\hline
\multirow{1}{*}{GREP-TS1}	&	\cellcolor{green!30}0.224 	&	\cellcolor{magenta!20}0.000 	&	\cellcolor{magenta!20}0.000 	&	\cellcolor{magenta!20}0.000 	&	\cellcolor{magenta!20}0.000 	&	NA	&	\cellcolor{yellow!50}0.285 	&	\cellcolor{blue!20}0.000 	&	\cellcolor{blue!20}0.000 	&	\cellcolor{blue!20}0.000 	&	\cellcolor{blue!20}0.000 	&	NA	\\
\hline 

\end{tabular}
}
\end{table*}

\begin{table*}[t!]
\centering
\scriptsize

\caption{$\mathcal{R}$ values of different strengths for Completeness(q=0.75) under RT and ART}
\label{TABLE:ap22}
\setlength{\tabcolsep}{2mm}{
\begin{tabular}{c|cccccc|cccccc}  \hline
\multirow{2}*{\textbf{Subject}} & \multicolumn{6}{c|}{\textbf{RT}} & \multicolumn{6}{c}{\textbf{ART}} \\
\cline{2-13}
& \textbf{$\lambda=1$} & \textbf{$\lambda=2$}   & \textbf{$\lambda=3$}  & \textbf{$\lambda=4$}   & \textbf{$\lambda=5$}   & \textbf{$\lambda=6$} & \textbf{$\lambda=1$} & \textbf{$\lambda=2$}   & \textbf{$\lambda=3$}  & \textbf{$\lambda=4$}   & \textbf{$\lambda=5$}   & \textbf{$\lambda=6$} 
   \\
\hline

\multirow{1}{*}{FLEX-TS1}	&	0.754 	&	\cellcolor{green!30}0.796 	&	0.590 	&	\cellcolor{magenta!20}0.000 	&	NA	&	NA	&	0.688 	&	\cellcolor{yellow!50}0.772 	&	0.626 	&	\cellcolor{blue!20}0.000 	&	NA	&	NA	\\	\hline
\multirow{1}{*}{FLEX-TS2}	&	0.744 	&	0.828 	&	\cellcolor{green!30}0.839 	&	0.730 	&	\cellcolor{magenta!20}0.000 	&	\cellcolor{magenta!20}0.000 	&	0.654 	&	0.734 	&	\cellcolor{yellow!50}0.795 	&	0.728 	&	\cellcolor{blue!20}0.000 	&	\cellcolor{blue!20}0.000 	\\	\hline
\multirow{1}{*}{FLEX-TS3}	&	0.837 	&	\cellcolor{green!30}0.885 	&	0.878 	&	0.688 	&	0.283 	&	\cellcolor{magenta!20}0.000 	&	0.801 	&	0.856 	&	\cellcolor{yellow!50}0.890 	&	0.788 	&	0.502 	&	\cellcolor{blue!20}0.000 	\\	\hline
\multirow{1}{*}{FLEX-TS4}	&	0.792 	&	\cellcolor{green!30}0.892 	&	0.874 	&	0.660 	&	\cellcolor{magenta!20}0.000 	&	\cellcolor{magenta!20}0.000 	&	0.749 	&	0.893 	&	\cellcolor{yellow!50}0.935 	&	0.885 	&	0.557 	&	\cellcolor{blue!20}0.000 	\\	\hline
\multirow{1}{*}{GREP-TS1}	&	\cellcolor{green!30}0.544 	&	\cellcolor{magenta!20}0.000 	&	\cellcolor{magenta!20}0.000 	&	\cellcolor{magenta!20}0.000 	&	\cellcolor{magenta!20}0.000 	&	NA	&	\cellcolor{yellow!50}0.909 	&	\cellcolor{blue!20}0.000 	&	\cellcolor{blue!20}0.000 	&	\cellcolor{blue!20}0.000 	&	\cellcolor{blue!20}0.000 	&	\cellcolor{blue!20}0.000 	\\
\hline 

\end{tabular}
}
\end{table*}

\FloatBarrier

\section{Correlation Values for Novelty}
\label{appendix:D}

\begin{table*}[h]
\centering
\scriptsize

\caption{$R^2$ values of different assignments for Novelty under RT and ART}
\label{TABLE:ap24}
\setlength{\tabcolsep}{0.8mm}{
\begin{tabular}{c|cccccc|cccccc}  \hline
\multirow{2}*{\textbf{Subject}} & \multicolumn{6}{c|}{\textbf{RT}} & \multicolumn{6}{c}{\textbf{ART}} \\
\cline{2-13}
& \textbf{$\varepsilon=2$} & \textbf{$\varepsilon=5$}   & \textbf{$\varepsilon=15$}  & \textbf{$\varepsilon=|T|/4$}   & \textbf{$\varepsilon=|T|/2$}   & \textbf{$\varepsilon=3|T|/4$} & \textbf{$\varepsilon=2$} & \textbf{$\varepsilon=5$}   & \textbf{$\varepsilon=15$}  & \textbf{$\varepsilon=|T|/4$}   & \textbf{$\varepsilon=|T|/2$}   & \textbf{$\varepsilon=3|T|/4$} 
   \\
\hline

\multirow{1}{*}{FLEX-TS1}	&	\cellcolor{green!30}0.504 	&	0.397 	&	\cellcolor{magenta!20}0.005 	&	0.154 	&	0.155 	&	0.111 	&	\cellcolor{yellow!50}0.614 	&	\cellcolor{yellow!50}0.614 	&	0.495 	&	\cellcolor{blue!20}0.325 	&	0.513 	&	0.553 	\\	\hline
\multirow{1}{*}{FLEX-TS2}	&	0.602 	&	\cellcolor{green!30}0.670 	&	0.487 	&	0.102 	&	0.117 	&	\cellcolor{magenta!20}0.093 	&	0.700 	&	\cellcolor{yellow!50}0.729 	&	0.702 	&	\cellcolor{blue!20}0.570 	&	0.576 	&	0.585 	\\	\hline
\multirow{1}{*}{FLEX-TS3}	&	0.600 	&	\cellcolor{green!30}0.633 	&	0.420 	&	\cellcolor{magenta!20}0.009 	&	0.035 	&	0.015 	&	0.849 	&	\cellcolor{yellow!50}0.855 	&	0.771 	&	0.758 	&	0.759 	&	\cellcolor{blue!20}0.754 	\\	\hline
\multirow{1}{*}{FLEX-TS4}	&	0.624 	&	\cellcolor{green!30}0.667 	&	0.503 	&	0.027 	&	0.046 	&	\cellcolor{magenta!20}0.026 	&	0.850 	&	\cellcolor{yellow!50}0.889 	&	0.857 	&	0.819 	&	0.692 	&	\cellcolor{blue!20}0.669 	\\	\hline
\multirow{1}{*}{GREP-TS1}	&	\cellcolor{green!30}0.183 	&	0.019 	&	0.019 	&	\cellcolor{magenta!20}0.001 	&	0.140 	&	0.055 	&	\cellcolor{yellow!50}0.502 	&	0.380 	&	0.380 	&	\cellcolor{blue!20}0.007 	&	0.373 	&	0.460 	\\	\hline
\multirow{1}{*}{GREP-TS2}	&	0.725 	&	\cellcolor{green!30}0.818 	&	0.570 	&	\cellcolor{magenta!20}0.011 	&	0.157 	&	0.189 	&	\cellcolor{yellow!50}0.916 	&	0.913 	&	\cellcolor{blue!20}0.793 	&	0.834 	&	0.843 	&	0.842 	\\
\hline 

\end{tabular}
}
\end{table*}

\begin{table*}[t!]
\centering
\scriptsize

\caption{$\mathcal{R}$ values of different assignments for Novelty under RT and ART}
\label{TABLE:ap25}
\setlength{\tabcolsep}{0.8mm}{
\begin{tabular}{c|cccccc|cccccc}  \hline
\multirow{2}*{\textbf{Subject}} & \multicolumn{6}{c|}{\textbf{RT}} & \multicolumn{6}{c}{\textbf{ART}} \\
\cline{2-13}
& \textbf{$\varepsilon=2$} & \textbf{$\varepsilon=5$}   & \textbf{$\varepsilon=15$}  & \textbf{$\varepsilon=|T|/4$}   & \textbf{$\varepsilon=|T|/2$}   & \textbf{$\varepsilon=3|T|/4$} & \textbf{$\varepsilon=2$} & \textbf{$\varepsilon=5$}   & \textbf{$\varepsilon=15$}  & \textbf{$\varepsilon=|T|/4$}   & \textbf{$\varepsilon=|T|/2$}   & \textbf{$\varepsilon=3|T|/4$} 
   \\
\hline

\multirow{1}{*}{FLEX-TS1}	&	\cellcolor{green!30}0.718 	&	0.676 	&	0.462 	&	\cellcolor{magenta!20}0.413 	&	0.545 	&	0.520 	&	\cellcolor{yellow!50}0.839 	&	0.838 	&	0.794 	&	\cellcolor{blue!20}0.654 	&	0.799 	&	0.827 	\\	\hline
\multirow{1}{*}{FLEX-TS2}	&	0.816 	&	\cellcolor{green!30}0.835 	&	0.745 	&	\cellcolor{magenta!20}0.514 	&	0.564 	&	0.572 	&	0.843 	&	\cellcolor{yellow!50}0.854 	&	0.828 	&	\cellcolor{blue!20}0.779 	&	0.824 	&	0.818 	\\	\hline
\multirow{1}{*}{FLEX-TS3}	&	0.808 	&	\cellcolor{green!30}0.819 	&	0.707 	&	\cellcolor{magenta!20}0.400 	&	0.448 	&	0.431 	&	0.898 	&	\cellcolor{yellow!50}0.903 	&	\cellcolor{blue!20}0.846 	&	0.856 	&	0.889 	&	0.882 	\\	\hline
\multirow{1}{*}{FLEX-TS4}	&	0.823 	&	\cellcolor{green!30}0.841 	&	0.763 	&	\cellcolor{magenta!20}0.345 	&	0.394 	&	0.377 	&	0.935 	&	\cellcolor{yellow!50}0.944 	&	0.920 	&	0.911 	&	0.896 	&	\cellcolor{blue!20}0.872 	\\	\hline
\multirow{1}{*}{GREP-TS1}	&	\cellcolor{green!30}0.522 	&	0.400 	&	0.400 	&	\cellcolor{magenta!20}0.176 	&	0.455 	&	0.469 	&	\cellcolor{yellow!50}0.700 	&	0.595 	&	0.595 	&	\cellcolor{blue!20}0.219 	&	0.589 	&	0.657 	\\	\hline
\multirow{1}{*}{GREP-TS2}	&	0.890 	&	\cellcolor{green!30}0.920 	&	0.805 	&	\cellcolor{magenta!20}0.445 	&	0.575 	&	0.621 	&	\cellcolor{yellow!50}0.945 	&	\cellcolor{yellow!50}0.945 	&	\cellcolor{blue!20}0.880 	&	0.911 	&	0.929 	&	0.924 	\\
\hline 

\end{tabular}
}
\end{table*}
\FloatBarrier

\section{Rankings of Correlation Values}
\label{appendix: B}
This appendix reports the rankings of correlation values for all investigated subjects and metrics across different strengths. A rank of "1" indicates the highest correlation, where smaller ranking values represent stronger correlations. The rank value of "0" is employed to denote cases that were not evaluated due to two primary reasons: 1) the subject's parameter constraints restricted the evaluation strength, or 2) the execution time exceeded our predefined 24-hour threshold.

\begin{table}[h]
\centering
\scriptsize

\caption{Rankings of $R^2$ values for VCC under RT and ART}
\label{TABLE:rankvccart}
\setlength{\tabcolsep}{0.8mm}{
\begin{tabular}{c|cccccc|cccccc}  \hline
\multirow{2}*{\textbf{Subject}} & \multicolumn{6}{c|}{\textbf{RT}} & \multicolumn{6}{c}{\textbf{ART}} \\
\cline{2-13}
 & \textbf{$\lambda=1$} & \textbf{$\lambda=2$}   & \textbf{$\lambda=3$}  & \textbf{$\lambda=4$}   & \textbf{$\lambda=5$} & \textbf{$\lambda=6$}    & \textbf{$\lambda=1$} & \textbf{$\lambda=2$}   & \textbf{$\lambda=3$}  & \textbf{$\lambda=4$}   & \textbf{$\lambda=5$} & \textbf{$\lambda=6$} \\
\hline
\multirow{1}{*}{FLEX-TS1}	&	4	&	1	&	2	&	3	&	0	&	0	&	4	&	2	&	1	&	3	&	0	&	0	\\	
\multirow{1}{*}{FLEX-TS2}	&	6	&	5	&	2	&	1	&	3	&	4	&	6	&	5	&	3	&	1	&	2	&	4	\\	
\multirow{1}{*}{FLEX-TS3}	&	6	&	3	&	1	&	2	&	4	&	5	&	5	&	3	&	1	&	1	&	2	&	4	\\	
\multirow{1}{*}{FLEX-TS4}	&	6	&	5	&	2	&	1	&	3	&	4	&	6	&	5	&	2	&	1	&	3	&	4	\\	
\multirow{1}{*}{GREP-TS1}	&	4	&	1	&	2	&	3	&	5	&	6	&	4	&	1	&	2	&	3	&	5	&	6	\\	
\multirow{1}{*}{GREP-TS2}	&	2	&	1	&	3	&	4	&	5	&	6	&	4	&	1	&	2	&	3	&	5	&	6	\\
\hline
\end{tabular}
}
\end{table}

\begin{table}[!t]
\centering
\scriptsize

\caption{Rankings of $\mathcal{R}$ values for VCC under RT and ART}
\label{TABLE:rankvccnonart}
\setlength{\tabcolsep}{0.8mm}{
\begin{tabular}{c|cccccc|cccccc}  \hline
\multirow{2}*{\textbf{Subject}} & \multicolumn{6}{c|}{\textbf{RT}} & \multicolumn{6}{c}{\textbf{ART}} \\
\cline{2-13}
 & \textbf{$\lambda=1$} & \textbf{$\lambda=2$}   & \textbf{$\lambda=3$}  & \textbf{$\lambda=4$}   & \textbf{$\lambda=5$} & \textbf{$\lambda=6$}    & \textbf{$\lambda=1$} & \textbf{$\lambda=2$}   & \textbf{$\lambda=3$}  & \textbf{$\lambda=4$}   & \textbf{$\lambda=5$} & \textbf{$\lambda=6$} \\
\hline
\multirow{1}{*}{FLEX-TS1}	&	4	&	2	&	1	&	3	&	0	&	0	&	4	&	2	&	1	&	3	&	0	&	0	\\	
\multirow{1}{*}{FLEX-TS2}	&	6	&	5	&	2	&	1	&	3	&	4	&	6	&	5	&	3	&	1	&	2	&	4	\\	
\multirow{1}{*}{FLEX-TS3}	&	6	&	3	&	1	&	2	&	4	&	5	&	6	&	4	&	1	&	2	&	3	&	5	\\	
\multirow{1}{*}{FLEX-TS4}	&	5	&	4	&	1	&	1	&	2	&	3	&	6	&	5	&	1	&	2	&	3	&	4	\\	
\multirow{1}{*}{GREP-TS1}	&	3	&	1	&	2	&	4	&	5	&	6	&	6	&	1	&	2	&	3	&	4	&	5	\\	
\multirow{1}{*}{GREP-TS2}	&	3	&	1	&	2	&	3	&	4	&	5	&	5	&	1	&	2	&	3	&	4	&	6	\\
\hline
\end{tabular}
}
\end{table}

\begin{table}[!t]
\centering
\scriptsize

\caption{Rankings of $R^2$ values for VID under RT and ART}
\label{TABLE:rankvidart}
\setlength{\tabcolsep}{0.8mm}{
\begin{tabular}{c|cccccc|cccccc}  \hline
\multirow{2}*{\textbf{Subject}} & \multicolumn{6}{c|}{\textbf{RT}} & \multicolumn{6}{c}{\textbf{ART}} \\
\cline{2-13}
 & \textbf{$\lambda=1$} & \textbf{$\lambda=2$}   & \textbf{$\lambda=3$}  & \textbf{$\lambda=4$}   & \textbf{$\lambda=5$} & \textbf{$\lambda=6$}    & \textbf{$\lambda=1$} & \textbf{$\lambda=2$}   & \textbf{$\lambda=3$}  & \textbf{$\lambda=4$}   & \textbf{$\lambda=5$} & \textbf{$\lambda=6$} \\
\hline
\multirow{1}{*}{FLEX-TS1}	&	1	&	2	&	3	&	4	&	0	&	0	&	1	&	2	&	3	&	4	&	0	&	0	\\	
\multirow{1}{*}{FLEX-TS2}	&	4	&	2	&	1	&	3	&	5	&	6	&	3	&	1	&	2	&	4	&	5	&	6	\\	
\multirow{1}{*}{FLEX-TS3}	&	4	&	2	&	1	&	3	&	5	&	6	&	2	&	1	&	3	&	4	&	5	&	6	\\	
\multirow{1}{*}{FLEX-TS4}	&	5	&	3	&	1	&	2	&	4	&	6	&	4	&	2	&	1	&	3	&	5	&	6	\\	
\multirow{1}{*}{GREP-TS1}	&	2	&	1	&	3	&	4	&	5	&	6	&	1	&	2	&	3	&	4	&	5	&	6	\\	
\multirow{1}{*}{GREP-TS2}	&	2	&	1	&	3	&	4	&	5	&	6	&	1	&	2	&	3	&	4	&	5	&	6	\\
\hline
\end{tabular}
}
\end{table}

\begin{table}[!t]
\centering
\scriptsize

\caption{Rankings of $\mathcal{R}$ values for VID under RT and ART}
\label{TABLE:rankvidnonart}
\setlength{\tabcolsep}{0.8mm}{
\begin{tabular}{c|cccccc|cccccc}  \hline
\multirow{2}*{\textbf{Subject}} & \multicolumn{6}{c|}{\textbf{RT}} & \multicolumn{6}{c}{\textbf{ART}} \\
\cline{2-13}
 & \textbf{$\lambda=1$} & \textbf{$\lambda=2$}   & \textbf{$\lambda=3$}  & \textbf{$\lambda=4$}   & \textbf{$\lambda=5$} & \textbf{$\lambda=6$}    & \textbf{$\lambda=1$} & \textbf{$\lambda=2$}   & \textbf{$\lambda=3$}  & \textbf{$\lambda=4$}   & \textbf{$\lambda=5$} & \textbf{$\lambda=6$} \\
\hline
\multirow{1}{*}{FLEX-TS1}	&	1	&	2	&	3	&	4	&	0	&	0	&	1	&	2	&	3	&	4	&	0	&	0	\\	
\multirow{1}{*}{FLEX-TS2}	&	2	&	1	&	1	&	2	&	3	&	4	&	2	&	1	&	3	&	4	&	5	&	6	\\	
\multirow{1}{*}{FLEX-TS3}	&	2	&	1	&	3	&	4	&	5	&	6	&	2	&	1	&	3	&	4	&	5	&	6	\\	
\multirow{1}{*}{FLEX-TS4}	&	3	&	1	&	2	&	4	&	5	&	6	&	4	&	2	&	1	&	3	&	5	&	6	\\	
\multirow{1}{*}{GREP-TS1}	&	1	&	2	&	3	&	4	&	5	&	6	&	1	&	2	&	3	&	4	&	5	&	6	\\	
\multirow{1}{*}{GREP-TS2}	&	1	&	2	&	3	&	4	&	5	&	6	&	1	&	2	&	3	&	4	&	5	&	6	\\
\hline
\end{tabular}
}
\end{table}

\begin{table}[!t]
\centering
\scriptsize

\caption{Rankings of $R^2$ values for WMVCC under RT and ART}
\label{TABLE:rankwmvccart}
\setlength{\tabcolsep}{0.8mm}{
\begin{tabular}{c|cccccc|cccccc}  \hline
\multirow{2}*{\textbf{Subject}} & \multicolumn{6}{c|}{\textbf{RT}} & \multicolumn{6}{c}{\textbf{ART}} \\
\cline{2-13}
 & \textbf{$\lambda=1$} & \textbf{$\lambda=2$}   & \textbf{$\lambda=3$}  & \textbf{$\lambda=4$}   & \textbf{$\lambda=5$} & \textbf{$\lambda=6$}    & \textbf{$\lambda=1$} & \textbf{$\lambda=2$}   & \textbf{$\lambda=3$}  & \textbf{$\lambda=4$}   & \textbf{$\lambda=5$} & \textbf{$\lambda=6$} \\
\hline
\multirow{1}{*}{FLEX-TS1}	&	0	&	3	&	1	&	2	&	0	&	0	&	0	&	3	&	2	&	1	&	0	&	0	\\	
\multirow{1}{*}{FLEX-TS2}	&	0	&	5	&	4	&	3	&	2	&	1	&	0	&	5	&	4	&	3	&	2	&	1	\\	
\multirow{1}{*}{FLEX-TS3}	&	0	&	5	&	4	&	3	&	1	&	2	&	0	&	5	&	4	&	3	&	1	&	2	\\	
\multirow{1}{*}{FLEX-TS4}	&	0	&	5	&	4	&	3	&	2	&	1	&	0	&	5	&	4	&	3	&	2	&	1	\\	
\multirow{1}{*}{GREP-TS1}	&	0	&	3	&	1	&	2	&	4	&	5	&	0	&	3	&	1	&	2	&	4	&	5	\\	
\multirow{1}{*}{GREP-TS2}	&	0	&	1	&	2	&	3	&	4	&	5	&	0	&	3	&	1	&	2	&	4	&	5	\\
\hline
\end{tabular}
}
\end{table}

\begin{table}[!t]
\centering
\scriptsize

\caption{Rankings of $\mathcal{R}$ values for WMVCC under RT and ART}
\label{TABLE:rankwmvccnonart}
\setlength{\tabcolsep}{0.8mm}{
\begin{tabular}{c|cccccc|cccccc}  \hline
\multirow{2}*{\textbf{Subject}} & \multicolumn{6}{c|}{\textbf{RT}} & \multicolumn{6}{c}{\textbf{ART}} \\
\cline{2-13}
 & \textbf{$\lambda=1$} & \textbf{$\lambda=2$}   & \textbf{$\lambda=3$}  & \textbf{$\lambda=4$}   & \textbf{$\lambda=5$} & \textbf{$\lambda=6$}    & \textbf{$\lambda=1$} & \textbf{$\lambda=2$}   & \textbf{$\lambda=3$}  & \textbf{$\lambda=4$}   & \textbf{$\lambda=5$} & \textbf{$\lambda=6$} \\
\hline
\multirow{1}{*}{FLEX-TS1}	&	0	&	3	&	2	&	1	&	0	&	0	&	0	&	3	&	2	&	1	&	0	&	0	\\	
\multirow{1}{*}{FLEX-TS2}	&	0	&	4	&	3	&	2	&	1	&	1	&	0	&	5	&	4	&	3	&	2	&	1	\\	
\multirow{1}{*}{FLEX-TS3}	&	0	&	4	&	3	&	1	&	1	&	2	&	0	&	5	&	4	&	1	&	2	&	3	\\	
\multirow{1}{*}{FLEX-TS4}	&	0	&	5	&	4	&	3	&	2	&	1	&	0	&	5	&	4	&	3	&	1	&	2	\\	
\multirow{1}{*}{GREP-TS1}	&	0	&	2	&	1	&	3	&	4	&	5	&	0	&	2	&	1	&	3	&	4	&	5	\\	
\multirow{1}{*}{GREP-TS2}	&	0	&	2	&	1	&	2	&	3	&	4	&	0	&	3	&	1	&	2	&	4	&	5	\\
\hline
\end{tabular}
}
\end{table}

\begin{table}[!t]
\centering
\scriptsize

\caption{Rankings of $R^2$ values for SVCC under RT and ART}
\label{TABLE:ranksvccart}
\setlength{\tabcolsep}{0.8mm}{
\begin{tabular}{c|cccccc|cccccc}  \hline
\multirow{2}*{\textbf{Subject}} & \multicolumn{6}{c|}{\textbf{RT}} & \multicolumn{6}{c}{\textbf{ART}} \\
\cline{2-13}
 & \textbf{$\lambda=1$} & \textbf{$\lambda=2$}   & \textbf{$\lambda=3$}  & \textbf{$\lambda=4$}   & \textbf{$\lambda=5$} & \textbf{$\lambda=6$}    & \textbf{$\lambda=1$} & \textbf{$\lambda=2$}   & \textbf{$\lambda=3$}  & \textbf{$\lambda=4$}   & \textbf{$\lambda=5$} & \textbf{$\lambda=6$} \\
\hline
\multirow{1}{*}{FLEX-TS1}	&	1	&	2	&	3	&	3	&	0	&	0	&	1	&	2	&	3	&	3	&	0	&	0	\\	
\multirow{1}{*}{FLEX-TS2}	&	3	&	1	&	2	&	4	&	5	&	5	&	3	&	1	&	2	&	4	&	5	&	5	\\	
\multirow{1}{*}{FLEX-TS3}	&	2	&	1	&	3	&	4	&	5	&	5	&	2	&	1	&	3	&	4	&	5	&	5	\\	
\multirow{1}{*}{FLEX-TS4}	&	2	&	1	&	3	&	4	&	5	&	5	&	3	&	1	&	2	&	4	&	5	&	5	\\	
\multirow{1}{*}{GREP-TS1}	&	1	&	3	&	3	&	3	&	2	&	2	&	1	&	3	&	3	&	3	&	2	&	2	\\	
\multirow{1}{*}{GREP-TS2}	&	1	&	2	&	4	&	4	&	3	&	3	&	1	&	2	&	4	&	4	&	3	&	3	\\
\hline
\end{tabular}
}
\end{table}

\begin{table}[!t]
\centering
\scriptsize

\caption{Rankings of $\mathcal{R}$ values for SVCC under RT and ART}
\label{TABLE:ranksvccnonart}
\setlength{\tabcolsep}{0.8mm}{
\begin{tabular}{c|cccccc|cccccc}  \hline
\multirow{2}*{\textbf{Subject}} & \multicolumn{6}{c|}{\textbf{RT}} & \multicolumn{6}{c}{\textbf{ART}} \\
\cline{2-13}
 & \textbf{$\lambda=1$} & \textbf{$\lambda=2$}   & \textbf{$\lambda=3$}  & \textbf{$\lambda=4$}   & \textbf{$\lambda=5$} & \textbf{$\lambda=6$}    & \textbf{$\lambda=1$} & \textbf{$\lambda=2$}   & \textbf{$\lambda=3$}  & \textbf{$\lambda=4$}   & \textbf{$\lambda=5$} & \textbf{$\lambda=6$} \\
\hline
\multirow{1}{*}{FLEX-TS1}	&	1	&	2	&	3	&	3	&	0	&	0	&	2	&	1	&	3	&	3	&	0	&	0	\\	
\multirow{1}{*}{FLEX-TS2}	&	3	&	1	&	2	&	4	&	5	&	5	&	3	&	1	&	2	&	4	&	5	&	5	\\	
\multirow{1}{*}{FLEX-TS3}	&	2	&	1	&	3	&	4	&	5	&	5	&	2	&	1	&	3	&	4	&	5	&	5	\\	
\multirow{1}{*}{FLEX-TS4}	&	2	&	1	&	3	&	4	&	5	&	5	&	3	&	1	&	2	&	4	&	5	&	5	\\	
\multirow{1}{*}{GREP-TS1}	&	1	&	3	&	3	&	3	&	2	&	2	&	1	&	3	&	3	&	3	&	2	&	2	\\	
\multirow{1}{*}{GREP-TS2}	&	1	&	2	&	4	&	4	&	3	&	3	&	1	&	2	&	4	&	4	&	3	&	3	\\
\hline
\end{tabular}
}
\end{table}

\begin{table}[!t]
\centering
\scriptsize

\caption{Rankings of $R^2$ values for Completeness(q=0.25) under RT and ART}
\label{TABLE:rankcomp25art}
\setlength{\tabcolsep}{0.8mm}{
\begin{tabular}{c|cccccc|cccccc}  \hline
\multirow{2}*{\textbf{Subject}} & \multicolumn{6}{c|}{\textbf{RT}} & \multicolumn{6}{c}{\textbf{ART}} \\
\cline{2-13}
 & \textbf{$\lambda=1$} & \textbf{$\lambda=2$}   & \textbf{$\lambda=3$}  & \textbf{$\lambda=4$}   & \textbf{$\lambda=5$} & \textbf{$\lambda=6$}    & \textbf{$\lambda=1$} & \textbf{$\lambda=2$}   & \textbf{$\lambda=3$}  & \textbf{$\lambda=4$}   & \textbf{$\lambda=5$} & \textbf{$\lambda=6$} \\
\hline
\multirow{1}{*}{FLEX-TS1}	&	4	&	3	&	1	&	2	&	0	&	0	&	4	&	3	&	1	&	2	&	0	&	0	\\	
\multirow{1}{*}{FLEX-TS2}	&	6	&	5	&	3	&	2	&	1	&	4	&	6	&	5	&	3	&	2	&	1	&	4	\\	
\multirow{1}{*}{FLEX-TS3}	&	6	&	5	&	3	&	1	&	2	&	4	&	6	&	5	&	4	&	1	&	2	&	3	\\	
\multirow{1}{*}{FLEX-TS4}	&	6	&	5	&	3	&	1	&	2	&	4	&	6	&	5	&	4	&	1	&	2	&	3	\\	
\multirow{1}{*}{GREP-TS1}	&	3	&	1	&	3	&	3	&	2	&	0	&	3	&	1	&	3	&	3	&	2	&	0	\\	
\multirow{1}{*}{GREP-TS2}	&	3	&	1	&	2	&	5	&	4	&	4	&	4	&	1	&	2	&	4	&	3	&	3	\\
\hline
\end{tabular}
}
\end{table}

\begin{table}[!t]
\centering
\scriptsize

\caption{Rankings of $\mathcal{R}$ values for Completeness(q=0.25) under RT and ART}
\label{TABLE:rankcomp25nonart}
\setlength{\tabcolsep}{0.8mm}{
\begin{tabular}{c|cccccc|cccccc}  \hline
\multirow{2}*{\textbf{Subject}} & \multicolumn{6}{c|}{\textbf{RT}} & \multicolumn{6}{c}{\textbf{ART}} \\
\cline{2-13}
 & \textbf{$\lambda=1$} & \textbf{$\lambda=2$}   & \textbf{$\lambda=3$}  & \textbf{$\lambda=4$}   & \textbf{$\lambda=5$} & \textbf{$\lambda=6$}    & \textbf{$\lambda=1$} & \textbf{$\lambda=2$}   & \textbf{$\lambda=3$}  & \textbf{$\lambda=4$}   & \textbf{$\lambda=5$} & \textbf{$\lambda=6$} \\
\hline
\multirow{1}{*}{FLEX-TS1}	&	4	&	3	&	1	&	2	&	0	&	0	&	4	&	3	&	1	&	2	&	0	&	0	\\	
\multirow{1}{*}{FLEX-TS2}	&	6	&	5	&	3	&	2	&	1	&	4	&	6	&	5	&	4	&	2	&	1	&	3	\\	
\multirow{1}{*}{FLEX-TS3}	&	6	&	5	&	4	&	2	&	1	&	3	&	6	&	5	&	4	&	2	&	1	&	3	\\	
\multirow{1}{*}{FLEX-TS4}	&	6	&	5	&	3	&	1	&	2	&	4	&	6	&	5	&	4	&	1	&	2	&	3	\\	
\multirow{1}{*}{GREP-TS1}	&	3	&	1	&	3	&	3	&	2	&	0	&	3	&	1	&	3	&	3	&	2	&	0	\\	
\multirow{1}{*}{GREP-TS2}	&	3	&	1	&	2	&	5	&	4	&	4	&	4	&	1	&	2	&	4	&	3	&	3	\\
\hline
\end{tabular}
}
\end{table}

\begin{table}[!t]
\centering
\scriptsize

\caption{Rankings of $R^2$ values for Completeness(q=0.50) under RT and ART}
\label{TABLE:rankcomp50art}
\setlength{\tabcolsep}{0.8mm}{
\begin{tabular}{c|cccccc|cccccc}  \hline
\multirow{2}*{\textbf{Subject}} & \multicolumn{6}{c|}{\textbf{RT}} & \multicolumn{6}{c}{\textbf{ART}} \\
\cline{2-13}
 & \textbf{$\lambda=1$} & \textbf{$\lambda=2$}   & \textbf{$\lambda=3$}  & \textbf{$\lambda=4$}   & \textbf{$\lambda=5$} & \textbf{$\lambda=6$}    & \textbf{$\lambda=1$} & \textbf{$\lambda=2$}   & \textbf{$\lambda=3$}  & \textbf{$\lambda=4$}   & \textbf{$\lambda=5$} & \textbf{$\lambda=6$} \\
\hline
\multirow{1}{*}{FLEX-TS1}	&	3	&	1	&	2	&	4	&	0	&	0	&	3	&	2	&	1	&	4	&	0	&	0	\\	
\multirow{1}{*}{FLEX-TS2}	&	5	&	3	&	2	&	1	&	4	&	6	&	5	&	4	&	1	&	2	&	3	&	6	\\	
\multirow{1}{*}{FLEX-TS3}	&	5	&	3	&	1	&	2	&	4	&	6	&	5	&	3	&	1	&	2	&	4	&	6	\\	
\multirow{1}{*}{FLEX-TS4}	&	5	&	3	&	1	&	2	&	4	&	6	&	5	&	3	&	1	&	2	&	4	&	6	\\	
\multirow{1}{*}{GREP-TS1}	&	1	&	3	&	3	&	3	&	2	&	0	&	1	&	3	&	3	&	3	&	2	&	0	\\	
\multirow{1}{*}{GREP-TS2}	&	2	&	1	&	4	&	4	&	3	&	3	&	2	&	1	&	4	&	4	&	3	&	3	\\
\hline
\end{tabular}
}
\end{table}

\begin{table}[!t]
\centering
\scriptsize

\caption{Rankings of $\mathcal{R}$ values for Completeness(q=0.50) under RT and ART}
\label{TABLE:rankcomp50nonart}
\setlength{\tabcolsep}{0.8mm}{
\begin{tabular}{c|cccccc|cccccc}  \hline
\multirow{2}*{\textbf{Subject}} & \multicolumn{6}{c|}{\textbf{RT}} & \multicolumn{6}{c}{\textbf{ART}} \\
\cline{2-13}
 & \textbf{$\lambda=1$} & \textbf{$\lambda=2$}   & \textbf{$\lambda=3$}  & \textbf{$\lambda=4$}   & \textbf{$\lambda=5$} & \textbf{$\lambda=6$}    & \textbf{$\lambda=1$} & \textbf{$\lambda=2$}   & \textbf{$\lambda=3$}  & \textbf{$\lambda=4$}   & \textbf{$\lambda=5$} & \textbf{$\lambda=6$} \\
\hline
\multirow{1}{*}{FLEX-TS1}	&	3	&	2	&	1	&	4	&	0	&	0	&	3	&	2	&	1	&	4	&	0	&	0	\\	
\multirow{1}{*}{FLEX-TS2}	&	5	&	3	&	2	&	1	&	4	&	6	&	5	&	4	&	2	&	1	&	3	&	6	\\	
\multirow{1}{*}{FLEX-TS3}	&	5	&	3	&	1	&	2	&	4	&	6	&	5	&	3	&	2	&	1	&	4	&	6	\\	
\multirow{1}{*}{FLEX-TS4}	&	5	&	3	&	1	&	2	&	4	&	6	&	5	&	4	&	2	&	1	&	3	&	6	\\	
\multirow{1}{*}{GREP-TS1}	&	1	&	3	&	3	&	3	&	2	&	0	&	1	&	3	&	3	&	3	&	2	&	0	\\	
\multirow{1}{*}{GREP-TS2}	&	2	&	1	&	4	&	4	&	3	&	3	&	2	&	1	&	4	&	4	&	3	&	3	\\
\hline
\end{tabular}
}
\end{table}

\begin{table}[!t]
\centering
\scriptsize

\caption{Rankings of $R^2$ values for Completeness(q=0.75) under RT and ART}
\label{TABLE:rankcomp75art}
\setlength{\tabcolsep}{0.8mm}{
\begin{tabular}{c|cccccc|cccccc}  \hline
\multirow{2}*{\textbf{Subject}} & \multicolumn{6}{c|}{\textbf{RT}} & \multicolumn{6}{c}{\textbf{ART}} \\
\cline{2-13}
 & \textbf{$\lambda=1$} & \textbf{$\lambda=2$}   & \textbf{$\lambda=3$}  & \textbf{$\lambda=4$}   & \textbf{$\lambda=5$} & \textbf{$\lambda=6$}    & \textbf{$\lambda=1$} & \textbf{$\lambda=2$}   & \textbf{$\lambda=3$}  & \textbf{$\lambda=4$}   & \textbf{$\lambda=5$} & \textbf{$\lambda=6$} \\
\hline
\multirow{1}{*}{FLEX-TS1}	&	2	&	1	&	3	&	4	&	0	&	0	&	2	&	1	&	3	&	4	&	0	&	0	\\	
\multirow{1}{*}{FLEX-TS2}	&	2	&	1	&	1	&	3	&	4	&	4	&	4	&	2	&	1	&	3	&	5	&	5	\\	
\multirow{1}{*}{FLEX-TS3}	&	3	&	1	&	2	&	4	&	5	&	6	&	3	&	2	&	1	&	4	&	5	&	6	\\	
\multirow{1}{*}{FLEX-TS4}	&	3	&	1	&	2	&	4	&	5	&	5	&	4	&	2	&	1	&	3	&	5	&	6	\\	
\multirow{1}{*}{GREP-TS1}	&	1	&	3	&	3	&	3	&	2	&	0	&	1	&	3	&	3	&	3	&	2	&	0	\\	
\multirow{1}{*}{GREP-TS2}	&	1	&	2	&	4	&	4	&	3	&	3	&	1	&	2	&	4	&	4	&	3	&	3	\\
\hline
\end{tabular}
}
\end{table}

\begin{table}[!t]
\centering
\scriptsize

\caption{Rankings of $\mathcal{R}$ values for Completeness(q=0.75) under RT and ART}
\label{TABLE:rankcomp75nonart}
\setlength{\tabcolsep}{0.8mm}{
\begin{tabular}{c|cccccc|cccccc}  \hline
\multirow{2}*{\textbf{Subject}} & \multicolumn{6}{c|}{\textbf{RT}} & \multicolumn{6}{c}{\textbf{ART}} \\
\cline{2-13}
 & \textbf{$\lambda=1$} & \textbf{$\lambda=2$}   & \textbf{$\lambda=3$}  & \textbf{$\lambda=4$}   & \textbf{$\lambda=5$} & \textbf{$\lambda=6$}    & \textbf{$\lambda=1$} & \textbf{$\lambda=2$}   & \textbf{$\lambda=3$}  & \textbf{$\lambda=4$}   & \textbf{$\lambda=5$} & \textbf{$\lambda=6$} \\
\hline
\multirow{1}{*}{FLEX-TS1}	&	2	&	1	&	3	&	4	&	0	&	0	&	2	&	1	&	3	&	4	&	0	&	0	\\	
\multirow{1}{*}{FLEX-TS2}	&	3	&	2	&	1	&	4	&	5	&	5	&	4	&	2	&	1	&	3	&	5	&	5	\\	
\multirow{1}{*}{FLEX-TS3}	&	3	&	1	&	2	&	4	&	5	&	6	&	3	&	2	&	1	&	4	&	5	&	6	\\	
\multirow{1}{*}{FLEX-TS4}	&	3	&	1	&	2	&	4	&	5	&	5	&	4	&	2	&	1	&	3	&	5	&	6	\\	
\multirow{1}{*}{GREP-TS1}	&	1	&	3	&	3	&	3	&	2	&	0	&	1	&	3	&	3	&	3	&	2	&	2	\\	
\multirow{1}{*}{GREP-TS2}	&	2	&	1	&	4	&	4	&	3	&	3	&	1	&	2	&	4	&	4	&	3	&	3	\\
\hline
\end{tabular}
}
\end{table}

\begin{table}[!t]
\centering
\scriptsize

\caption{Rankings of $R^2$ values for ICBM metrics under RT and ART(best)}
\label{TABLE:rankrt}
\setlength{\tabcolsep}{0.8mm}{
\begin{tabular}{c|ccccc|ccccc}  \hline
\multirow{2}*{\textbf{Subject}} & \multicolumn{5}{c|}{\textbf{RT}} & \multicolumn{5}{c}{\textbf{ART}} \\
\cline{2-11}
 & \textit{VCC} & \textit{VID}   & \textit{WMVCC}  & \textit{SVCC}   & \textit{Completeness} & \textit{VCC} & \textit{VID}   & \textit{WMVCC}  & \textit{SVCC}   & \textit{Completeness}  \\
\hline
\multirow{1}{*}{FLEX-TS1}	&	2	&	3	&	1	&	5	&	4	&	1	&	3	&	2	&	5	&	4	\\	
\multirow{1}{*}{FLEX-TS2}	&	2	&	3	&	1	&	4	&	5	&	3	&	1	&	2	&	4	&	5	\\	
\multirow{1}{*}{FLEX-TS3}	&	1	&	4	&	2	&	5	&	3	&	3	&	1	&	2	&	4	&	5	\\	
\multirow{1}{*}{FLEX-TS4}	&	1	&	4	&	2	&	5	&	3	&	1	&	3	&	2	&	4	&	5	\\	
\multirow{1}{*}{GREP-TS1}	&	2	&	3	&	1	&	4	&	4	&	2	&	3	&	1	&	4	&	4	\\	
\multirow{1}{*}{GREP-TS2}	&	3	&	1	&	2	&	4	&	5	&	3	&	2	&	1	&	4	&	5	\\
\hline
\end{tabular}
}
\end{table}

\begin{table}[!t]
\centering
\scriptsize

\caption{Rankings of $\mathcal{R}$ values for ICBM metrics under RT and ART(best)}
\label{TABLE:rankrtnon}
\setlength{\tabcolsep}{0.8mm}{
\begin{tabular}{c|ccccc|ccccc}  \hline
\multirow{2}*{\textbf{Subject}} & \multicolumn{5}{c|}{\textbf{RT}} & \multicolumn{5}{c}{\textbf{ART}} \\
\cline{2-11}
 & \textit{VCC} & \textit{VID}   & \textit{WMVCC}  & \textit{SVCC}   & \textit{Completeness} & \textit{VCC} & \textit{VID}   & \textit{WMVCC}  & \textit{SVCC}   & \textit{Completeness}  \\
\hline
\multirow{1}{*}{FLEX-TS1}	&	1	&	2	&	4	&	5	&	3	&	1	&	2	&	5	&	4	&	3	\\	
\multirow{1}{*}{FLEX-TS2}	&	1	&	2	&	3	&	4	&	5	&	2	&	1	&	5	&	3	&	4	\\	
\multirow{1}{*}{FLEX-TS3}	&	1	&	3	&	4	&	5	&	2	&	2	&	1	&	3	&	4	&	5	\\	
\multirow{1}{*}{FLEX-TS4}	&	1	&	3	&	4	&	5	&	2	&	1	&	2	&	5	&	4	&	3	\\	
\multirow{1}{*}{GREP-TS1}	&	1	&	3	&	2	&	4	&	4	&	1	&	2	&	3	&	4	&	4	\\	
\multirow{1}{*}{GREP-TS2}	&	2	&	1	&	3	&	4	&	5	&	2	&	1	&	3	&	4	&	5	\\
\hline
\end{tabular}
}
\end{table}

\begin{table}[!t]
\centering
\scriptsize

\caption{Rankings of $R$ values and $\mathcal{R}$ values for the best Completeness metric under RT and ART}
\label{TABLE:rankfcomp}
\setlength{\tabcolsep}{0.8mm}{
\begin{tabular}{c|ccc|ccc|ccc|ccc}  \hline
\multirow{2}*{\textbf{Subject}} & \multicolumn{3}{c|}{\textbf{$R^2$ values under RT}} & \multicolumn{3}{c|}{\textbf{$R^2$ values under ART}} & \multicolumn{3}{c|}{\textbf{$\mathcal{R}$ values under RT}} & \multicolumn{3}{c}{\textbf{$\mathcal{R}$ values under ART}}  \\
\cline{2-13}
 & $q=0.25$ & $q=0.50$   & $q=0.75$  &  $q=0.25$ & $q=0.50$   & $q=0.75$  & $q=0.25$ & $q=0.50$   & $q=0.75$  & $q=0.25$ & $q=0.50$   & $q=0.75$    \\
\hline
\multirow{1}{*}{FLEX-TS1}	&	3	&	2	&	1	&	3	&	1	&	2	&	3	&	2	&	1	&	2	&	1	&	3	\\
\multirow{1}{*}{FLEX-TS2}	&	2	&	3	&	1	&	1	&	3	&	2	&	2	&	3	&	1	&	2	&	3	&	1	\\
\multirow{1}{*}{FLEX-TS3}	&	3	&	1	&	2	&	3	&	1	&	2	&	2	&	1	&	3	&	3	&	2	&	1	\\
\multirow{1}{*}{FLEX-TS4}	&	2	&	1	&	3	&	1	&	2	&	3	&	1	&	2	&	3	&	3	&	2	&	1	\\
\multirow{1}{*}{GREP-TS1}	&	1	&	1	&	1	&	1	&	1	&	1	&	1	&	1	&	1	&	1	&	1	&	1	\\
\multirow{1}{*}{GREP-TS2}	&	2	&	2	&	1	&	2	&	2	&	1	&	2	&	2	&	1	&	1	&	1	&	1	\\
\hline
\end{tabular}
}
\end{table}

\begin{table}[t!]
\centering
\scriptsize

\caption{Ranking scores of ICBM metrics (best)}
\label{TABLE:rank2}
\setlength{\tabcolsep}{2mm}{
\begin{tabular}{lcccccc}  \hline

\textbf{Metric} & \textit{VCC} & \textit{VID}   & \textit{WMVCC}  & \textit{SVCC}   & \textit{Completeness}  \\
\hline
$R^2$ values under RT 
&	2.09 	&	3.19 	&	1.44 	&	4.00 	&	3.56 	
 \\ 

$R^2$ values under ART &	2.22 	&	2.81 	&	1.72 	&	3.53 	&	3.66 		
\\ 

$\mathcal{R}$ values under RT
&	1.53 	&	2.56 	&	3.97 	&	3.03 	&	3.00 	
 \\ 

$\mathcal{R}$ values under ART &	1.75 	&	2.19 	&	4.19 	&	3.00 	&	3.38
 \\ 

\hline \textit{Average} &	1.90 	&	2.69 	&	2.83 	&	3.39 	&	3.40 

 \\
\hline
\end{tabular}
}
\end{table}

\begin{table*}[!t]
\centering
\scriptsize

\caption{Rankings of $R^2$ values for Novelty under RT and ART}
\label{TABLE:ranknovelrt}
\setlength{\tabcolsep}{0.8mm}{
\begin{tabular}{c|cccccc|cccccc}  \hline
\multirow{2}*{\textbf{Subject}} & \multicolumn{6}{c|}{\textbf{RT}} & \multicolumn{6}{c}{\textbf{ART}} \\
\cline{2-13}
& \textbf{$\varepsilon=2$} & \textbf{$\varepsilon=5$}   & \textbf{$\varepsilon=15$}  & \textbf{$\varepsilon=|T|/4$}   & \textbf{$\varepsilon=|T|/2$}   & \textbf{$\varepsilon=3|T|/4$} & \textbf{$\varepsilon=2$} & \textbf{$\varepsilon=5$}   & \textbf{$\varepsilon=15$}  & \textbf{$\varepsilon=|T|/4$}   & \textbf{$\varepsilon=|T|/2$}   & \textbf{$\varepsilon=3|T|/4$} 
   \\
\hline
\multirow{1}{*}{FLEX-TS1}	&	1	&	2	&	6	&	4	&	3	&	5	&	1	&	1	&	4	&	5	&	3	&	2	\\	
\multirow{1}{*}{FLEX-TS2}	&	2	&	1	&	3	&	5	&	4	&	6	&	3	&	1	&	2	&	6	&	5	&	4	\\	
\multirow{1}{*}{FLEX-TS3}	&	2	&	1	&	3	&	6	&	4	&	5	&	2	&	1	&	3	&	5	&	4	&	6	\\	
\multirow{1}{*}{FLEX-TS4}	&	2	&	1	&	3	&	5	&	4	&	6	&	3	&	1	&	2	&	4	&	5	&	6	\\	
\multirow{1}{*}{GREP-TS1}	&	1	&	4	&	4	&	5	&	2	&	3	&	1	&	3	&	3	&	5	&	4	&	2	\\	
\multirow{1}{*}{GREP-TS2}	&	2	&	1	&	3	&	6	&	5	&	4	&	1	&	2	&	6	&	5	&	3	&	4	\\
\hline
\end{tabular}
}
\end{table*}

\begin{table}[!t]
\centering
\scriptsize

\caption{Rankings of $\mathcal{R}$ values for Novelty under RT and ART}
\label{TABLE:ranknovelrtnon}
\setlength{\tabcolsep}{0.8mm}{
\begin{tabular}{c|cccccc|cccccc}  \hline
\multirow{2}*{\textbf{Subject}} & \multicolumn{6}{c|}{\textbf{RT}} & \multicolumn{6}{c}{\textbf{ART}} \\
\cline{2-13}
& \textbf{$\varepsilon=2$} & \textbf{$\varepsilon=5$}   & \textbf{$\varepsilon=15$}  & \textbf{$\varepsilon=|T|/4$}   & \textbf{$\varepsilon=|T|/2$}   & \textbf{$\varepsilon=3|T|/4$} & \textbf{$\varepsilon=2$} & \textbf{$\varepsilon=5$}   & \textbf{$\varepsilon=15$}  & \textbf{$\varepsilon=|T|/4$}   & \textbf{$\varepsilon=|T|/2$}   & \textbf{$\varepsilon=3|T|/4$} 
   \\
\hline
\multirow{1}{*}{FLEX-TS1}	&	1	&	2	&	5	&	6	&	3	&	4	&	1	&	2	&	5	&	6	&	4	&	3	\\	
\multirow{1}{*}{FLEX-TS2}	&	2	&	1	&	3	&	6	&	5	&	4	&	2	&	1	&	3	&	6	&	4	&	5	\\	
\multirow{1}{*}{FLEX-TS3}	&	2	&	1	&	3	&	6	&	4	&	5	&	2	&	1	&	6	&	5	&	3	&	4	\\	
\multirow{1}{*}{FLEX-TS4}	&	2	&	1	&	3	&	6	&	4	&	5	&	2	&	1	&	3	&	4	&	5	&	6	\\	
\multirow{1}{*}{GREP-TS1}	&	1	&	4	&	4	&	5	&	3	&	2	&	1	&	3	&	3	&	5	&	4	&	2	\\	
\multirow{1}{*}{GREP-TS2}	&	2	&	1	&	3	&	6	&	5	&	4	&	1	&	1	&	5	&	4	&	2	&	3	\\
\hline
\end{tabular}
}
\end{table}

\begin{table}[!t]
\centering
\scriptsize

\caption{Ranking scores of different assignments for Novelty}
\label{TABLE:ranknovel}
\setlength{\tabcolsep}{1.6mm}{
\begin{tabular}{lcccccc}  \hline
\textbf{Parameter} & \textbf{$\varepsilon=2$} & \textbf{$\varepsilon=5$}   & \textbf{$\varepsilon=15$}  & \textbf{$\varepsilon=|T|/4$}   & \textbf{$\varepsilon=|T|/2$}   & \textbf{$\varepsilon=3|T|/4$}     \\
\hline
$R^2$ values under RT &	1.56 	&	1.69 	&	2.97 	&	5.25 	&	4.03 	&	4.97 \\	

$R^2$ values under ART &	2.16 	&	2.19 	&	3.53 	&	4.90 	&	3.47 	&	3.84 \\ 
$\mathcal{R}$ values under RT &	1.72 	&	1.59 	&	3.19 	&	5.65 	&	3.97 	&	4.38 \\	

$\mathcal{R}$ values under ART &	2.03 	&	2.13 	&	3.84 	&	5.19 	&	3.25 	&	3.72  
\\ 
\hline
\textit{Average} & 1.87 	&	1.90 	&	3.38 	&	5.25 	&	3.68 	&	4.23 
\\
\hline
\end{tabular}
}
\end{table}

\begin{table}[!t]
\centering
\scriptsize

\caption{Rankings of $R^2$ values for DBM metrics under RT and ART(best)}
\label{TABLE:rankrtdbm}
\setlength{\tabcolsep}{0.8mm}{
\begin{tabular}{c|cccccc|cccccc}  \hline
\multirow{2}*{\textbf{Subject}} & \multicolumn{6}{c|}{\textbf{RT}} & \multicolumn{6}{c}{\textbf{ART}} \\
\cline{2-13}
& \textit{Disc.} & \textit{Disp.}   & \textit{Dive.}  & \textit{Simi.}   & \textit{Nove.}   & \textit{Divg.}  & \textit{Disc.} & \textit{Disp.}   & \textit{Dive.}  & \textit{Simi.}   & \textit{Nove.}   & \textit{Divg.} \\
\hline
\multirow{1}{*}{FLEX-TS1}	&	1	&	4	&	3	&	5	&	2	&	2	&	2	&	4	&	3	&	5	&	1	&	6	\\	
\multirow{1}{*}{FLEX-TS2}	&	1	&	4	&	5	&	6	&	2	&	3	&	1	&	3	&	4	&	5	&	2	&	6	\\	
\multirow{1}{*}{FLEX-TS3}	&	1	&	4	&	5	&	6	&	2	&	3	&	2	&	4	&	3	&	5	&	1	&	6	\\	
\multirow{1}{*}{FLEX-TS4}	&	1	&	5	&	4	&	6	&	2	&	3	&	1	&	4	&	3	&	5	&	2	&	6	\\	
\multirow{1}{*}{GREP-TS1}	&	1	&	5	&	3	&	6	&	4	&	2	&	1	&	6	&	5	&	4	&	2	&	3	\\	
\multirow{1}{*}{GREP-TS2}	&	1	&	4	&	5	&	6	&	2	&	3	&	2	&	5	&	3	&	4	&	1	&	6	\\
\hline
\end{tabular}
}
\end{table}

\begin{table}[!t]
\centering
\scriptsize

\caption{Rankings of $\mathcal{R}$ values for DBM metrics under RT and ART(best)}
\label{TABLE:rankrtdbmnon}
\setlength{\tabcolsep}{0.8mm}{
%
}
\end{table}

\begin{table}[!t]
\centering
\scriptsize

\caption{Rankings of $R^2$ values and $\mathcal{R}$ values for VCC ($\lambda=1$) and Discrepancy under RT and ART}
\label{TABLE:rankfc1}
\setlength{\tabcolsep}{0.8mm}{
%
}
\end{table}

\begin{table}[!t]
\centering
\scriptsize

\caption{Rankings of $R^2$ values and $\mathcal{R}$ values for VCC ($\lambda=2$) and Discrepancy under RT and ART}
\label{TABLE:rankfc2}
\setlength{\tabcolsep}{0.8mm}{
%
}
\end{table}

\begin{table}[!t]
\centering
\scriptsize

\caption{Rankings of $R^2$ values and $\mathcal{R}$ values for VCC ($\lambda=3$) and Discrepancy under RT and ART}
\label{TABLE:rankf2}
\setlength{\tabcolsep}{0.8mm}{
%
}
\end{table}

\begin{table}[!t]
\centering
\scriptsize

\caption{Rankings of $R^2$ values and $\mathcal{R}$ values for VCC ($\lambda=4$) and Discrepancy under RT and ART}
\label{TABLE:rankfc4}
\setlength{\tabcolsep}{0.8mm}{
%
}
\end{table}

\FloatBarrier

\section{Computation Time of ICBM Metrics}
\label{appdendix:C}
{
This appendix reports the computational time of the ICBM metrics across different strengths. The notation "NA" indicates that the subject was unevaluated due to strength constraints, while "TO" represents a timeout where the execution exceeded the predefined time threshold.
}
\begin{table*}[h]
\centering
\scriptsize

\caption{Computation time of ICBM metrics for $\lambda=1$}
\label{TABLE:time1}
\setlength{\tabcolsep}{2mm}{
%
}
\end{table*}

\begin{table*}[t!]
\centering
\scriptsize

\caption{Computation time of ICBM metrics for $\lambda=2$}
\label{TABLE:time2}
\setlength{\tabcolsep}{2mm}{
%
}
\end{table*}

\begin{table*}[t!]
\centering
\scriptsize

\caption{Computation time of ICBM metrics for $\lambda=3$}
\label{TABLE:time3}
\setlength{\tabcolsep}{2mm}{
%
}
\end{table*}

\begin{table*}[t!]
\centering
\scriptsize

\caption{Computation time of ICBM metrics for $\lambda=4$}
\label{TABLE:time4}
\setlength{\tabcolsep}{2mm}{
%
}
\end{table*}


\vfill

\end{document}